\RequirePackage{amsmath}
\documentclass{article}
\usepackage[a4paper, total={6in, 8in}]{geometry}
\usepackage{booktabs} 
\usepackage{fullpage}
\usepackage{graphicx}
\usepackage{pgfplots}
\usepackage[utf8]{inputenc}
\usepackage{tikz}
\usepackage{dblfloatfix}
\usepackage{float}
\usepackage{textgreek}
\usepackage{multicol}
\usepackage{algorithm}
\usepackage{relsize}
\usepackage{mathtools, bm, amsfonts, amssymb, lmodern}

\DeclareMathOperator*{\argmin}{argmin}
\usepackage{hyperref}
\usepackage{subfig}
\usepackage{verbatim}
\usepackage{algorithmic}
\usepackage[inline]{enumitem} 
\usepackage{setspace}

\usepackage{natbib}

\definecolor{blue}{HTML}{1F77B4}
\definecolor{orange}{HTML}{FF7F0E}
\definecolor{green}{HTML}{2CA02C}

\pgfplotsset{compat=1.14}

\newcommand\MEETtitle[1]{\Large \bf \hskip2.25pc \parbox{.8\textwidth}{ \noindent%
   \large \bf \begin{center} #1 \end{center}\rm } \vskip.1in \rm\normalsize }

\newcommand\MEETauthor[1]{\hskip2.25pc \parbox{.8\textwidth}{ \noindent%
   \normalsize \bf \begin{center} #1 \end{center}\rm } \vskip-1pc }

\let\title\MEETtitle
\let\author\MEETauthor

\let\address\MEETaddress
\let\email\MEETemail

\begin{document}	

\title{Image Reconstruction by Splitting Expectation Propagation Techniques from Iterative Inversion}

	
	\author{Robert G. Aykroyd}
	\address{Department of Statistics, School of Mathematics,  \\ University of Leeds, UK}
	\email{R.G.Aykroyd@leeds.ac.uk}
		\author{Kehinde Olobatuyi}
	\address{Department of Statistics and Mathematical Finance, \\ University of Milan-Bicocca, Italy}
	\email{k.olobatuyi@campus.unimib.it}
	
\date{}
	%
	%

\begin{abstract}
\noindent
Reconstructing images from downsampled and noisy measurements, such as MRI and low dose Computed Tomography (CT), is a mathematically ill-posed inverse problem. We propose an easy-to-use reconstruction method based on Expectation Propagation (EP) techniques. We incorporate the Monte Carlo (MC) method, Markov Chain Monte Carlo (MCMC), and Alternating Direction Method of Multiplier (ADMM) algorithm into EP method to address the intractability issue encountered in EP. We demonstrate the approach on complex Bayesian models for image reconstruction. Our technique is applied to images from Gamma-camera scans. We compare EPMC, EP-MCMC, EP-ADMM methods with MCMC only. The metrics are the better image reconstruction, speed, and parameters estimation. Experiments with Gamma-camera imaging in real and simulated data show that our proposed method is convincingly less computationally expensive than MCMC and produces relatively a better image reconstruction.

\medskip

\end{abstract}	


\section{Introduction}

The Bayesian approach to modeling data provides a flexible mathematical frameworks for incorporating uncertainty of unknown quantities within complex statistical models. The Bayesian posterior distribution encodes the probabilistic uncertainty in the model parameters and can be used for predictions of new unobserved data. In general, the posterior distribution is intractable and it is therefore necessary to approximate it. Deterministic approximations, such as Laplace approximation (see Bishpop 2006), variational Bayes (Blei, Kucukelbir, and McAuliffe 2017), and expectation-propagation (Minka 2001) aim to approximate the posterior with a simpler tractable distribution (e.g., a normal distribution). The deterministic approaches are often fit using fast optimization techniques and trade-off exact posterior inference for computational efficiency.
\par Markov chain Monte Carlo (MCMC) algorithms (Brooks et al. 2011) approximate the posterior distributions with a discrete set of samples generated from a Markov chain whose invariant distribution is the posterior distribution. Simple MCMC algorithms, such as random-walk Metropolis (Metropolis et al. 1953), are easy to apply and only require that the unnormalized density of the posterior can be evaluated point-wise. Under mild conditions, the samples generated from the Markov chain converge to the posterior distribution (Roberts and Rosenthal 2004) and for many popular MCMC algorithms, rates of convergence based on geometric ergodicity have been established (Meyn and Tweedie 1994; Roberts and Rosenthal 1997)
\par While MCMC algorithms have the advantage of providing asymptotic exact posterior samples, this comes at the cost of beign computationally slow to apply in practice. This issue is further exacerbated by the demand to store in the memory and analyze large-scale datasets and to fit increasingly sophisticated and complex models to the high-dimensional data. For example, brain imaging (Anderson et al. 2018), natural language processing (Yogatama et al. 2014), popular genetics (Raj, Stephens, and Pritchard 2014), commonly use a Bayesian approach to data analysis, but the continual growth in the size of the dataset in these fields prevents the use of traditional MCMC methods. Computational challenges such as these have led to recent research interest in scalable Monte Carlo algorithms. Broadly speaking, the new developments in Monte Carlo algorithm achieve computational efficiency by either parallelizing the MCMC scheme, or by subsampling the data. \par If the data can be distributed across multiple computer cores then the computational challenge of inference can be shared, with MCMC algorithm run on each core to draw samples from a partial posterior that is conditional on the subset of the full data. The major challenge is to combine these posterior samples from each computer to generate an approximation to the full posterior distribution. It is then possible to merge the sample that are exact if the partial posteriors are Gaussian (Scott et al. 2016) with the update rules that just depend on the mean and variance for each partial posterior. However, it is difficult to quantify the level of approximation of the rules due to non-Gaussianity of the partial posterior. Alternative merging procedures that are more robust to non-Gaussianity, have been proposed (Neiswanger, Wang, and Xing 2014; Rabionovich, Angelino, and Jordan 2015; Minsker et al. 2017; Nemeth and Sherlock 2018; Srivastava, Li, and Dunson 2018) but it is difficult to measure the level of approximation accuracy. Bespoke methods are also needed when interest in the joint posterior of the parameters relates to subsets of the data, or individual data points (Zuanetti et al. 2019).

We briefly review the standard EP algorithm as it forms the basis for the novel proposed methods presented in this paper.

\section{Expectation Propagation}\label{sec:2} 

Suppose a dataset $\mathcal{D}$ is observed which consists of $n$ independent and identically distributed components such that $\mathcal{D} = \{\mathbf{x}_i\}_{i=1}^n$.
Further, each component is from the parametric probabilistic model 
$p(\mathbf{x}|\bm{\theta})$ parameterized by an unknown 
$p$-dimensional vector $\bm{\theta}$ whose prior distribution is 
$p_0(\bm{\theta})$. 
Given the data, the posterior distribution is formulated as follows;
\begin{equation}
	p(\bm{\theta}|\mathcal{D}) 
	=  \frac{ p_0(\bm{\theta})}{p(\mathcal{D})} \; \mathlarger \, \prod_{i=1}^n p(\mathbf{x}_i|\bm{\theta}). 
	\label{equ:1}
\end{equation}
Exact Bayesian inference then involves the computation of typically intractable posterior distribution over the parameters. 
Note that in what follows, to simplify notation, the posterior distribution is denoted $p(\bm{\theta})$ rather than the more usual $p(\bm{\theta}|\mathcal{D})$.

The goal of EP is to capture the contribution of each data point, through the likelihood term, to the posterior by refining approximate terms, that is;
\begin{equation}
p(\bm{\theta}) 
	\propto  p_0(\bm{\theta}) \mathlarger \prod_{i=1}^n t_i(\bm{\theta}) 
\, \approx \, q(\bm{\theta} ) \propto p_0(\bm{\theta}) \mathlarger \prod_{i=1}^n \tilde{t}_i(\bm{\theta}),
	\label{equ:1b}
\end{equation}
where  $t_i(\bm{\theta}) \propto p(\mathbf{x}_i|\bm{\theta})$ is the exact term,
$\tilde{t}_i(\bm{\theta})$ is the corresponding approximating term for $t_i(\bm{\theta})$, and $q(\bm{\theta})$ is the approximate posterior distribution that will be refined by EP. 
One approach is to search for the approximate posterior distribution which minimizes the Kullback-Leibler (KL) divergence between the exact posterior distribution and the distribution formed by replacing, in turn, one of the likelihood components by the approximate term $\tilde{t}_i(\bm{\theta})$, that is $\text{KL}[p(\bm{\theta}|\mathcal{D}) ~||~ p(\bm{\theta}|\mathcal{D})\tilde{t}_i(\bm{\theta})/p(\mathbf{x}_i|\bm{\theta})]$;
%
%
%
Even with this approach, however, the update remains intractable in high dimensional space. 
Instead, EP approximates this procedure by removing the observations one-by-one from the posterior distribution,
that is 
$p_{-i}(\bm{\theta}) 
\propto p(\bm{\theta}|\mathcal{D})/t_i(\bm{\theta})$, 
which is called the exact leave-one-out posterior,
and the corresponding  approximate leave-one-out posterior, that is
$q_{-i}(\bm{\theta}) \propto q(\bm{\theta})/\tilde{t}_i(\bm{\theta})$, which is called the cavity distribution. 
Since this couples the updates for the approximating factors, the updates must now be iterated until convergence. 

In a more general perspective, EP first selects a component to update and then removes it from the approximate posterior distribution to produce the cavity distribution, and 
the corresponding exact likelihood term is combined with the cavity distribution to produce the tilted posterior distribution, 
$\tilde{p}_i(\bm{\theta}) \propto q_{-i}(\bm{\theta})t_i(\bm{\theta})$. 
Then, EP updates the approximate posterior distribution by minimizing the tilted distribution, 
that is;
%
%
\begin{equation}
\tilde{t}_i(\bm{\theta})= \argmin_{q(\bm{\theta})} \text{KL}[\tilde{p}(\bm{\theta}) \; || \; q(\bm{\theta})], 
\quad i=1,\ldots, n.
\end{equation}
This establishes that the measure of contribution the exact likelihood term makes on both exact posterior distribution and the tilted posterior distribution is the same. 
The KL minimization often reduces to moment matching when the approximate distribution is in the exponential family. 
Finally, we update each approximate likelihood term by dividing the new approximate posterior distribution by the old approximate posterior distribution. 
\par In practice, EP often performs well due to this local updating but it can become intractable again in high dimensional problems due to the need to compute normalizing factors caused by marginalizing the parameters.

Many attempts have been made to solve instability problem in Expectation Propagation (EP) technique [\citep{akietal2019}; \citep{gelman2014way}]. 
Several research have focused on the problem of the energy function in expectation propagation.
EP may have stability issues even with one-dimensional tilted distributions and moment computations are more challenging if the tilted distribution is multimodal or has long tails \citep{jylanki2011ep}. 
\cite{matthias2014propagation} studies expectation propagation method and demonstrates its significant potentials on the nonlinear inverse problem. They propose the coupling of the EP with an iterative linearization strategy. A direct integral was used for computing the marginal likelihood. However, this is only tractable in a low dimensional space as the work assumed.
\cite{jose2015propagation} works on a hierarchical problem as a layered Bayesian Neural Network. However, their approach can lead to underestimation of variance of the approximate posterior. \cite{gelman2014way} revisit expectation propagation as a prototype for scalable algorithm that partition big datasets into many parts and analyze each part in parallel to perform inference of shared parameters. \cite{minka2004ex} proposes an extension of EP called a fractional EP to improve the robustness of EP algorithm when the approximation family is not enough \citep{minka2005ep} or when the propagation of information is difficult due to vague prior information \citep{seeger2008pro}. \cite{yin2015se} presents an extension of expectation propagation called a Stochastic Expectation Propagation that maintains a global posterior approximation like variational inference but updates the posterior in a local way. The work used the expectation propagation strategy but differs only at the updating stage. \cite{grave2011vi} proposes a Monte Carlo approximation to compute the lower bound of the variational inference optimized by stochastic gradient descent (SDG). \cite{john2011vi} focuses on the problem of intractable lower bound on the marginal likelihood in variational inference at the updates of the parameters. They propose an alternative approach based on stochastic optimization that allows for direct optimization of the variational lower bound.

\par Image reconstruction problems are common to medical imaging area such as fast MRI and low dose CT are generally mathematically ill-posed inverse problems. 
Often times, linear imaging system are considered with a forward operator $G$, e.g. a Fourier transform for MRI and X-ray transform for CT. The measurement $y$ is given as $y = Gx$, where $x$ is the underlying image in the noise-free state. The linear operator $G$ is ill-posed for most applications; therefore, some statistical priors are necessary to make these problems invertible. Consequently, intractability becomes inevitable due to high-dimensional data. Moreover, when the number of hidden variables to be estimated grows bigger and results in multiple integration, intractability poses a problem in Bayesian inference. In this paper, we generalize EP to the mixture of Gaussian and non-Gaussian distributions. We propose a new hybrid algorithms called \textit{Splitting Expectation Propagation} (SEP). We incorporate the Monte Carlo approximation, Markov Chain Monte Carlo method (MCMC), and Alternating Direction Method of Multiplier (ADMM) into the EP. 
SEP has the ability to handle non-Gaussian factors in hierarchical Bayesian models in high dimensional space.

The paper is organized as follows; 
In Sect.\ \eqref{sec:2}, we give a detailed explanation of the standard EP technique whereas
the proposed techniques is fully derived in Sect.\ \eqref{sec:3}. 
Furthermore, to illustrate the methods, in Sect.\ \eqref{sec:4}, we present an experiment on synthetic data sets and finally, in Sect.\ \eqref{sec:5}, we draw some future direction and the main conclusions of this work.

\section{Splitting Expectation Propagation}\label{sec:3}
In this section, we introduce a new hybrid algorithm motivated by EP, called Splitting Expectation Propagation (SEP). SEP incorporates Monte Carlo (MC), Markov Chain Monte Carlo (MCMC), and Alternating Direction method of multiplier (ADMM) at the EP update. The algorithm can be interpreted as a version of EP that is capable of handling the marginal likelihood with mixture of exponential family in a stochastic and deterministic views. 

\subsection{Model Specification}
We describe a Bayesian model for the data based on an inverse problem. Given data $\mathcal{D} = \lbrace\mathcal{X}_n, \mathcal{Y}_n\rbrace_{n=1}^N$, made up of $D-$dimensional vector $\mathcal{X}_n \in \mathcal{R}^D$ and corresponding target variables $\mathcal{Y}_n \in \mathcal{R}^D$. In the context of inverse problems, we only have access to a noisy version $\mathcal{Y}$ of the exact data $\mathcal{X}$. We assume that $\mathcal{Y}$ is obtained as $\mathcal{Y} = G\mathcal{X} + \xi$, where the $\xi \in \mathcal{R}^{\mathrm{N} \times \mathrm{D}}$ represents the noise in the data.  
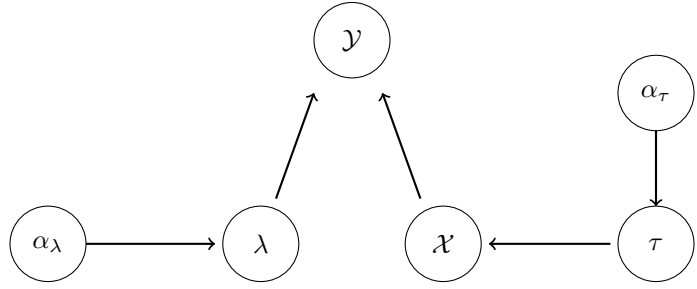
\begin{figure}[H]
	\begin{center}
		\begin{tikzpicture}
			\shadedraw [inner color = white] (0.8,0.5) circle (0.5cm) node {$\lambda$};
			\draw[thick,->] (-2,0.5) -- (0.2, 0.5);
			\filldraw [draw = black, fill = white] (-2,0.5) circle (0.5cm) node {$\alpha_\lambda$};
			\filldraw [draw = black, fill = white] (2,3.2) circle (0.5cm) node {$\mathcal{Y}$};
			\draw[thick,<-] (1.5,2.5) -- (1,1.1);
			\draw[thick,<-] (2.4,2.5) -- (2.9,1.1);
			\shadedraw [inner color = white] (3.2,0.5) circle (0.5cm) node {$\mathcal{X}$};
			\filldraw [draw = black, fill = white] (6,0.5) circle (0.5cm) node {$\tau$};
			\draw[thick,->] (5.4,0.5) -- (3.8, 0.5);
			\filldraw [draw = black, fill = white] (6,2.5) circle (0.5cm) node {$\mathbf{\alpha_\tau}$};
			\draw[thick,->] (6,2.) -- (6, 1);
		\end{tikzpicture}
	\end{center}
	\caption{Graph networks connecting priors and their hyper-priors} \label{fig:1}
\end{figure}
\bigskip
\par \noindent Figure \eqref{fig:1} shows graphical representation of the Bayesian hierarchical model which is mathematically represented as follows;
\begin{equation}
	p(\mathcal{X}, \lambda, \tau | \mathcal{Y}) = \frac{p(\mathcal{Y}|\mathcal{G}\mathcal{X}, \lambda)p(\mathcal{X}|\tau)p(\lambda)p(\tau)}{\mathlarger \int ...\mathlarger \int p(\mathcal{Y}|\mathcal{G}\mathcal{X}, \lambda)p(\mathcal{X}|\tau)p(\lambda)p(\tau) d\mathcal{X}d\lambda d\tau} \label{equ:500},
\end{equation}
Equation \eqref{equ:500} has a large size of the integration which grows with the size of the reconstructed image $\mathcal{X}$. The standalone or independent variable is directly connected to both its mean $\mathcal{X}$ and variance $\lambda$. The priors are disconnected from each other but directly connected to their respective hyper-priors.
We consider a linear equation
\begin{equation}
	\mathcal{F}(\mathcal{X}) = \mathcal{Y} \label{equ:50}
\end{equation}
where the map $\mathcal{F}:\mathcal{R}^{\mathrm{N} \times \mathrm{D}} \rightarrow \mathcal{R}^{\mathrm{N} \times \mathrm{D}}$, matrices $\mathcal{X} \in \mathcal{R}^{\mathrm{N} \times \mathrm{D}}$ and $\mathcal{Y} \in \mathcal{R}^{\mathrm{N} \times \mathrm{D}}$ refer to data formation mechanism, unknown parameter and the given data respectively.
\par \noindent Let $\mathcal{Y}$ be an $N \times D$ matrix and $\mathcal{X}$ be an $N \times D$ matrix also. The likelihood for the model and the noise variance $\lambda$, with data $\mathcal{D} = (\mathcal{X}, \mathcal{Y})$ is then 
\begin{equation}
	p(\mathcal{Y}|G\mathcal{X}, \lambda) = \prod_i\prod_j \mathcal{N}(\mathcal{Y}_{ij} \big | \mathcal{G}\mathcal{X}_{ij}, \lambda),
	\label{equ:6.43}
\end{equation}
Also, we specify a Gaussian prior distribution for each entry $\mathcal{L}\mathcal{X}$ which is used as a matrix with zero mean and $\tau^2$ as its variance. The Gaussian prior for $\mathcal{L}\mathcal{X}$ is as follows; 
\begin{equation}
	p(\mathcal{L}\mathcal{X}|\tau) = \mathlarger \prod_i \mathlarger \prod_j \mathcal{N}(\mathcal{L}\mathcal{X}_{ij} \big | 0, \tau),
	\label{equ:6.44}
\end{equation}
The prior distributions for $\lambda$ and $\tau$ are chosen to be an exponential distributions, i.e., $p(\tau) = \exp(\tau|\alpha_\tau)$ and also the noise variance to be $p(\lambda) = \exp(\lambda|\alpha_\lambda)$. The posterior distribution for the parameters $\mathcal{X}, \lambda, \tau$ can then be obtained by
\begin{equation}
	p(\mathcal{X}, \lambda, \tau|\mathcal{Y}) = \frac{p(\mathcal{Y}|\mathcal{G}\mathcal{X}, \lambda)p(\mathcal{L}\mathcal{X}|\tau)p(\lambda)p(\tau)}{p(\mathcal{Y})},
	\label{equ:6.45}
\end{equation}
where $\lambda$, $\tau^2$ are unknown variances and $\mathcal{G}$ is a known matrix, $p(\mathcal{Y})$ is the normalizing factor and it will be denoted as $Z$ throughout this work and in particular
\begin{equation}
	Z = \mathlarger \int .... \mathlarger \int p(\mathcal{Y}| \mathcal{G}\mathcal{X}, \lambda) p(\mathcal{L}\mathcal{X}|\tau)p(\lambda)p(\tau) d\mathcal{X} d\lambda d\tau,
	\label{equ:6.46}
\end{equation} 
where $\mathcal{X}$ is $\mathcal{N} \times D$ dimension and $\mathcal{Y}$ is $\mathcal{N} \times D$ dimension. Next, we choose an approximating families. Here, the approximate distributions are mixture of different distributions of the exponential family.
For instance, 
\begin{equation}
	q(\mathcal{X}, \lambda, \tau) = \bigg[\mathlarger \prod_{i=1}^m \mathlarger \prod_{j=1}^n \mathcal{N}(\mathcal{X}_{ij} \big | mx_{i,j}, vx_{i,j}) \bigg] \exp(\lambda|\alpha_\lambda) \exp(\tau|\alpha_\tau),
	\label{equ:6.47}
\end{equation}
$\mathcal{X} \sim \mathcal{N}(mx_{i,j}, vx_{i,j})$, $\lambda \sim \exp(\alpha_\lambda)$ and $\tau \sim  \exp(\alpha_\tau)$.
\subsubsection{EP-MC Technique}
Here, the approximation parameters $mx_{i,j}, vx_{i,j}, \alpha_\tau$ and $\alpha_\lambda$ are determined by applying a stochastic search on the posterior in Equation (\ref{equ:6.45}). 
Finally, we sequence through and incorporate the terms $t_i$ into the approximate posterior in Equation (\ref{equ:6.47}). At each step, we move from an old $q_{-i}(\mathcal{X},\lambda,\tau)$ to a new $q(\mathcal{X},\lambda,\tau)$. 
\bigskip
\par \noindent The update rule is given in Equation (\ref{equ:6.48}) below;
\begin{equation}
	\alpha_\tau^{new} \approx \alpha_{-ij} + v_{\alpha_{-ij}}\zeta_{\alpha_{-ij}},
	\label{equ:6.48}
\end{equation}
Where the mean $\frac{1}{K}\sum_{k=1}^{K}\tau^{(k)}$ is denoted as $\zeta_{\alpha_{-ij}}$ and variance as $v_{\alpha_{-ij}}$. 
\par \noindent The update rule for $\lambda$ is given in Equation (\ref{equ:6.491}) below;
\begin{equation}
	\alpha_\lambda^{new} \approx \alpha^{\lambda}_{-ij} + v_{\alpha^{\lambda}_{-ij}}\frac{1}{K}\mathlarger \sum_{k=1}^{K}\lambda^{(k)},
	\label{equ:6.491}
\end{equation}
We Take the derivative of the normalizing factor with respect to $\omega$ concerning $\mathcal{X}$ as follows;
\begin{equation}
	\nabla_\omega \log{Z} = \frac{1}{Z} \int_{\mathcal{X}_{i,j}, \tau} \frac{t_{i,j}(L\mathcal{X}_{i,j}, \tau)}{\tilde{t}_{i,j}(\mathcal{X}_{i,j}, \tau)}q_{i,j}(\mathcal{X}_{i,j}, \tau)\nabla_\omega \log{q_{i,j}(\mathcal{X}_{i,j}, \tau)} d \mathcal{X}_{i,j}d\tau,
	\label{equ:6.831}
\end{equation}
At this juncture, we proceed with the use of Monte Carlo Integration
\begin{equation}
	\frac{\delta_1}{\delta_0} = \sum_{n = 1}^N \frac{t_{i,j}(L\mathcal{X}_{i,j}, \tau)}{\tilde{t}_{i,j}(\mathcal{X}_{i,j}, \tau)}\nabla_\omega \log{q_{i,j}(\mathcal{X}_{i,j}, \tau)} \bigg / \sum_{n = 1}^N \frac{t_{i,j}(L\mathcal{X}_{i,j}, \tau)}{\tilde{t}_{i,j}(\mathcal{X}_{i,j}, \tau)},
	\label{equ:6.841}
\end{equation}
when taking $\mathcal{X}$ into consideration then we fix $\tau$ and vice versa. So we update the parameter by taking the gradient step
\begin{equation}
	\omega^{t+1} = \omega^t + lr * \frac{\delta_1}{\delta_0},
	\label{equ:6.851}
\end{equation}
One of the peculiarities of EP is the ability to minimize Kullback-Leibler divergence between the tilted posterior and the approximate posterior.
The full derivatives are provided in the Appendix.
\subsubsection{EP-ADMM technique}
We introduce the Alternating direction method of multiplier (ADMM) algorithm to update the approximate posterior parameters.
\begin{align}
	\begin{split}
		\text{minimize} \hspace*{0.2in}\text{KL}(\tilde{p}(x_{ij}||q(x)_{ij}))
		\label{equ:6.862}
	\end{split} \\
	&\ \text{subject to}: m_{ij} \geq a; v_{ij} \geq b \notag
\end{align}
where $a$ and $b$ are constants. Then updating according to \cite{minka2001ep} is as follows;
\begin{equation}
	m_x = \nabla_{m_{-ij}} \log Z_x + \alpha + \rho(m_{-ij} - a) 
	\label{equ:6.872}
\end{equation}
and 
\begin{equation}
	v_x = \nabla_{v_{-ij}} \log Z_x + \beta + \rho(v_{-ij} - b) 
	\label{equ:6.882}
\end{equation}
where 
\begin{equation}
	q(x) = \mathcal{N}(\mathcal{X};m_x,v_x) \label{equ:6.892}
\end{equation}
then the tilted distribution is 
\begin{equation}
	\tilde{p}(x_{ij}) = \frac{t_{ij}(x)q_{-ij}(x)}{\mathlarger \int_{\mathcal{X}_{ij}}t_{ij}(x)q_{-ij}(x) dx_{ij}}
	\label{equ:6.902}
\end{equation}
Now the normalizing factor is 
\begin{equation}
	Z_x = \mathlarger \int_{x_{ij}} \mathcal{N}(\mathcal{Y}_{ij};G\mathcal{X}_{ij},\lambda)\mathcal{N}(\mathcal{X}_{ij};m_{-ij},v_{-ij}) dx_{ij}
	\label{equ:6.912}
\end{equation}
The update rule for the parameters $m_x$ and $v_x$ of $q(\mathcal{X})$ according to \cite{minka2001ep} with a method of multipliers are
\begin{equation}
	m^{new}_x = m_{-ij} + v_{-ij} \frac{y_{ij} - gm_{-ij}}{v_{-ij}g^2 + \lambda}g + \alpha + \rho(m_i - a) 
	\label{equ:6.971}
\end{equation}
\begin{equation}
	v_x^{new} = \frac{v_{-ij}\lambda}{v_{-ij}g^2 + \lambda} + \beta + \rho(v_i - b) 
	\label{equ:6.981}
\end{equation}
\begin{equation}
	\alpha^{new} = \alpha^k + \rho(m_x^{new} - a) \hspace*{0.2in} \text{and} \hspace{0.2in} \beta^{new} = \beta^k + \rho(v_x^{new} - b)
	\label{equ:6.991}
\end{equation}
for full derivatives, Appendix provides details.
\subsubsection{EP-MCMC Technique}
The EP-MCMC method provides an approach when the model is complex or the number of parameters is large and deterministic gradient-based techniques are infeasible and most importantly when the posteriors are not Gaussian. The Kullback-Leibler divergence between the tilted posterior and the approximated posterior can be addressed using the MCMC technique. In brief, the transitions in the Markov chain are designed so that an equilibrium distribution exists and is equal to the target distribution (tilted posterior). According to EP's strategy of moment matching, this is also possible but instead we update the moments of the approximated posterior with the moment of the tilted posterior from the Markov chain. We update the approximate term and compute the new cavity distribution from the new approximated posterior and thereafter compute the new tilted posterior. With this cycle in mind, the advantage of EP-MCMC over the MCMC is that at each iteration in EP-MCMC the tilted posterior is updated with new cavity distribution thereby generating samples that are independent at each step of the algorithm. Unlike EP-MCMC, the major problem of MCMC is the dependence among samples generated at different iterations. This is due to a static target posterior. 
The EP setup for the parameters $\tau$ and $\lambda$ is approached via Markov Chain Monte Carlo method.
\begin{algorithm}[H]
	\caption[Algorithm for EP-MCMC]{The General EP-MCMC Algorithm}
	\begin{algorithmic}[a]
		\begin{spacing}{0.02}
			\STATE Set an initial value for $\tau$ and $\lambda$
			\vspace*{0.1in}
			\begin{enumerate}
				\item Initialize all of the approximating factors $\tilde{t_i}(\tau)$ and $\tilde{t_i}(\lambda)$.
				\item Compute the initial approximation $q(\tau)$ and $q(\lambda)$ from the product of the approximating factors: 
				\begin{equation}
					q(\tau) = \frac{\mathlarger \prod_i \tilde{t}_i}{\mathlarger \int \mathlarger \prod_i \tilde{t}_i \hspace*{0.05in} d\tau} \hspace*{0.2in} \text{and} \hspace*{0.2in}  q(\lambda) = \frac{\mathlarger \prod_i \tilde{t}_i}{\mathlarger \int \mathlarger \prod_i \tilde{t}_i \hspace*{0.05in}d\lambda}
					\notag
				\end{equation}
				\item Until all $\tilde{t}_i$ converge:
				\begin{enumerate}
					\item Choose a $\tilde{t}_i$ to refine
					\item Remove $\tilde{t}_i$ from the approximation $q(\tau)$ and $q(\lambda)$ 
					by division:
					\begin{equation}
						q_{-i}(\tau) \propto \frac{q(\tau)}{\tilde{t}_i(\tau)} \hspace{0.2in} \text{and} \hspace{0.2in} q_{-i}(\lambda) \propto \frac{q(\lambda)}{\tilde{t}_i(\lambda)}\notag
					\end{equation}
					\item Compute the tilted distribution $\tilde{p}_i$ from $q_{-i}$ and the exact factor $t_i$
					\begin{equation}
						\tilde{p}_i(\tau) \propto t_i(\tau) \hspace{.1cm} q_{-i}(\tau) \hspace{0.2in} \text{and} \hspace{0.2in} \tilde{p}_i(\lambda) \propto t_i(\lambda) \hspace{.1cm} q_{-i}(\lambda)\notag
					\end{equation}
					\begin{enumerate}
						\item Generate $\epsilon_\tau$ and $\epsilon_\lambda$ from a Gaussian distribution 
						\item Generate a proposed new value $\tau^* = \tau + \epsilon_\tau$ and $\lambda^* = \lambda + \epsilon_\lambda$
						\item Evaluate
						\begin{equation}
							\alpha_{\tau} = \frac{\pi(\tau^*|\mathcal{X})}{\pi(\tau|\mathcal{X})} \hspace{0.2in} \text{and} \hspace{0.2in} \alpha_{\lambda} = \frac{\pi(\lambda^*|\mathcal{Y})}{\pi(\lambda|\mathcal{Y})}\notag
						\end{equation}
					\item Generate $u_\tau$ and $u_\lambda$ from a uniform distribution $U(0,1)$
					\item If $\alpha_{\tau} > u_\tau$ and $\alpha_\lambda > u_\lambda$ then accept the proposals $\tau^*$ and $\lambda^*$ and set $\tau = \tau^*$ and $\lambda = \lambda^*$  
					\end{enumerate}
				    \item Update the approximate posterior from the inference of tilted posterior, this is similar to a moment matching in EP. $q^{new}(\tau)$ and $q^{new}(\lambda)$
					\item Compute the new approximate term:
					\begin{equation}
						\tilde{t}^{new}_i(\tau) \propto \frac{q^{new}(\tau)}{q_{-i}(\tau)} \hspace{0.2in} \text{and} \hspace{0.2in} \tilde{t}^{new}_i(\lambda) \propto \frac{q^{new}(\lambda)}{q_{-i}(\lambda)}
						\notag
					\end{equation}
				\end{enumerate}
				\item Evaluate the approximation to the model evidence:
				\begin{equation}
					p(\mathcal{\mathcal{Y}}) \approxeq \mathlarger \int \mathlarger \prod_i \tilde{t}_i(\tau)\hspace{0.05in} d\tau \hspace{0.2in} \text{and} \hspace{0.2in} p(\mathcal{\mathcal{Y}}) \approxeq \mathlarger \int \mathlarger \prod_i \tilde{t}_i(\lambda)\hspace{0.05in} d\lambda
				\end{equation}
				  
			\end{enumerate}
		\end{spacing}
	\end{algorithmic}
\end{algorithm}
\section{Implementation details}\label{sec:4}
After incorporating the factors in Equation \eqref{equ:6.902} for the first time, we constrained only mean and variance parameters of approximate posterior $q$ of $\mathcal{X}$ for the purpose of this work using Alternating direction method of multipliers. Because of the structure of the data presented in Figure \eqref{fig:6.4} which is sparse, EP often breaks down when the algorithm produces variance parameter that is negative after incorporating one likelihood factor, especially at the update step of approximate term which involves division. A similar operation was reported in \cite{minka2001ep} when negative variances arise in Gaussian approximation factors.
\subsection{Monitoring Convergence of EP-MCMC}
To monitor convergence of the embedded Metropolis Hastings in EP, we run EP-MCMC $m = 10$ replications and $n = 1000$ to compute the Brook-Gelman statistic; \cite{brookandgelman1998} which compare the between-chain variance and within-chain variance. However, EP is generally known to be deterministic which is still preserved in EP-ADMM for the prior on $\mathcal{X}$ but the priors on $\lambda$ and $\tau$ are randomly tuned. The reconstruction images presented are mean of the approximate posterior $m_{ij}$. To know if a convergence has been reached, \cite{gelmanandrubin1992a} suggests comparing $m$ inferences computed from the $m$ chains to the inferences computed by mixing together the $mn$ draws from all the sequences. 
\noindent To compute the between-chain and within-chain variances, let $s_{jt}$ where $j = 1,..., m$ and $t = 1, ..., n$ be the $t$th iterations of $s$ in sequence $j$, the between-chain variance $B/n$ and within-chain variance are given by
\begin{equation}
	B/n = \frac{1}{m-1}\mathlarger \sum_{j=1}^{m} (\bar{s}_{j.} - \bar{s}_{..})^2
	\label{equ:6.102}
\end{equation}
and 
\begin{equation}
	W = \frac{1}{m(n-1)}\mathlarger \sum_{j=1}^{m} \mathlarger\sum_{t=1}^{n}(s_{jt} - \bar{s}_{j.})^2
	\label{equ:6.103}
\end{equation}
We also computed the \textit{potential scale reduction factor} PSRF which is the variance of the pooled and within-chain inferences
\begin{equation}
	R = \frac{\hat{V}}{\sigma^2}\notag
\end{equation}
R will be estimated by $\hat{R}$ because its denominator is unknown 
\begin{equation}
	\hat{R} = \frac{\hat{V}}{W} = \frac{m+1}{m} \frac{\hat{\sigma}^2_+}{W} - \frac{n-1}{mn}
	\label{equ:6.104}
\end{equation}
where the pooled variance is given by
\begin{equation}
	\hat{V} = \hat{\sigma}^2_+ + B/(mn) 
	\label{equ:6.195}
\end{equation}
and the estimated variance of the $\sigma^2$ given by the weighted average of $B$ and $W$
\begin{equation}
	\hat{\sigma}^2_+ = \frac{(n-1)}{n}W + \frac{B}{n}
	\label{equ:6.106}
\end{equation}
Equation \eqref{equ:6.106} is the unbiased estimate of the true variance $\sigma^2$. And $\hat{V}$ accounts for the sampling variability of the estimator which yields a pooled posterior variance estimate given in Equation \eqref{equ:6.106}. The PSRF is expected to be close to $1$ for convergence to be guaranteed and each of the $m$ sets of $n$ simulated observations is close to the target distribution, however large value of $\hat{R}$ indicates no convergence or divergence and further simulations can be taken into consideration or the proposal distribution should be critically looked into.
\subsection{Splitting EP on Clutter Problem}
The clutter problem has been addressed in the work by \cite{minka2001ep}. This work replicates this process to compare the proposed algorithm with the EP algorithm for a one-dimensional space. Suppose we have observations from a Gaussian distribution embedded in a sea of unrelated clutter where $w$ is the clutter ratio, so that the density observation is a mixture of two Gaussians:
\begin{equation}
	p(\textbf{y}|\theta) = (1 - w) \mathcal{N}(\textbf{y}; \theta, \textbf{I}) + w \mathcal{N}(\textbf{y}; \textbf{0}, 10\textbf{I})
	\label{equ:6.1}
\end{equation}
Let the \textit{d}-dimensional vector $\vec{\theta}$ with a Gaussian prior distribution:
\begin{equation}
	p(\theta) \sim \mathcal{N}(\textbf{0}, 100\textbf{I})
	\label{equ:6.2}
\end{equation}
The joint distribution of $\vec{\theta}$ and $n$ observation $D = \lbrace \bf{y}_1, ..., \bf{y}_n \rbrace$
\begin{equation}
	p(D, \theta) = p(\theta)\prod_i p(\textbf{y}_i|\theta)
	\label{equ:6.3}
\end{equation}
\begin{figure}[H]
	\centering
	\begin{minipage}[b]{0.45\textwidth}
		\includegraphics[width = 3in, height=2in]{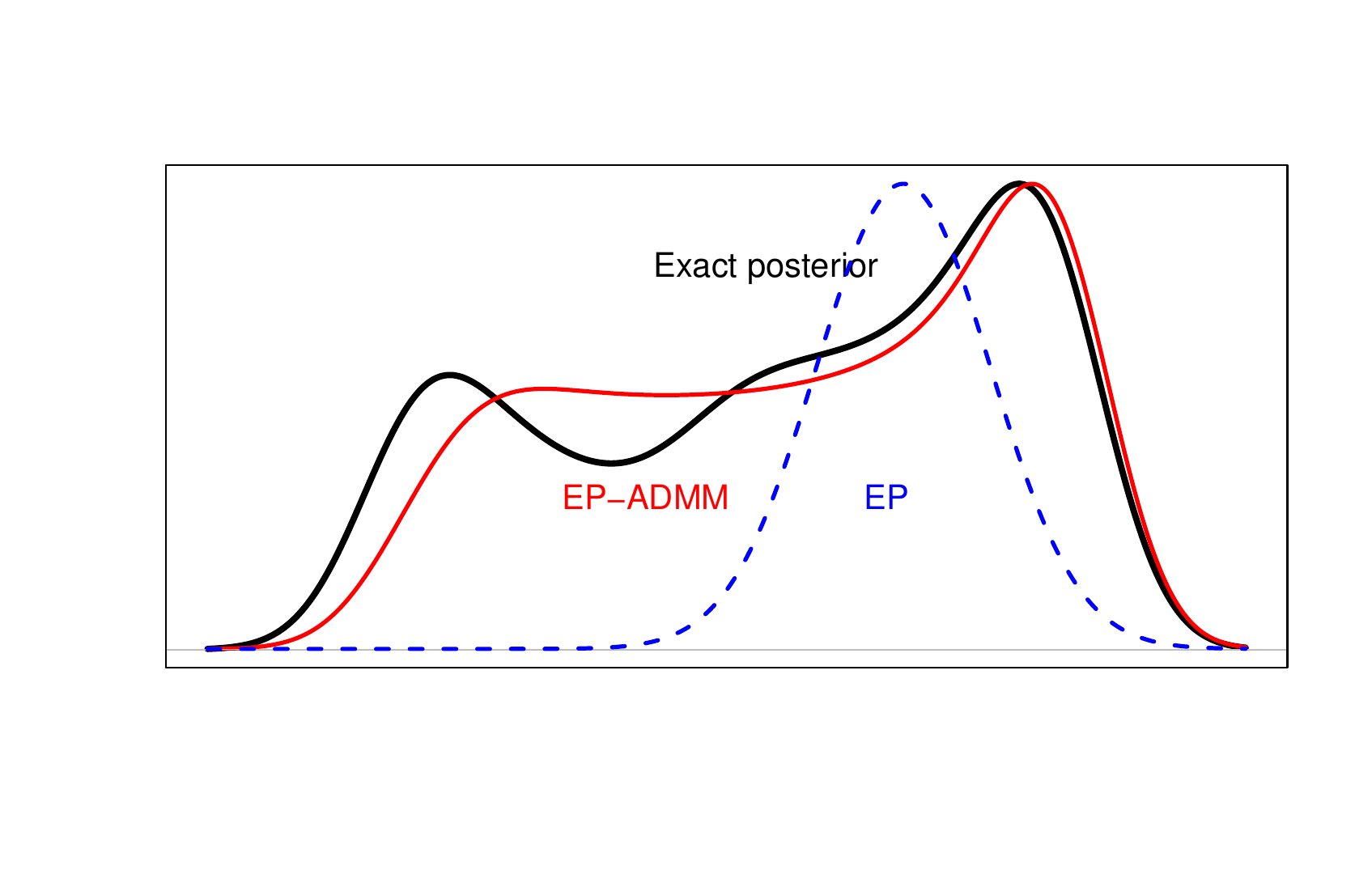}
		\centering\caption[A approximate posterior of clutter problem by EP-ADMM compared to EP]{A complex posterior in the clutter problem produced by EP with ADMM and EP compared with the exact posterior.} 
		\label{fig:6.1}
	\end{minipage}
	\hfil
	\begin{minipage}[b]{0.45\textwidth}
		\includegraphics[width = 2.5in, height=1.98in]{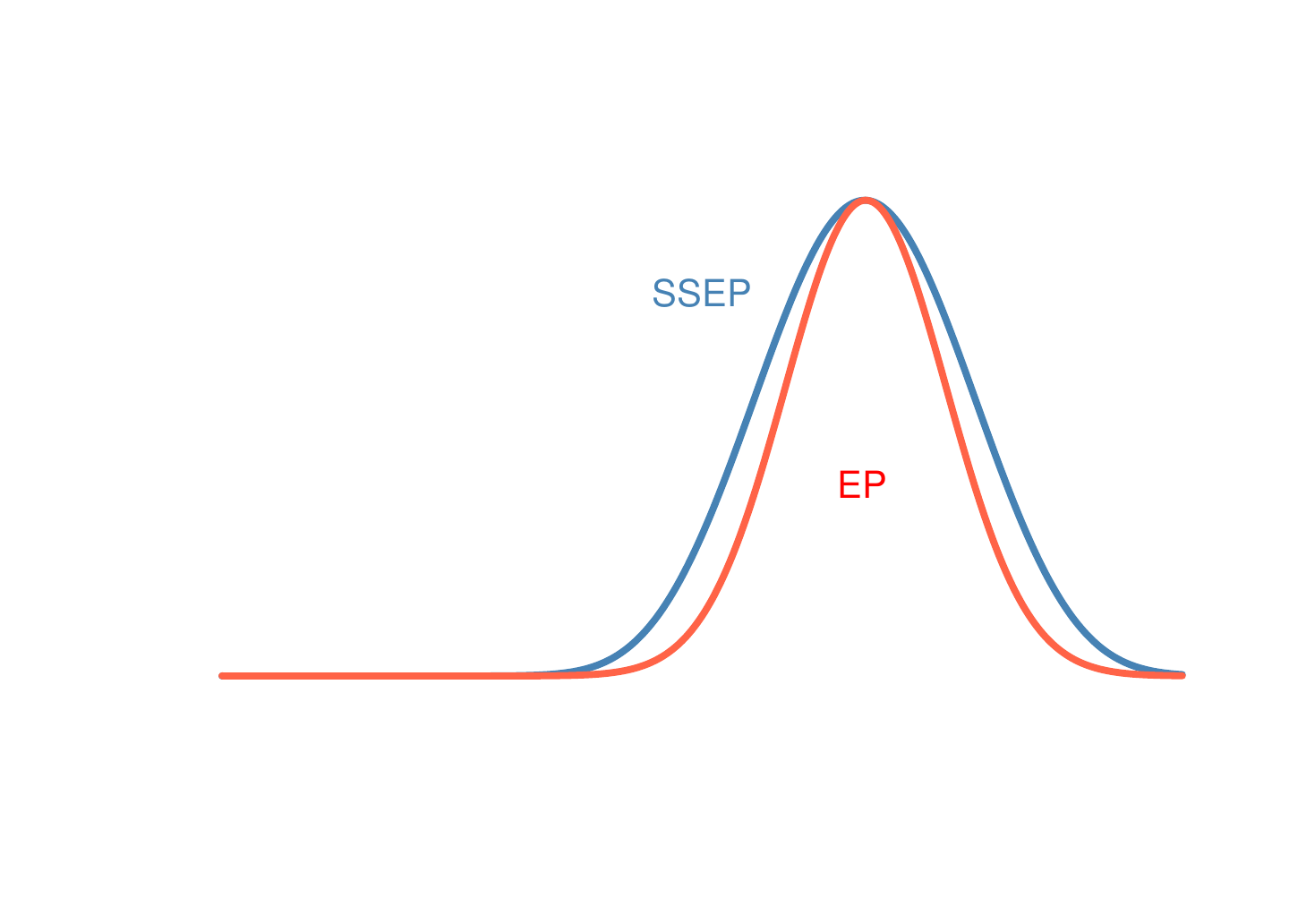}
		\caption[A approximate posterior of clutter problem by SSEP compared with EP]{A complex posterior in the clutter problem produced by Stochastic search EP with Monte Carlo integration compared with EP.} 
		\label{fig:6.2}
	\end{minipage}
\end{figure}
\bigskip
The goal is to approximate the posterior distribution
\begin{equation}
	p(\vec{\theta} | D) = \frac{p(D,\theta)}{\int p(D,\theta)\hspace{0.05cm} d\hspace{0.05cm}\theta}
	\label{equ:6.4}
\end{equation}
The first component in Equation \ref{equ:6.1} contains the parameter to be estimated which is $\theta$ and the second component describes the clutter where $w$ is the known ratio of the clutter. The Bayesian network for this problem is $\theta$ point to the $y_i$ [Check \cite{minka2001ep} for more detail].
\par The strength of Splitting EP over expectation propagation is the use of stochastic approximation and method of multiplier within EP. The learning rate (lr) determines how stable the Splitting EP algorithm is, unlike expectation propagation which fails at the refining step. Learning rate controls the erratic nature of the expectation propagation algorithm, and this seems to be a viable improvement over EP. Moreover, it forces the variance not to reduce to negative value which often leads to EP's stability issue. EP was compared to four algorithms for approximate inference in \cite{minka2001ep} such as the Laplace's method, variational Bayes, Importance sampling where the prior is used as the importance sampling, and Gibbs sampling by introducing the hidden variables that determines if a data point is clutter or not. According to \cite{minka2001ep}, EP competed well with other deterministic algorithms by approximating the posterior with a Gaussian. However, their performance improved substantially with more data i.e. the posterior is more Gaussian with more data. Figure (\ref{fig:6.1}) shows true posterior with multi-modal shape (Black line). Here, EP-ADMM (Red line) captures at least two modes while EP captures only one of the modes. This justifies that EP is a unimodal algorithm. Figure (\ref{fig:6.2}) shows how Stochastic search expectation propagation competes well with EP.
\subsection{Splitting EP on Synthetic Data}
\begin{figure}[H]
	\begin{center}
		\includegraphics[width=3in]{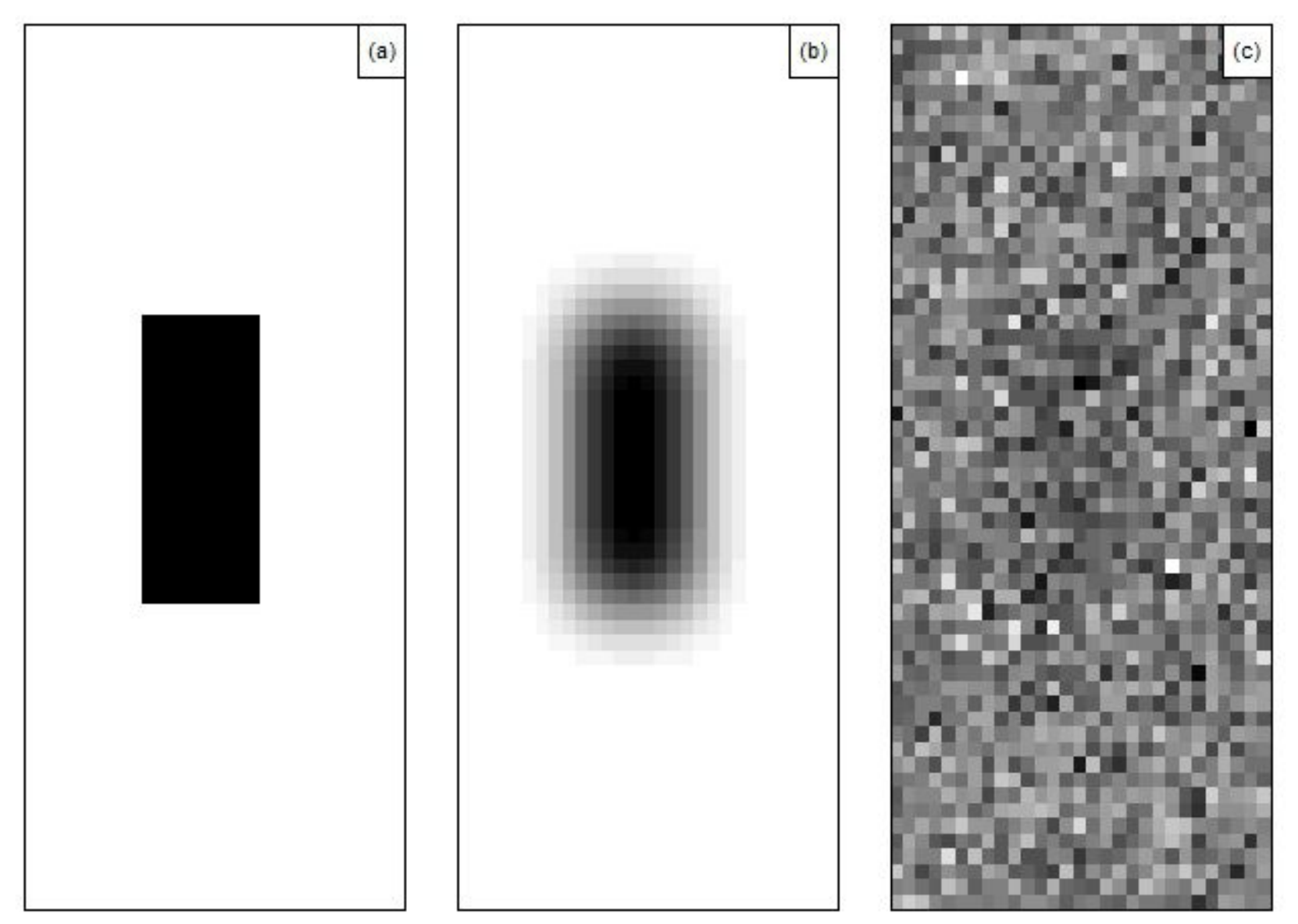}
	\end{center}
	\caption{$(a)$ True data: $\mathcal{X}$, $(b)$ the mean: $\mathcal{G}\mathcal{X}$ and $(c)$ Noisy data:$\mathcal{Y}$ \label{fig:6.4}}
\end{figure}

\begin{figure}[H]
	\centering
	\begin{minipage}[b]{0.45\textwidth}
		\includegraphics[width = 3in, height = 2.5in]{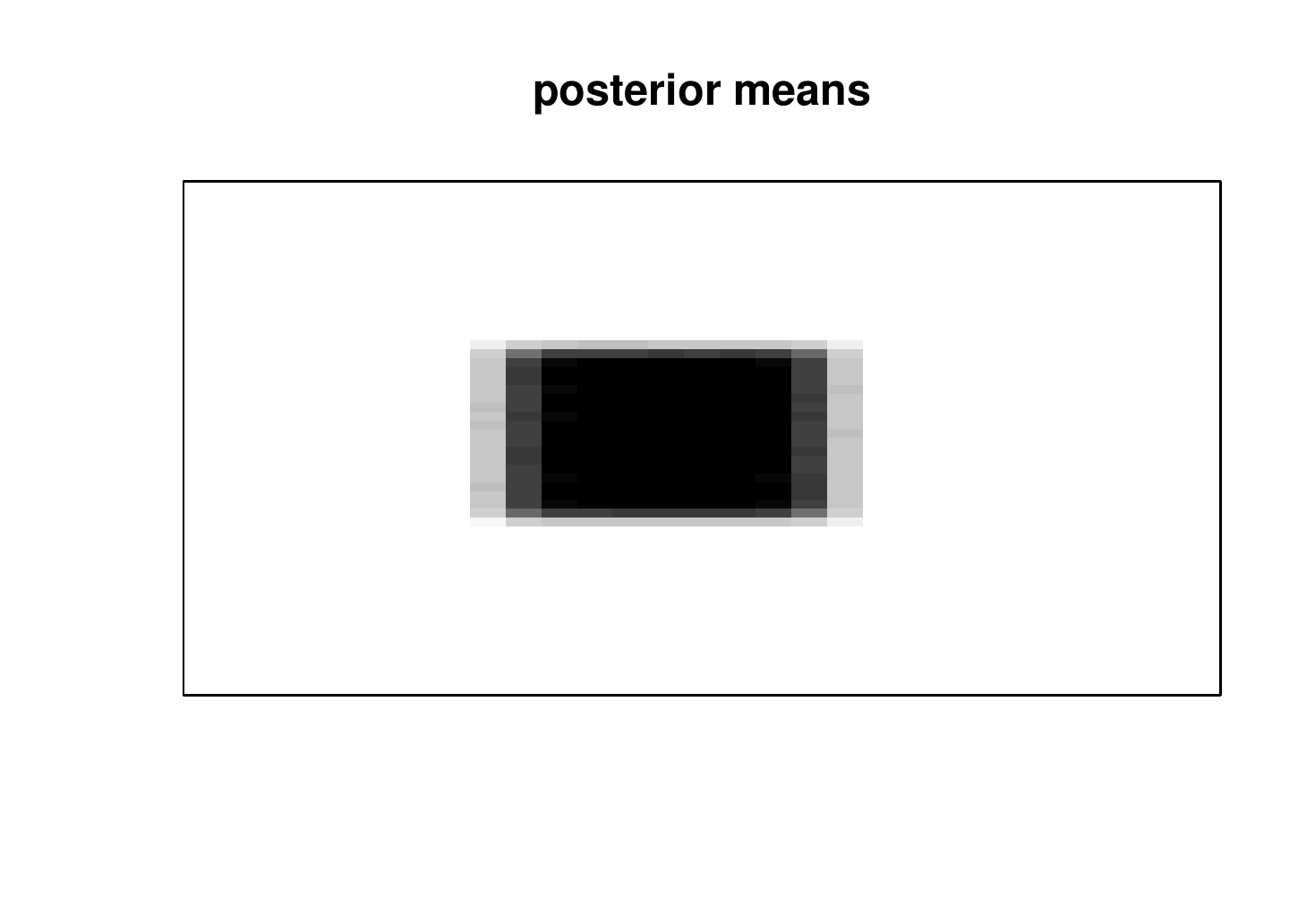}
		\centering\caption{The reconstruction of the cylindrical image produced by EP-MC} 
		\label{fig:6.5}
	\end{minipage}
	\hfil
	\begin{minipage}[b]{0.45\textwidth}
		\includegraphics[width = 2.8in, height = 1.5in]{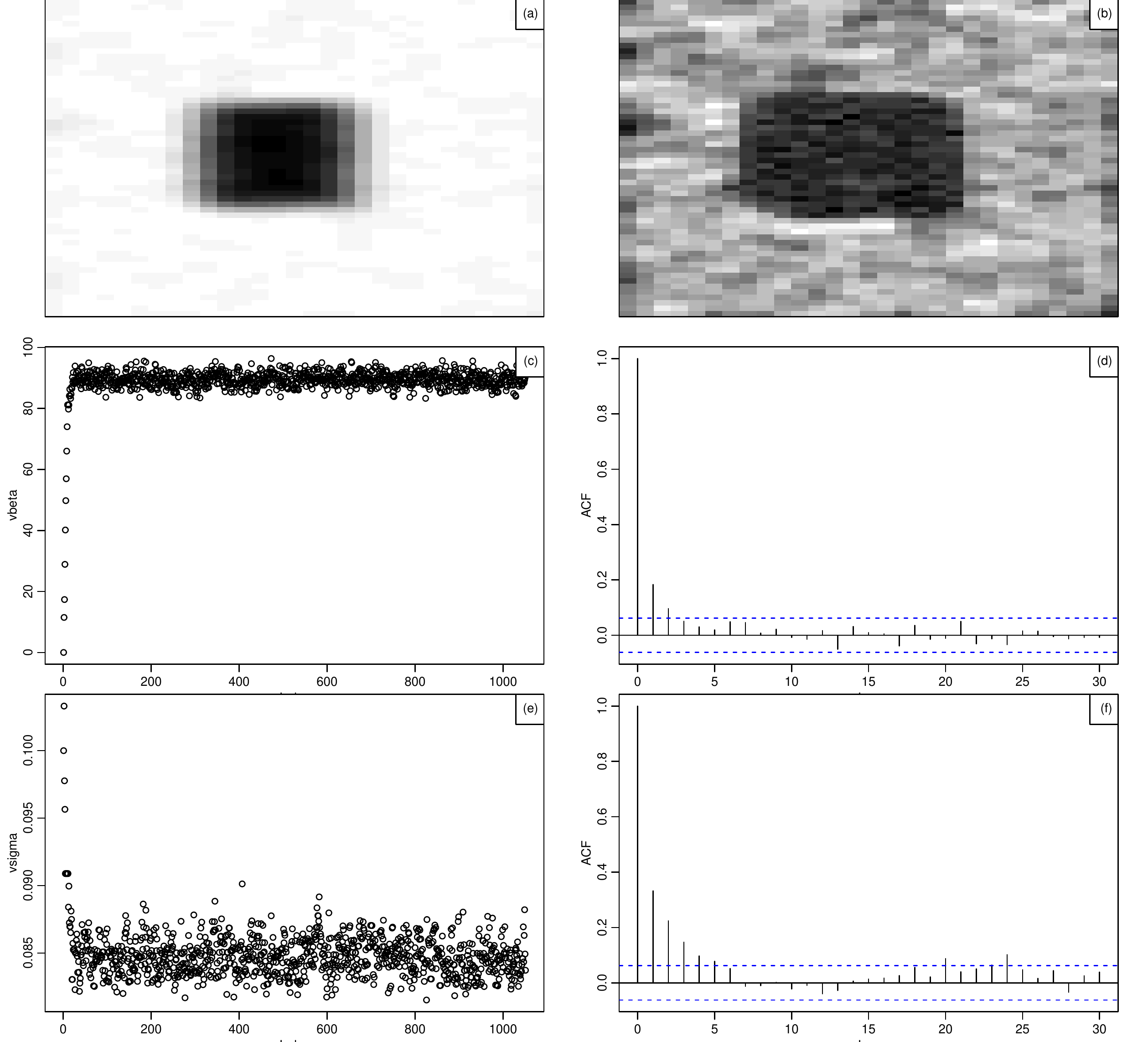}
		\caption[The result of the cylindrical image produced by MCMC]{(a): The reconstruction image of the cylindrical image produced by MCMC.(b): relative error (c): The estimates of precision converge at about $85$, (d): The estimates of standard deviation converges at $0.085$.}
		\label{fig:6.6}
	\end{minipage}
\end{figure}

A cylindrical image produced by a gamma camera is presented in Figure \eqref{fig:6.4}. The true image to be reconstructed is presented in Figure \eqref{fig:6.4}(c) mixed with a noise. We evaluate EP-Monte Carlo in inverse problem with hierarchical Bayesian models, with data of cylindrical image from gamma camera. In EP-Monte Carlo, we retain the originality of the EP. However, we improve on the intractability of the normalizing factor by following the idea of \cite{john2011vi}. Due to the randomness introduced to EP, we ran EP-MC $1000$ iterations and output the mean of the posterior. The true parameters are the following; the precision is $100$ and the true standard deviation is $0.1$.
We hereby present the reconstruction results of the EP-MC and MCMC. Figure \eqref{fig:6.5} shows the final output of EP-MC after $1000$ iterations while Figure \eqref{fig:6.6} shows the outcome of the MCMC. The reconstruction result of EP-Monte Carlo outperforms that of MCMC in terms of sharpness of the image. 
\begin{figure}[H]
	\centering
	\begin{minipage}[b]{0.45\textwidth}
		\includegraphics[width = 2.5in, height = 1.5in]{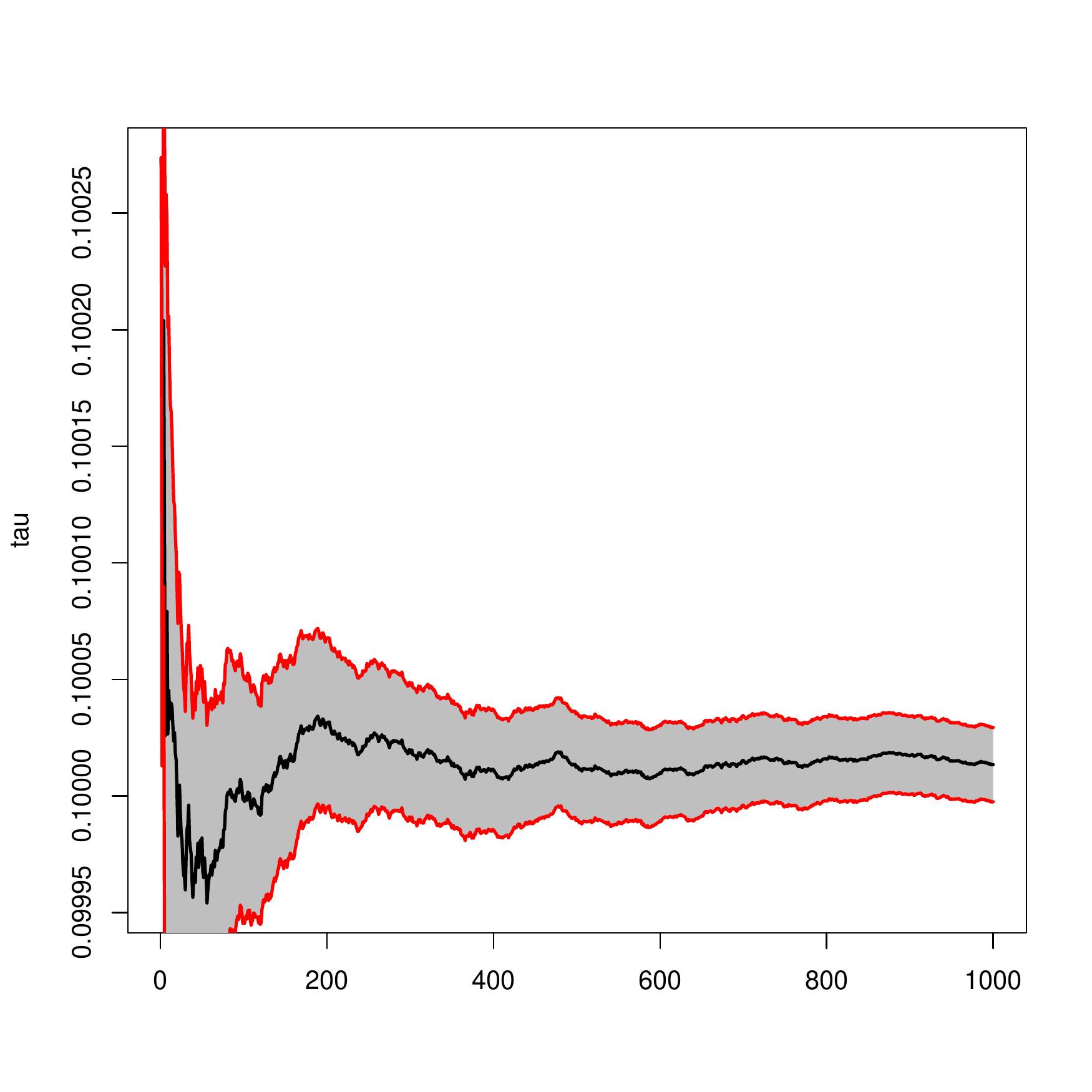}
		\centering\caption{The precision is calculated from the plot as $\tau^{-2} = 100$. Since the variance is $\tau^2 = 0.01$} 
		\label{fig:6.7}
	\end{minipage}
	\hfil
	\begin{minipage}[b]{0.45\textwidth}
		\includegraphics[width = 2.5in, height = 1.5in]{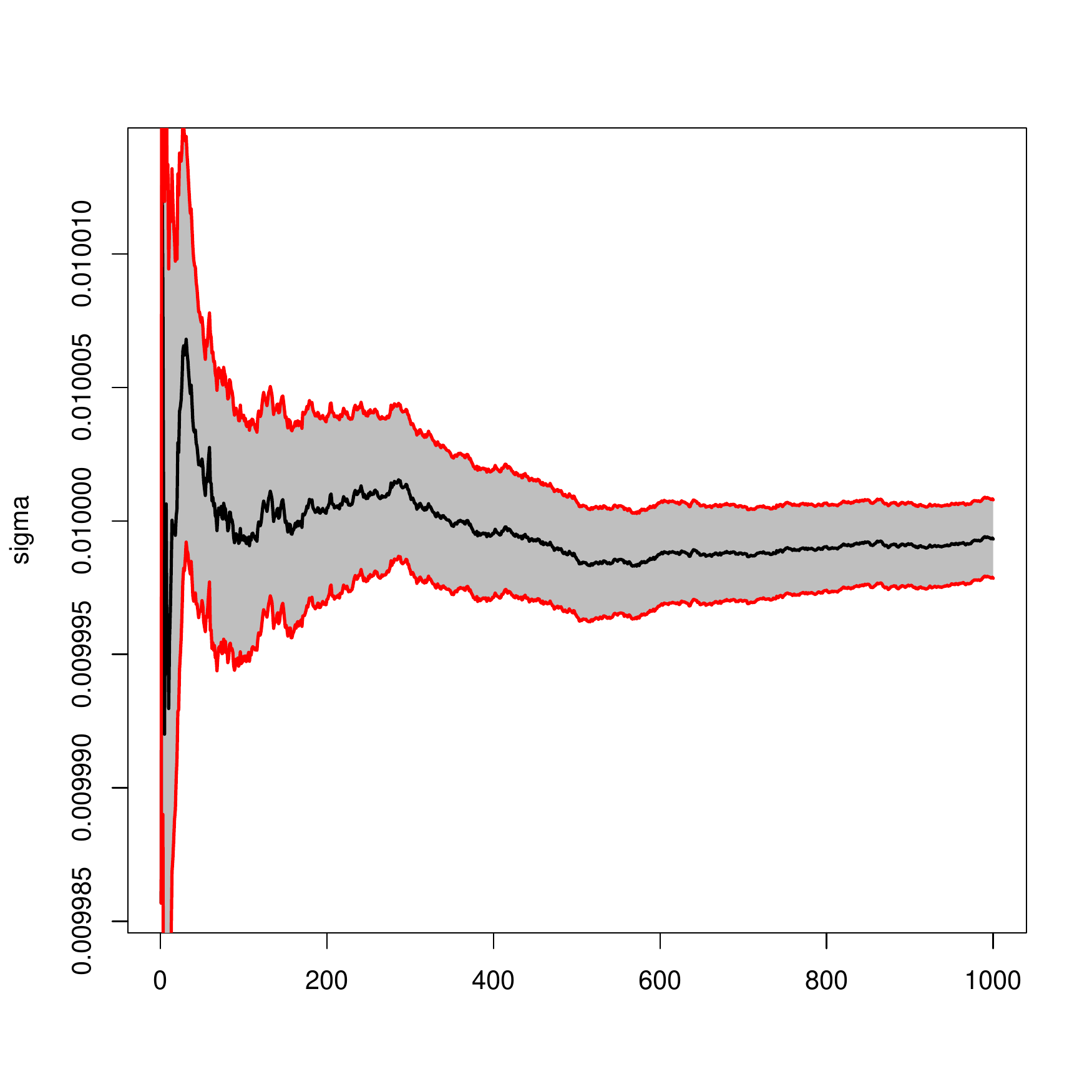}
		\caption{The variance output from the EP-MC is $0.01$ then the estimated standard deviation of EP-MC is $0.1$.} 
		\label{fig:6.8}
	\end{minipage}
\end{figure}
\bigskip
\begin{table}[H]
	\centering\caption{The Accuracy of the Standard deviation and precision produced \\ \centering by EP-MC and MCMC for prior $\tau$ and $\lambda$ compared to the True values provided}
	\begin{tabular}{rrrrrrrrr}
		\hline
		Parameter&&True-Value&&EP-MC&&MCMC&&\\
		\hline
		$\tau^{-2}$&&$100$&&$100$&&$85$&&\\
		$\lambda$&&$0.1$&&$0.1$&&$0.085$&&\\
		\hline
	\end{tabular}
	\label{tab:0}
\end{table}
\bigskip
\noindent Compared to the true image in Figure \eqref{fig:6.4}(a), we observe that the posterior of $\mathcal{X}$ is well approximated by the EP-MC method. In terms of the convergent time, i.e. time for both methods to reach convergence, 
\begin{figure}[H]
	\begin{center}
		\includegraphics[height=1.5in]{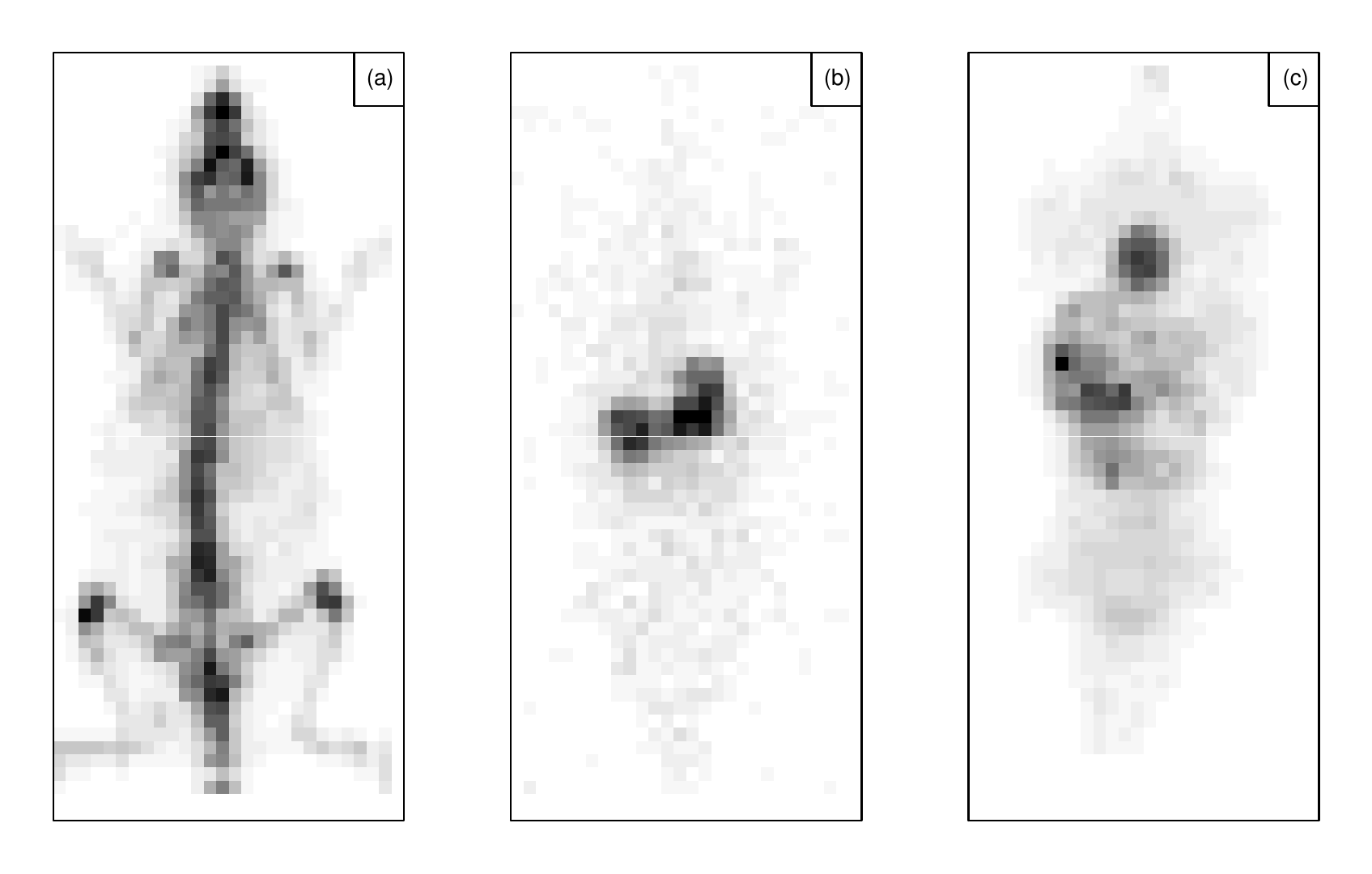}
		\centering\caption[The injected Mouse with MDP, DMSA, and MIBI]{\textit{(a)}: Mouse injected with $[^{99m}Tc]MDP$ at $4$ h pi (15-min) acquisition time. (\textit{b}): A mouse injected with $^{99m}Tc$ DMSA at $1 h$ (10-min scan), 3 h (10-min scan), 5 h (10-min scan), 6h (10-min scan), and 24 h pi (30-min scan), (\textit{c}):  A static image of the mouse injected with the $[^{99m} Tc]$ MIBI at different intervals of time.}
		\label{fig:6.9}
	\end{center}
\end{figure}
\bigskip
\noindent EP-MC is faster than MCMC. Unlike MCMC that needs about $500$ chains as a burn-in period, EP-MC does not require any thin-in or burn-in period. We observed a clear pattern in the residual plot shown in Figure \eqref{fig:6.6}(b). This non-randomness may be due to the choice of the starting value of hyperparameters. The estimated precision for both EP-MC and MCMC are provided in Figure \eqref{fig:6.7} and Figure \eqref{fig:6.6}(c) respectively. 
\noindent It is not surprising to observe that EP-MC produces the exact precision parameter value because of the high accuracy in the approximation of $\mathcal{X}$. Similarly, the estimate of variance parameter value are provided in Figure \eqref{fig:6.8} and Figure \eqref{fig:6.6}(d) for EP-MC and MCMC respectively. Table \eqref{tab:0} shows the comparison of the estimate of precision and variance parameter values for EP-MC and MCMC with the true parameters. 
We observe that EP-MC provides a closer estimate of both precision and variance to the true parameters.
\subsection{Splitting EP on Real Data}
The full description of the study on the $\gamma$-eye system evaluated in a proof-of-concept animal study using normal Webster Swiss Albino mice with average weights of $25g$ can be found in \cite{maria2016}. The author conducted the study on three different clinical radio-pharmaceuticals which was radio-labeled with Tc-99m and injected the mice via their tail vein. The first mouse presented in Figure \eqref{fig:6.9}(a) was injected with $100 \mu 1/7.5$ MBq $[^{99m}Tc]$ MDP, which is a suitable agent for bone imaging and static images were obtained at $1 h$ post-injection $(pi)$, $2 h$ pi, $3 h$ pi, and $4 h$ pi. The second mouse was injected by $100 \mu1/7.5$ MBq $[^{99m}Tc]$ DMSA, which is a tracer for imaging the anatomical structure and functional process of the kidneys. The author performed dynamic imaging for the first hour after injection and then static images for $10$ min were acquired every $1 h$ up to $24 h$ pi. The image presented in Figure \eqref{fig:6.9}(b) is a 1-h dynamic study of a mouse injected with $100 \hspace{.2mm}\mu l$ of $[^{99m} Tc]$ DMSA radio-tracer. Then, static images were acquired at different time intervals. Figure \eqref{fig:6.9}(c) shows a third mouse injected with $100 \mu1/5.6$ MBq $[^{99mTc}]$ MIBI for heart perfusion. Dynamic imaging was performed for $2 h$ pi and $10$ min static images were obtained up to $5$-h pi. 
\subsubsection{Parameter Setup for Splitting EP}
The Splitting EP algorithm was initialized with the following parameters; $\tau = \lambda = 0.01$ as a starting point for Metropolis Hasting algorithm embedded in EP to solve the intractability of the normalizing factor. We fixed the hyper-priors to $\tilde{\alpha}_{ij}^\tau = 2$, $\tilde{\alpha}_{ij}^\lambda = 1$ for the approximate term distribution and $\alpha^\tau_{ij} = \alpha^\lambda_{ij} = 10$ for the approximate posterior. According to the update of splitting EP, we computed the cavity distribution parameters $\alpha_{-ij}^\tau$ and $\alpha_{-ij}^\lambda$ just once to initialize the target distribution $p(\tau)$ and $p(\lambda)$. The outer loop consists of the EP-MCMC, i.e. tilted posteriors of $\tau$ and $\lambda$ are hereby assumed to be intractable and thereby approximated using MCMC technique. We note that EP-MCMC has some advantages over ordinary MCMC in terms of speed to convergence, and accuracy. Also, we observed that due to different values of cavity distribution computed at different iterations of the algorithm, EP-MCMC does not need either burn-in period or thinning to achieve stationarity. To illustrate the accuracy of EP-ADMM for the image reconstruction and EP-MCMC for parameter recovery, we present also the results by a standard Metropolis Hasting MCMC with a sample of $1000$ and burn-in period of $500$. 
\begin{figure}[H]
	\centering
	\begin{minipage}[b]{0.45\textwidth}
		\includegraphics[width = 3in, height=2in]{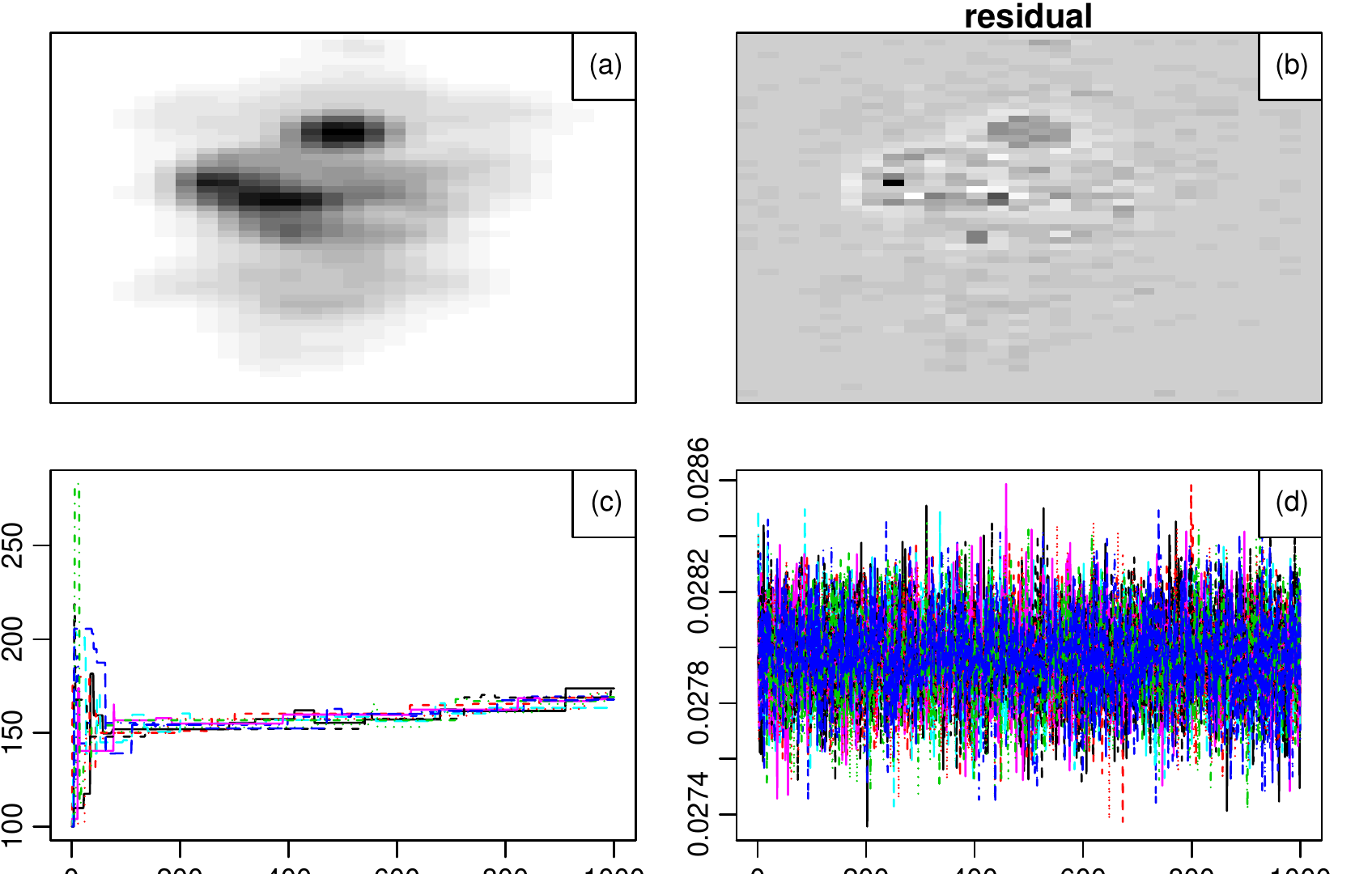}
		\centering\caption[The Reconstruction of Mibi image by EP-ADMM]{(a): 3h static\_mibi\_results from EP-ADMM, (b): relative error (c): estimates of $\tau$ converges at $158.48$ from EP-MCMC, (d): estimates of $\lambda$ converges at $0.028$ from EP-MCMC.} 
		\label{fig:6.10}
	\end{minipage}
	\hfil
	\begin{minipage}[b]{0.45\textwidth}
		\includegraphics[width = 2.5in, height=1.98in]{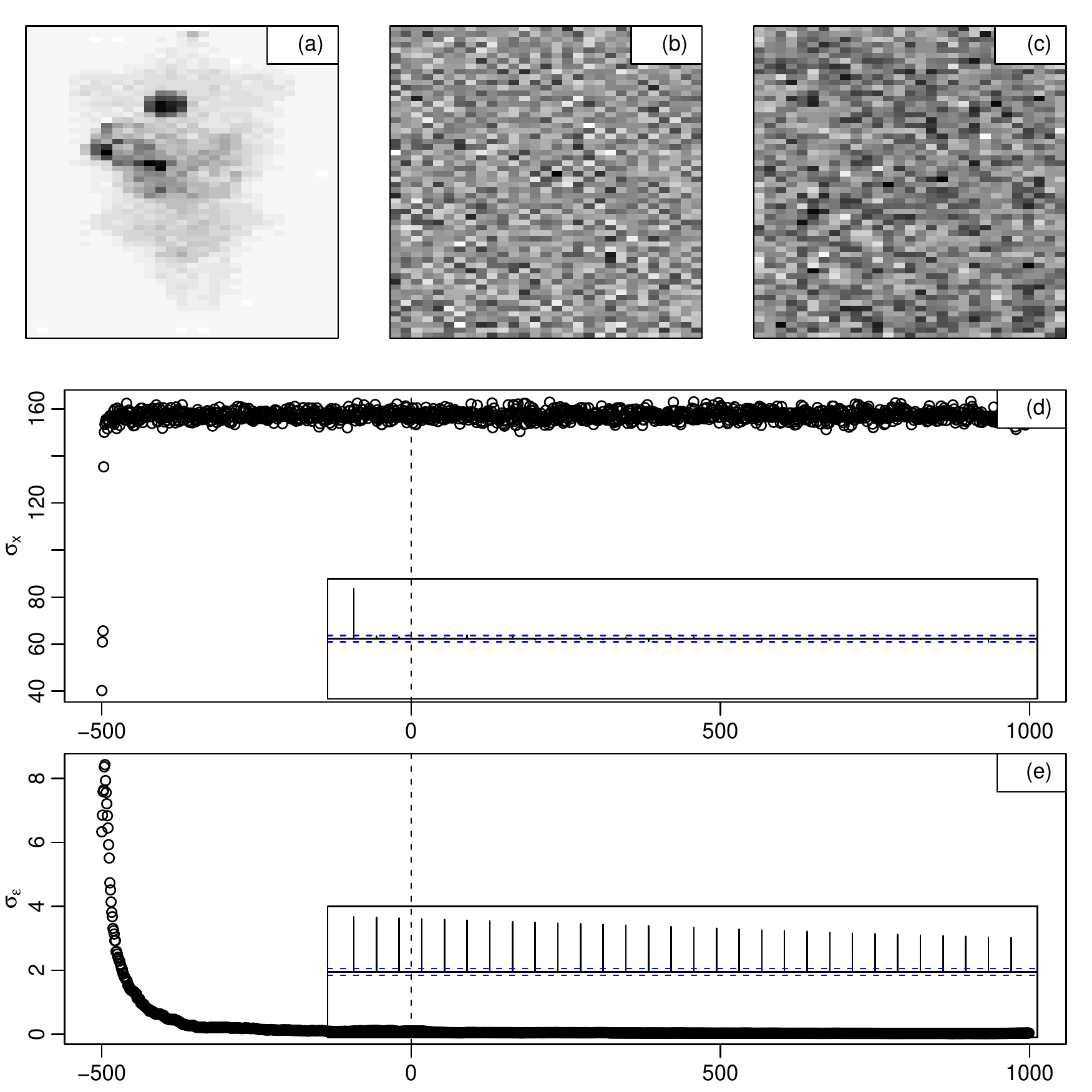}
		\caption[The reconstruction of Mibi image by MCMC]{(a): 3h static\_mibi\_results from MCMC, (b)\&(c): error \& residual, (d): the estimate of $\sigma_x$ converges at $159$, (e): the estimates of $\sigma_\epsilon$ converges to about $0.9$.} 
		\label{fig:6.11}
	\end{minipage}
\end{figure}
\bigskip
\begin{figure}[H]
	\centering
	\begin{minipage}[b]{0.45\textwidth}
		\includegraphics[width = 3in, height=1.5in]{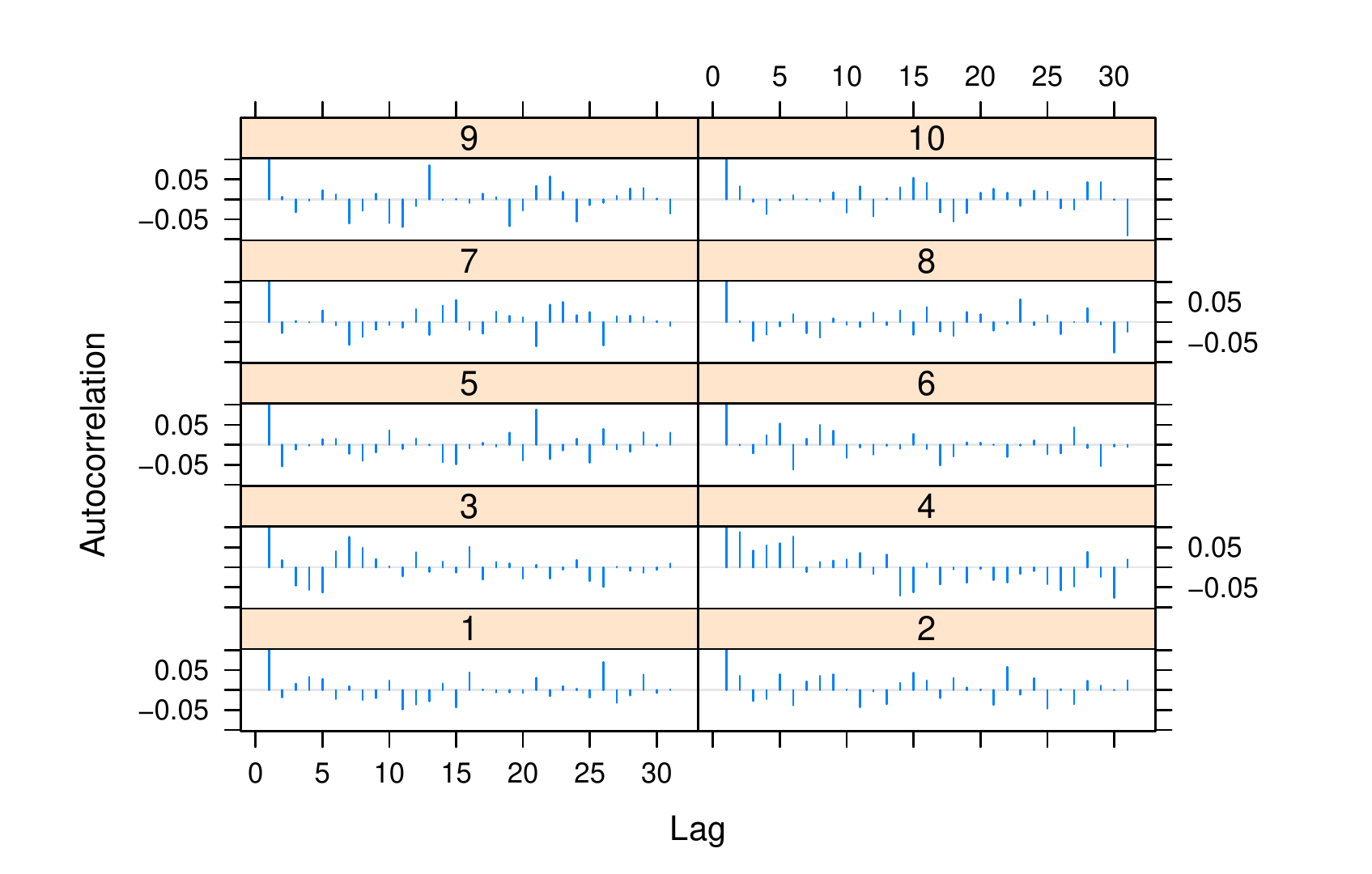}
		\centering\caption[The Autocorrelation plot of Mibi for $\sigma$ of EP-MCMC algorithm]{The Autocorrelation plot of the mouse injected with Mibi reagent for sigma; the EP-MCMC algorithm was runs $10$ time each of length $1000$. The plot shows independence of chains at each replication.} 
		\label{fig:6.12}
	\end{minipage}
	\hfil
	\begin{minipage}[b]{0.45\textwidth}
		\includegraphics[width = 2.5in, height=1.5in]{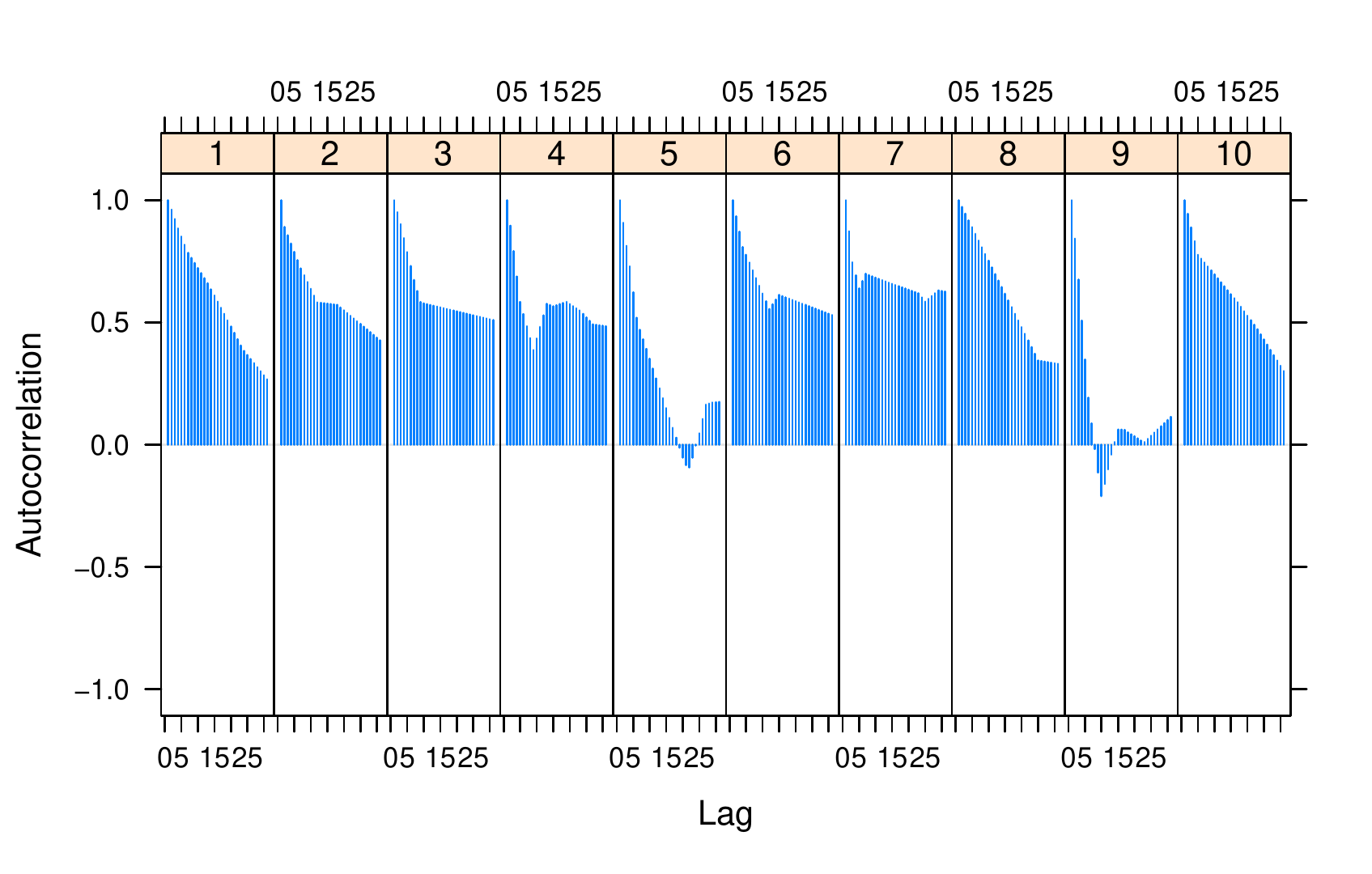}
		\caption[The Autocorrelation plot of Mibi for $\tau$ of EP-MCMC algorithm]{For the $\tau$, the plot shows a strong dependence of chains at first four replications but at the fifth and ninth there is no correlation and the last replication shows a decline in correlation of the chains.} 
		\label{fig:6.13}
	\end{minipage}
\end{figure}
\bigskip 
\noindent The random walk stepsize is chosen so that the acceptance rate is close to $0.234$, which is widely known to be optimal \citep{gelmanetal1996}. It is well known that the convergence of MCMC is very crucial. 
\subsubsection{Reconstruction result for Mibi}
First, we present the reconstruction result for the mouse injected with $100 \mu1/5.6$ MBq $[^{99mTc}]$ MIBI for heart perfusion. The estimate based on the Gaussian prior is shown in Figure \eqref{fig:6.10}. It can be seen that a well reconstruction is obtained with a substantive reduction in the noise. The result produced from ordinary MCMC is presented in Figure \eqref{fig:6.11}(a). The reconstructions produced by both algorithms have a clear distinctions. First off, the reconstruction by EP-ADMM has a white background and smooth edges while the reconstruction produced by MCMC has a gray background and rough edges around the image reconstructed. This is a very important distinction because expectation propagation is well known for its fast computational speed. In contrast, MCMC is known for its slow convergence and it requires more computational time to reach a stationary distribution. Figure \eqref{fig:6.10}(b) shows the error comparing the estimates with the data $\mathcal{Y}$. Comparing the MCMC results with the reconstruction produced in EP-MCMC, the parameter estimates of $\sigma_x$ in MCMC is $159$ while $\tau$ in EP-MCMC is about $158.48$, similarly the parameter estimates of $\sigma_\epsilon$ is about $0.9$ while that of EP-MCMC is $0.028$. The residual error shown in Figure \eqref{fig:6.10}(b) comparing the estimates with the data shows a foggy pattern. It is possible to produce similar random residual as shown in Figure \eqref{fig:6.11}(b\&c) by adjusting the value of $\alpha_\lambda$ and $\alpha_\tau$ in the hyperparameter $\lambda$ and $\tau$ respectively. This may be achieved by tweaking the values of the hyperparameters $\lambda$ and $\tau$. However, given our primary aim to produce a constant reconstruction, our focus is on the appropriate values of hyperparameters that produce smoother reconstruction.
\begin{figure}[H]
	\centering
	\begin{minipage}[b]{0.45\textwidth}
		\includegraphics[width = 2in]{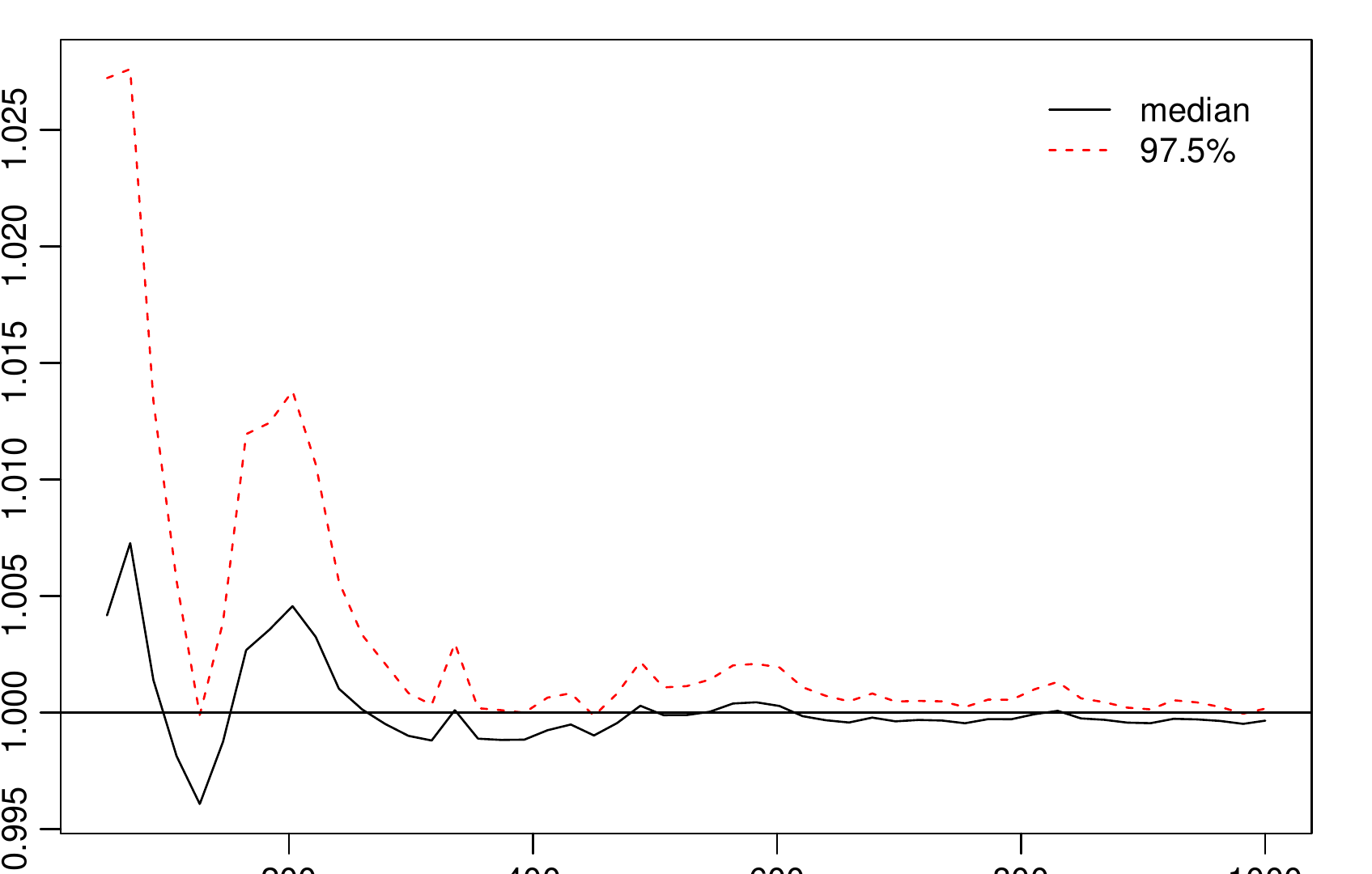}
		\centering\caption{Iterative PSRF Plot for $\lambda$ in Mibi image data (from $m = 10$ parallel sequence and $n = 1000$). The convergence starts at about $300$ iterations till the end of the iteration.}
		\label{fig:6.14}
	\end{minipage}
	\hfil
	\begin{minipage}[b]{0.45\textwidth}
		\includegraphics[width = 2.5in]{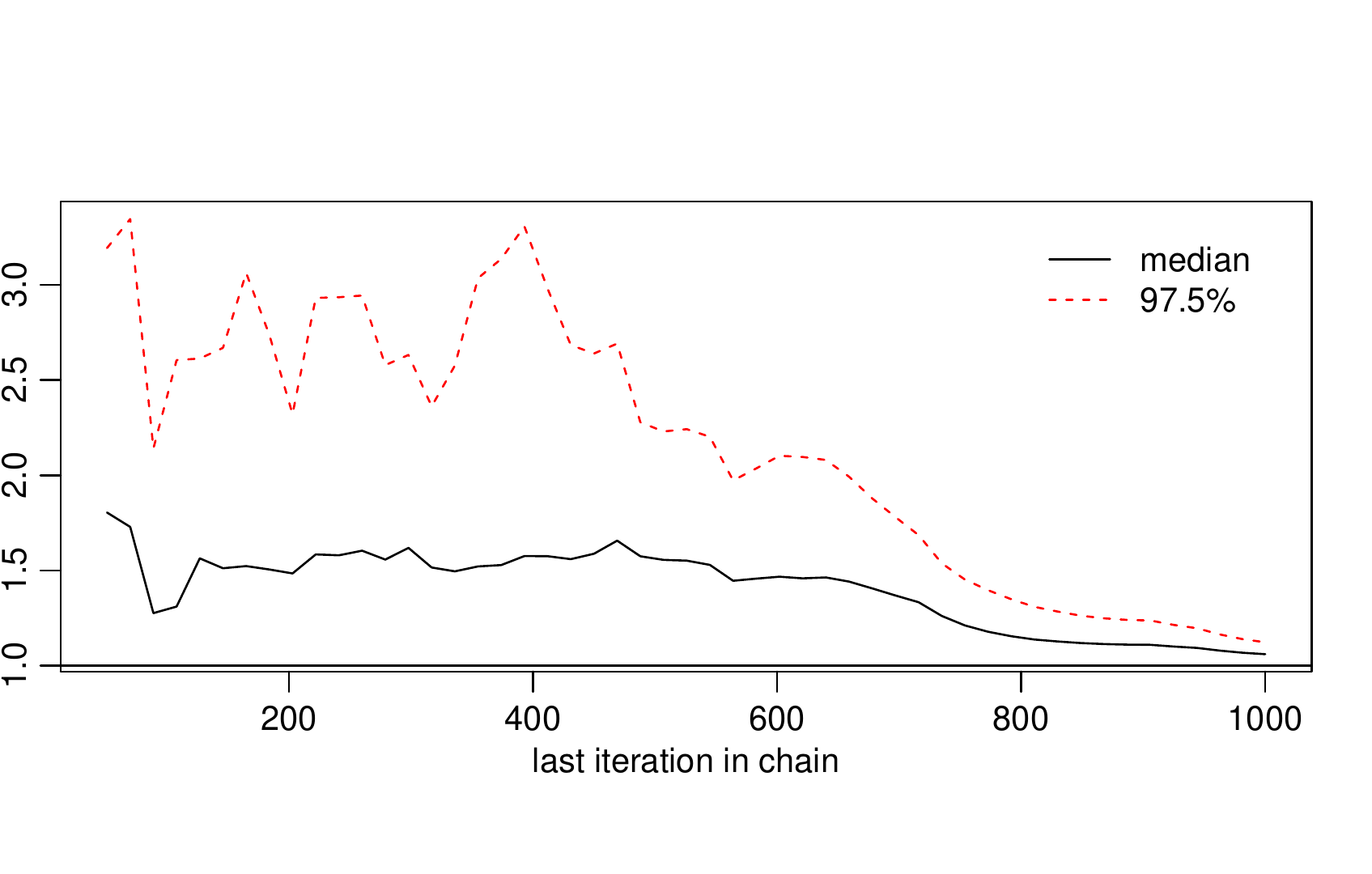}
		\caption[Iterative psrf Plot for $\tau$ in Mibi image data]{Iterative PSRF Plot for $\tau$ in Mibi image data (from $m = 10$ parallel sequence and $n = 1000$). The convergence starts at About $800$ iteration till the end of the iteration.} 
		\label{fig:6.15}
	\end{minipage}
\end{figure}
\bigskip
\begin{figure}[H]
	\centering
	\begin{minipage}[b]{0.45\textwidth}
		\includegraphics[width = 2.5in, height=2in]{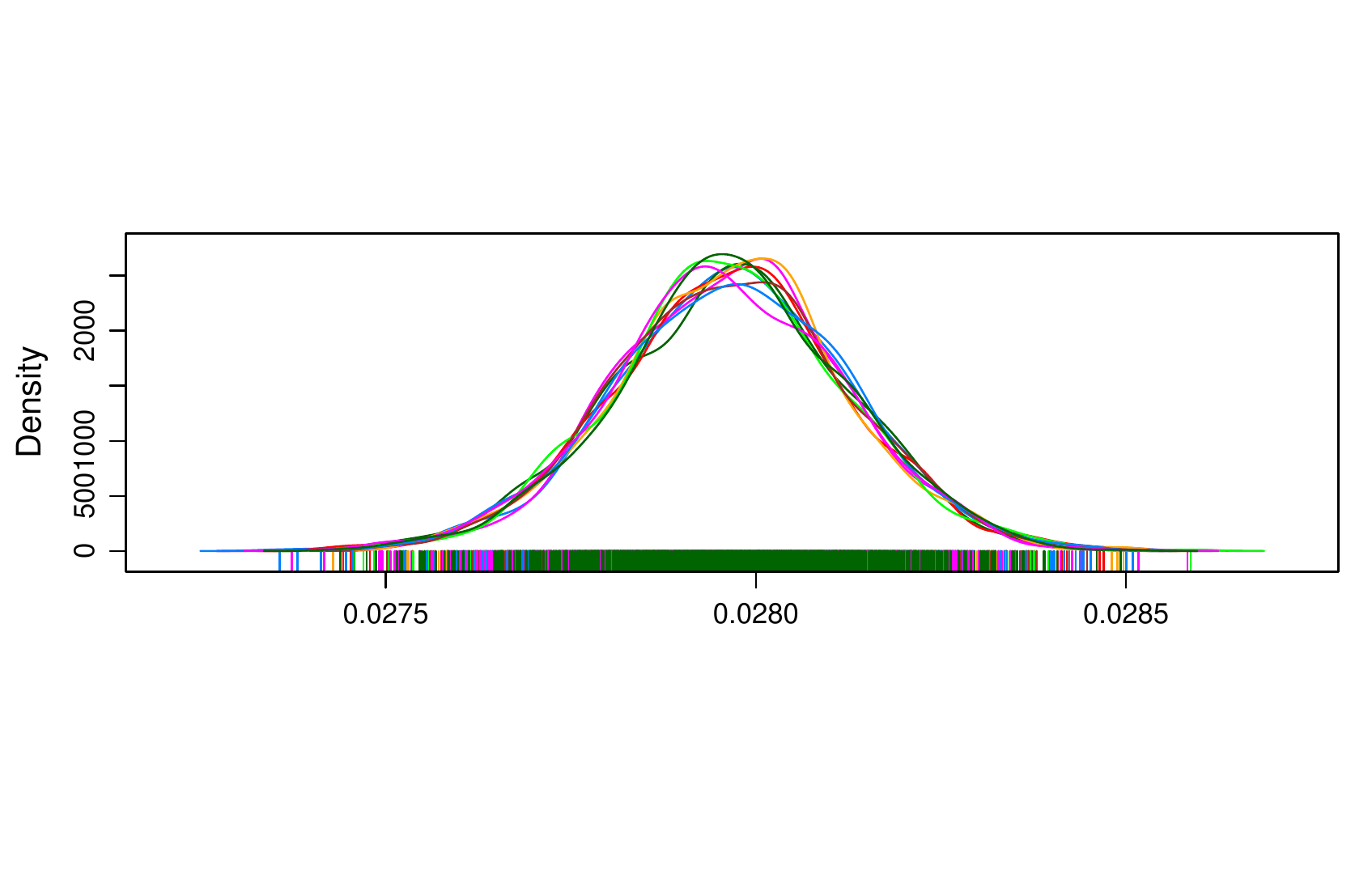}
		\centering\caption[Marginal posterior distribution of the parameter lambda for Mibi data]{Marginal posterior distribution of the parameter $\lambda$ for Mibi data. The tick horizontal bar signifies the posterior $95\%$ credible interval and the maximum value at $0.028$ is the posterior mean.}
		\label{fig:6.141}
	\end{minipage}
	\hfil
	\begin{minipage}[b]{0.45\textwidth}
		\includegraphics[width = 2.5in, height=1.9in]{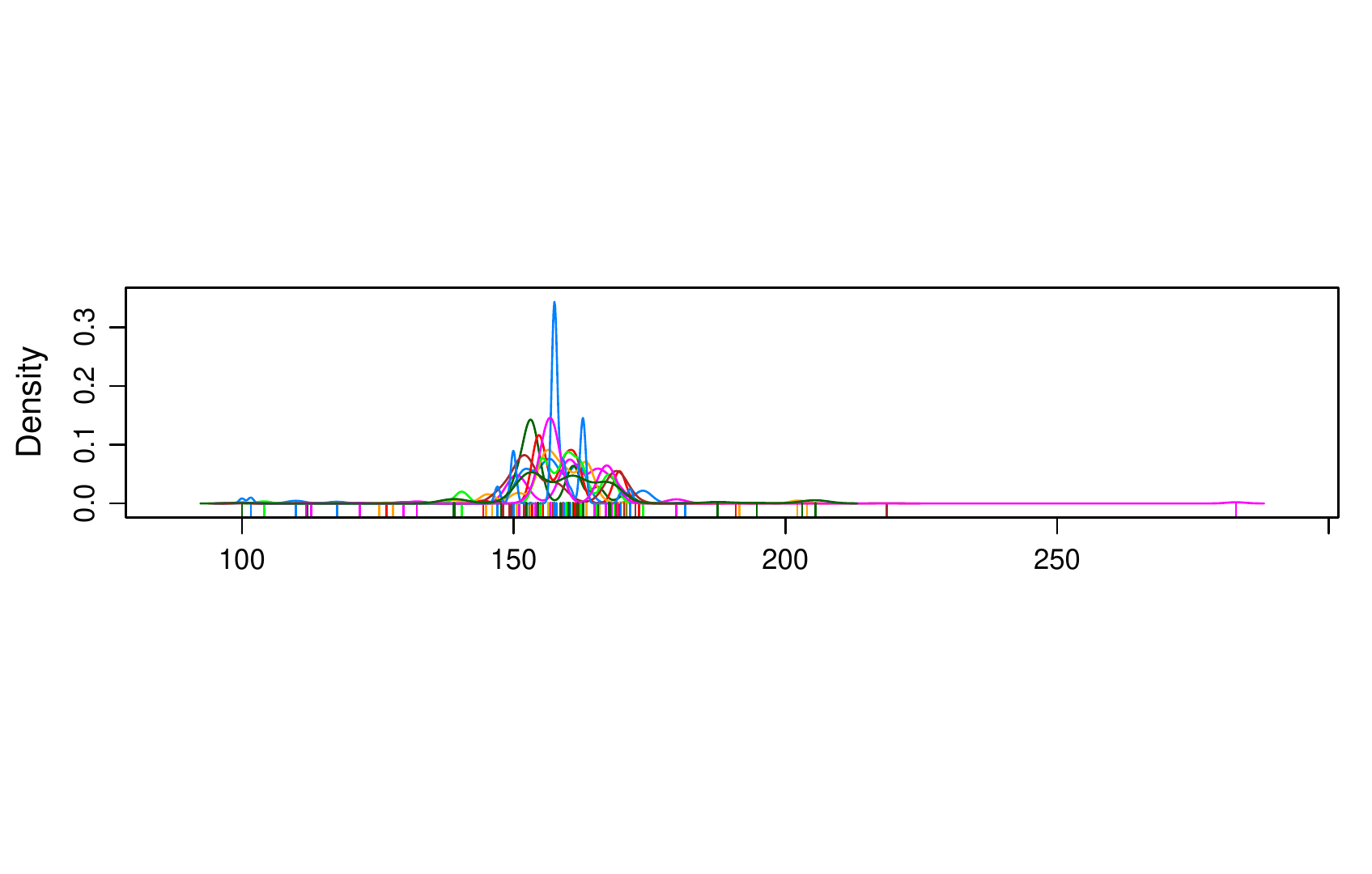}
		\caption[Marginal posterior distribution of the parameter tau for Mibi data]{Marginal posterior distribution of the parameter $\tau$ for Mibi data. The tick horizontal bar signifies the posterior $95\%$ credible interval and the maximum value at about $160$ is the posterior mean.} 
		\label{fig:6.151}
	\end{minipage}
\end{figure}
\bigskip
\begin{table}[H]
	\centering\caption{The mean estimates of the ten EP-MCMC chains for priors $\tau$ and $\lambda$\\ of Mibi data}
	\begin{tabular}{rrrrrrrrr}
		\hline
		parameter&Posterior-Mean&$\hat{R}$&$\hat{V}$&$W$&$B/n$&$\sigma_+^2$&\\
		\hline
		$\tau$&$158.48$&$1.06$&$173.58$&$98.53$&$68.32$&$166.75$&\\
		$\lambda$&$0.028$&$1.00$&$2.98e^{-08}$&$2.49e^{-08}$&$2.48e^{-08}$&$2.73e^{-08}$&\\
		\hline
	\end{tabular}
	\label{tab:1}
\end{table}
\bigskip
\par \noindent These parameter estimates are very significant in producing the clearer and smooth image. Given our primary aim to produce a clear and smoother image, which means larger $\tau$ in EP-MCMC and $\sigma_x$ in MCMC would be preferred and lower values of $\sigma_\epsilon$ in MCMC and $\lambda$ in EP-MCMC would also be preferred. The autocorrelations of the estimates produced by MCMC are nested in their respective plot and it can be seen that the autocorrelations have been controlled by the well set-up of MCMC such as burn-in of $500$ samples and thin-in of $1$-of-$10$ samples. However, there seems to be a high correlation in Figure \eqref{fig:6.11}(e) which might be reduced over time with large iteration. The computed potential scale reduction factors $\hat{R}$ for $\tau$ is $1.06$, with upper C.I to be $1.12$ and for $\lambda$ is $1$ with upper C.I to be $1$ as shown in Table \eqref{tab:1}. Here, the potential scale reduction factor for $\lambda$ is exactly $1$ which indicates strong convergence across the $10$ runs. According to \cite{gelmanandrubin1992a}, large $\hat{R}$ can be interpreted as a need for more simulations to further reduce the estimate of the variance $\hat{\sigma}^2$ or to increase the $W$. The closeness to $1$ indicates that each $m$ sets of $n$ simulated samples is close to the target distribution. However, potential scale reduction factor for $\tau$ is a bit more than one which is in tandem with further plots below. The simulation result agrees with the result from MCMC. Figure \eqref{fig:6.12} shows the correlation between the chains of $\lambda$ for each run. It can be seen that there is no autocorrection in the chains for 
\begin{figure}[H]
	\centering
	\begin{minipage}[b]{0.45\textwidth}
		\includegraphics[width = 3in, height=2in]{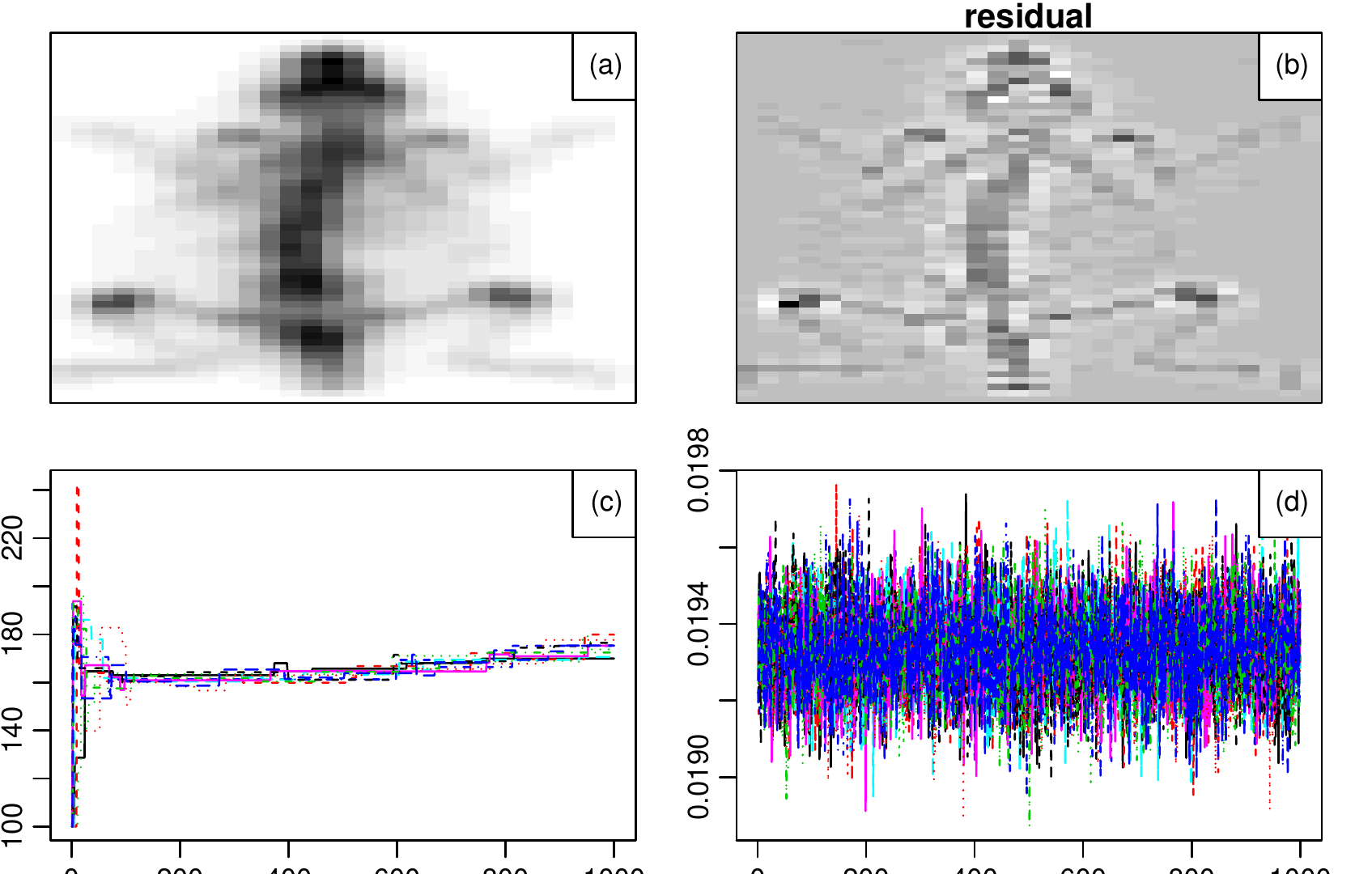}
		\centering\caption[mdp\_1h\_mouse\_results from EP-ADMM, relative error, and estimates of $\tau$, $\lambda$]{(a): mdp\_1h\_mouse\_results from EP-ADMM, (b): relative error (c): estimates of $\tau$ which converges at $165.443.$, (d): estimates of $\lambda$ converges at $0.019$.} 
		\label{fig:6.16}
	\end{minipage}
	\hfil
	\begin{minipage}[b]{0.45\textwidth}
		\includegraphics[width = 2.5in, height=1.98in]{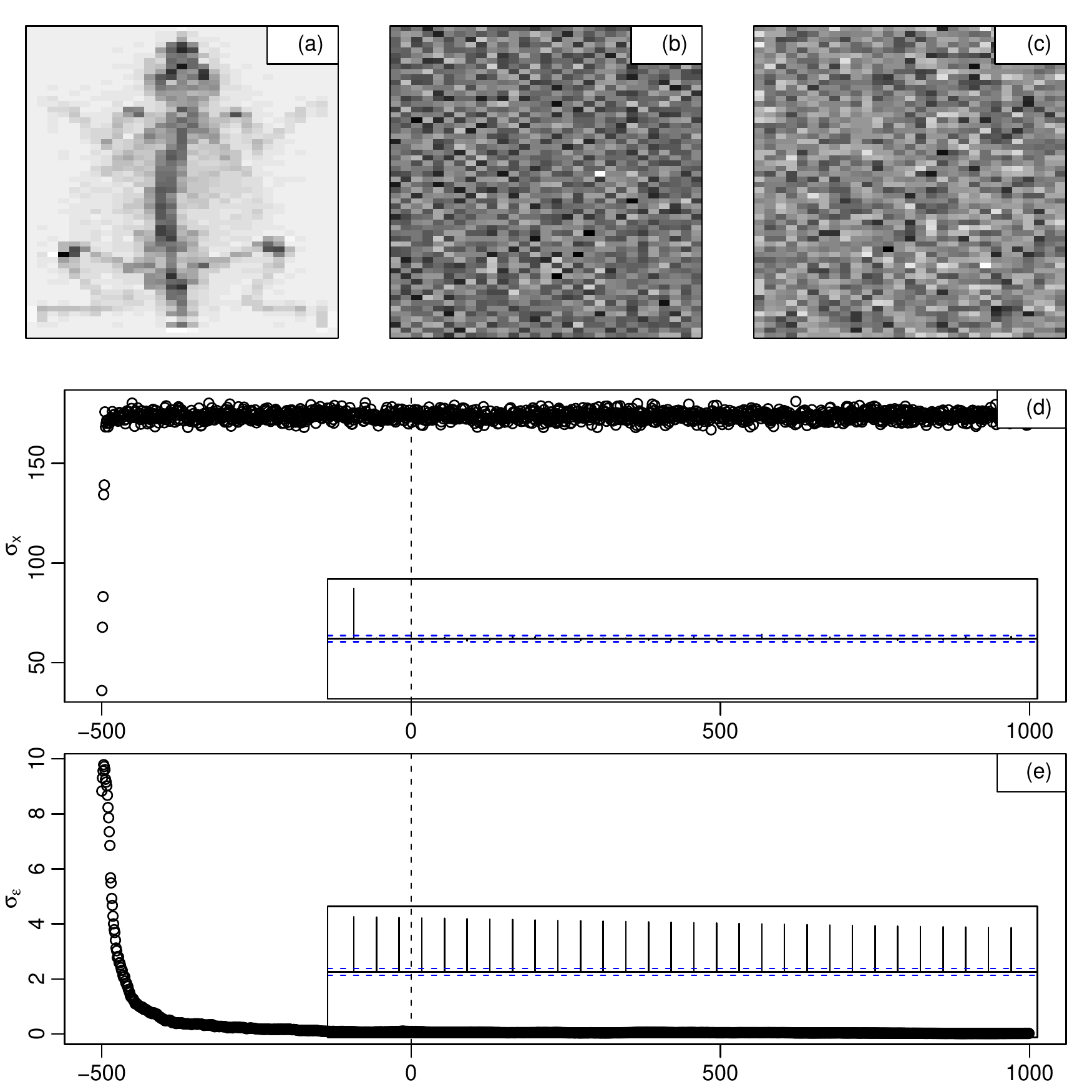}
		\caption[mdp\_1h\_mouse\_results from MCMC, error \& residual, estimate of $\sigma_x$ and $\sigma_\epsilon$]{(a): mdp\_1h\_mouse\_results from MCMC (b)\&(c): error \& residual, (d): the estimate $\sigma_x$ converges at $160$, (e): the estimates $\sigma_\epsilon$ converges to about $0.9$.} 
		\label{fig:6.17}
	\end{minipage}
\end{figure}
\bigskip
\noindent each of the runs, this indicates that there is a convergence at every run which also contributes to the potential scale reduction factor of $1$. Similarly, Figure \eqref{fig:6.13} shows an autocorrelation plot of $\tau$ for the $10$ runs. There is a high correlation among the chain of the first four runs, but at the fifth and ninth runs there is a sharp reduction in the correlation while the tenth run shows a continuous decline in the correlation. Figure \eqref{fig:6.141} and Figure \eqref{fig:6.151} show the density of the marginal posterior distributions for both parameter $\lambda$ and $\tau$ respectively. In each plot, the density summarizes the posterior sample. The thick horizontal bar at the bottom shows the posterior $95\%$ credible interval and the maximum value is the posterior mean. The posterior estimate for $\lambda$ is $\hat{\lambda} = 0.028$, with the credible interval of $(0.0275, 0.0285)$. The posterior estimate for $\tau$ is $\hat{\tau} = 158.48$ with the credible interval of $(150, 170)$.
\subsubsection{Reconstruction result for Mouse Data}
Now, the reconstruction result for the mouse injected with $100 \mu 1/7.5$ MBq $[^{99m}Tc]$ MDP, to X-ray the bone imaging is presented in Figure \eqref{fig:6.16} and Figure \eqref{fig:6.17} for splitting EP and MCMC respectively. Figure \eqref{fig:6.16}(a) shows the estimate of the true image reconstructed from the noise by EP-ADMM. It can be seen that the image has a clear background with slightly rough edges on the image. The residual error shown in Figure \eqref{fig:6.16}(b) comparing the estimates with the data, shows no randomness. Similar comment can be made about the residual shown in Figure \eqref{fig:6.17}(c\&d). It is possible to reduce the patterns in the residual by adjusting the value of $\alpha_\lambda$ in the hyperparameter $\lambda$. This may lead to introducing another noise into the posterior mean estimate. Figure \eqref{fig:6.16}(c) and Figure \eqref{fig:6.16}(d) show the estimates of the parameter which converges to around $165.44$ and $0.019$ produced by EP-MCMC. Figure \eqref{fig:6.17} on the other hand, shows a result produced by MCMC which is seen to have slightly similar results with EP-ADMM with a difference of blurry background. This might be as a result of long chain needed to reach convergence. Also, in Figure \eqref{fig:6.17}(e), it can be seen that the samples are highly correlated even at $1500$ iteration. To reduce the correlation, the number of $n$ samples to thin might be increased that is, the number of samples to be discarded for each kept sample.
Figure \eqref{fig:6.16}(c) shows the trace plot for the parameter $\tau$. It can be seen that the acceptance rate is quite low, this might be due to the choice of proposal distribution. Also, in Figure \eqref{fig:6.17}(e) the $\sigma_\epsilon$ chains produced by MCMC have high correlation which doesn't show any sign of reduction. On the contrary, the observations $\sigma_x$ of MCMC exhibit independence in the chains. However, all the parameters from EP-MCMC and MCMC converge to a very similar value with slight differences. We further establish the convergence of the $\tau$ and $\lambda$ for EP-MCMC.
\begin{figure}[H]
	\centering
	\begin{minipage}[b]{0.45\textwidth}
		\includegraphics[width = 2.5in]{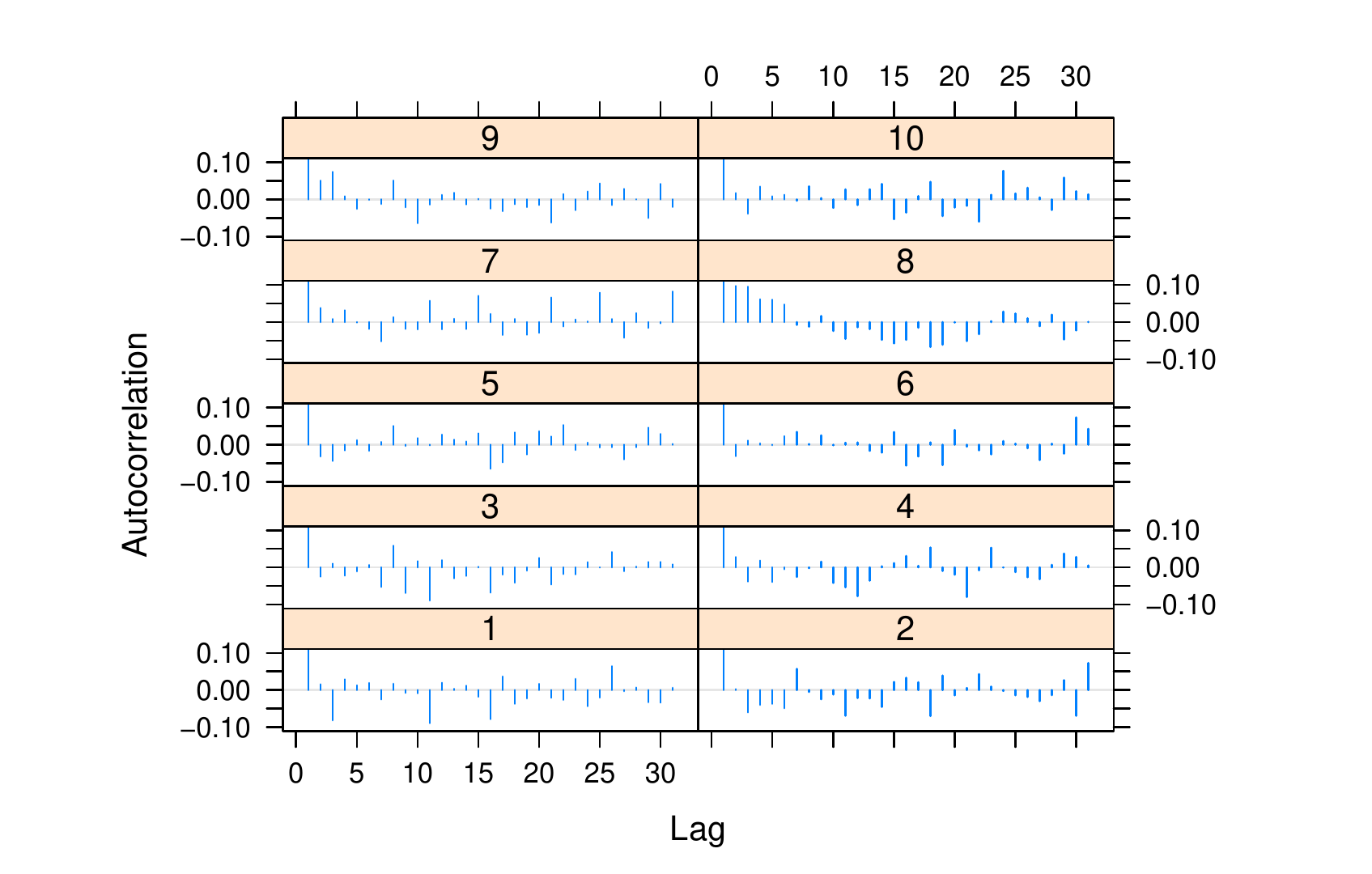}
		\centering\caption[The Autocorrelation plot of $\lambda$ for EP-MCMC algorithm on Mdp data]{The Autocorrelation plot of the mouse injected with Mdp reagent; the EP-MCMC algorithm was run $10$ time each of length $1000$. The plot shows independence of chains at each replication.}
		\label{fig:6.18}
	\end{minipage}
	\hfil
	\begin{minipage}[b]{0.45\textwidth}
		\includegraphics[width = 2.5in]{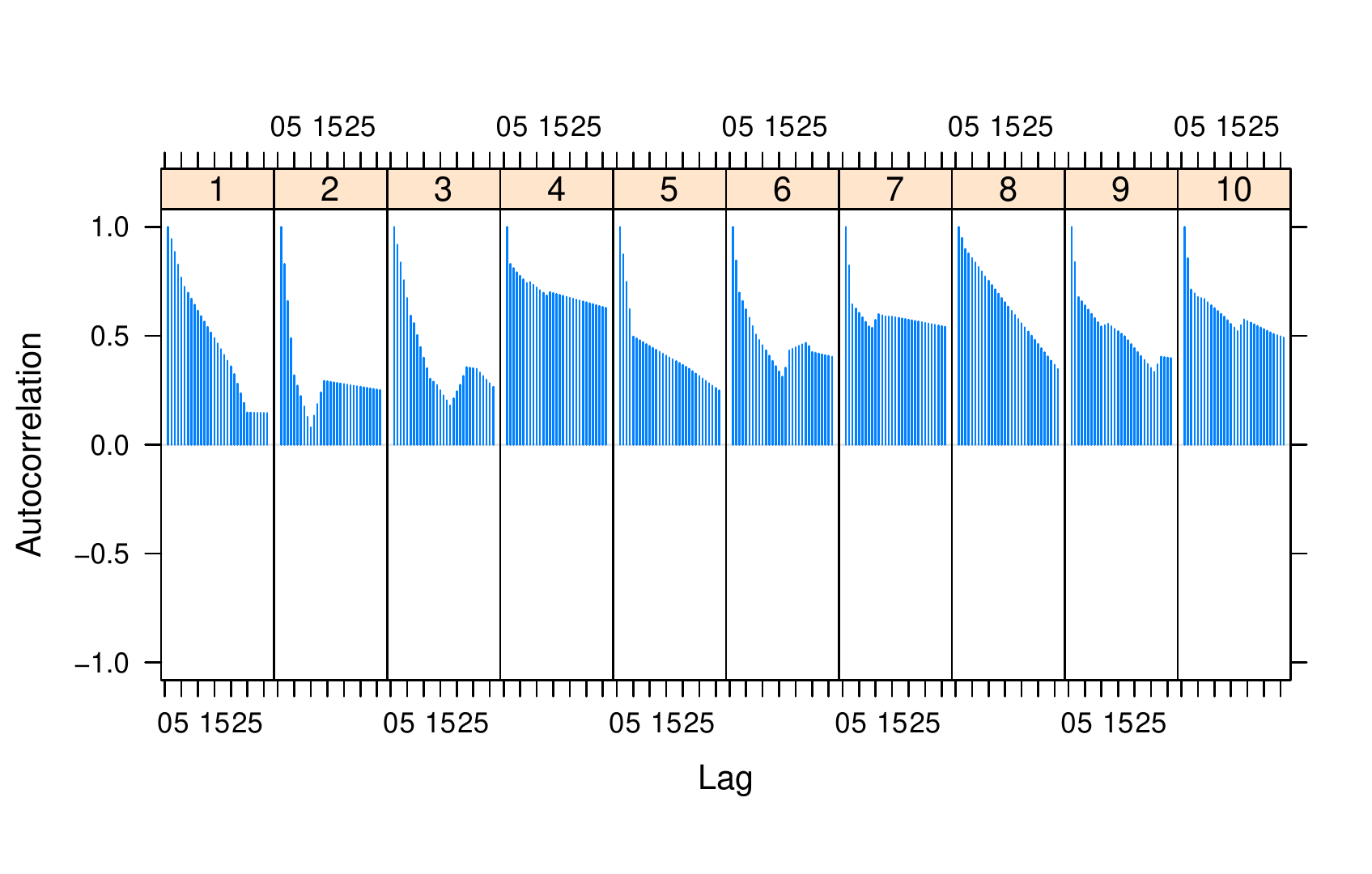}
		\caption{The plot shows a reduction in dependence of chains at first three replications but at the fifth and ninth there is no correlation and the last replication shows a decline in correlation.} 
		\label{fig:6.19}
	\end{minipage}
\end{figure}
\bigskip
\begin{figure}[H]
	\centering
	\begin{minipage}[b]{0.45\textwidth}
		\includegraphics[width = 2.5in]{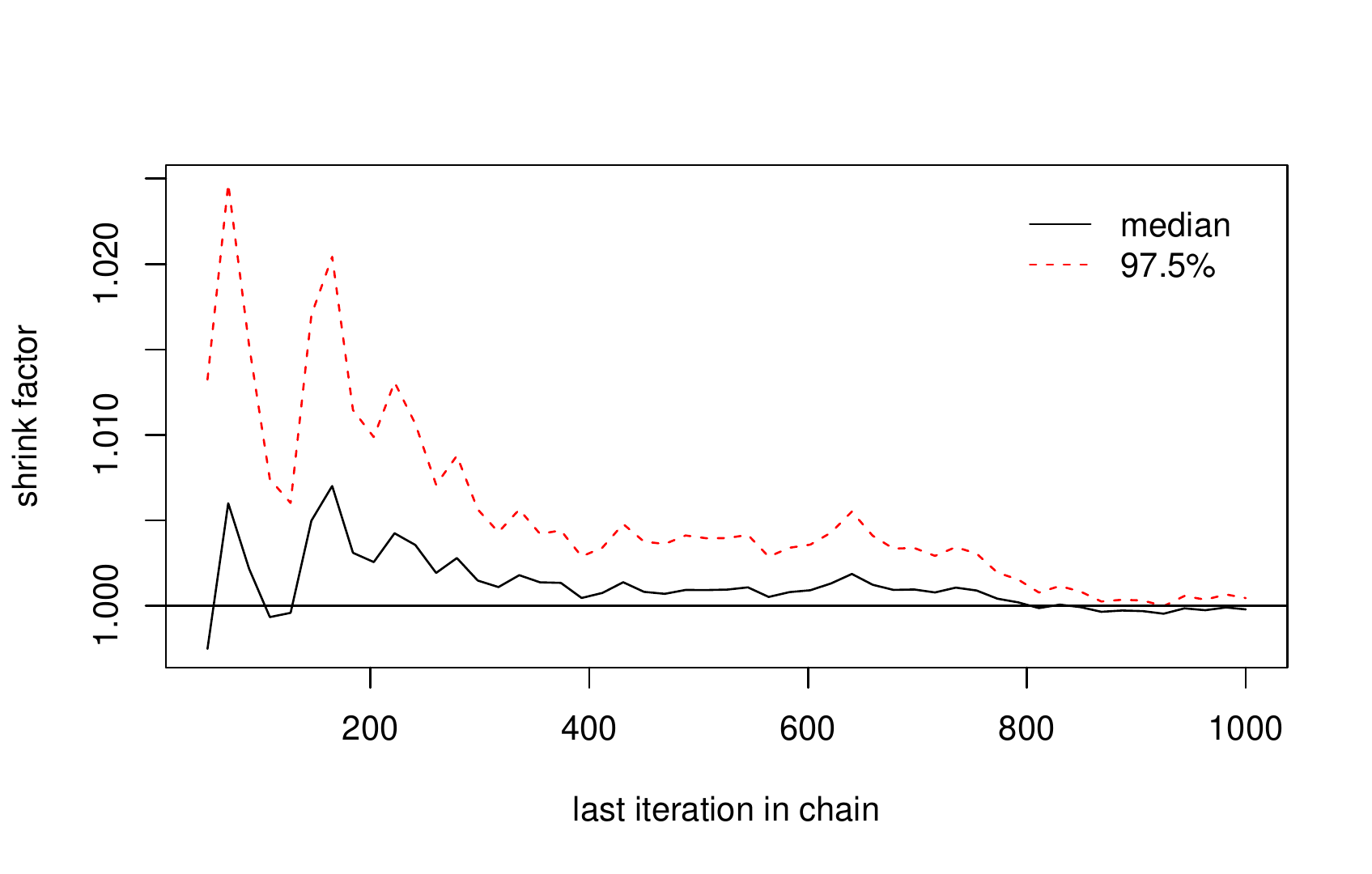}
		\centering\caption{Iterative PSRF Plot for $\lambda$ in Mouse image data (from $m = 10$ parallel sequence and $n = 1000$). About $800$ iterations, the convergence starts till the end of the iteration.}
		\label{fig:6.20}
	\end{minipage}
	\hfil
	\begin{minipage}[b]{0.45\textwidth}
		\includegraphics[width = 2.5in]{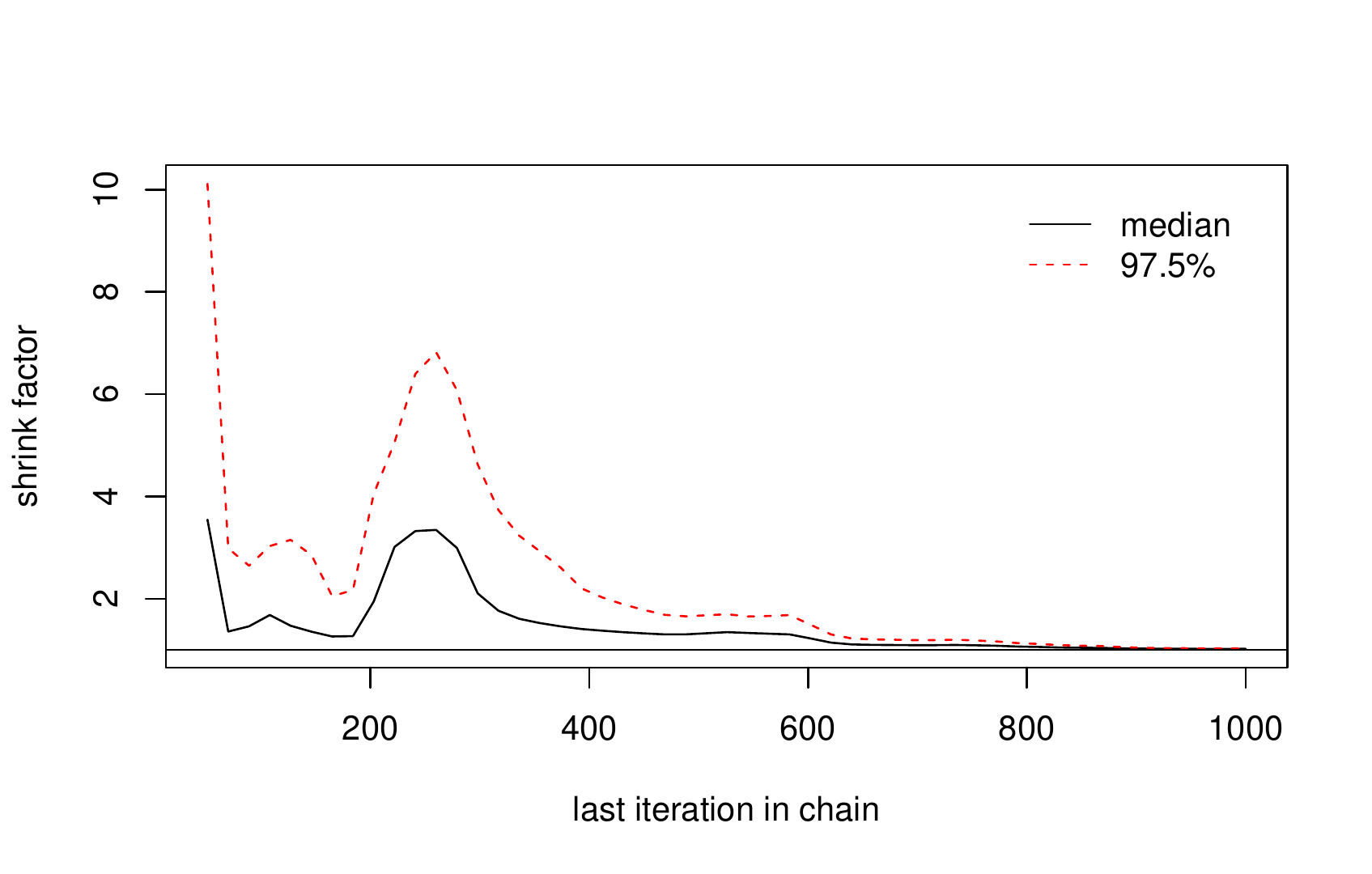}
		\caption[Iterative psrf Plot for $\tau$ in MDP data]{Iterative PSRF Plot for $\tau$ in Mouse image data (from $m = 10$ parallel sequence and $n = 1000$). About $600$ iterations, the convergence starts till the end of the iteration} 
		\label{fig:6.21}
	\end{minipage}
\end{figure}
\bigskip
\begin{table}[H]
	\centering\caption{The mean estimate of the ten EP-MCMC chains for prior $\tau$ and $\lambda$ of mouse data}
	\begin{tabular}{rrrrrrrrr}
		\hline
		parameter&Posterior-Mean&$\hat{R}$&$\hat{V}$&$W$&$B/n$&$\sigma_+^2$&\\
		\hline
		$\tau$&$165.44$&$1.02$&$56.84$&$56.58$&$0.00031$&$56.83$&\\
		$\lambda$&$0.019$&$1.00$&$1.21e^{-08}$&$1.21e^{-08}$&$1.14e^{-11}$&$1.21e^{-08}$&\\
		\hline
	\end{tabular}
	\label{tab:6.3}
\end{table}
\bigskip
\noindent Figure \eqref{fig:6.20} and Figure \eqref{fig:6.21} give a visual interpretation of the values presented. The shrinking factor for $\lambda$ in Figure \eqref{fig:6.21} was above $1$ at the initial stage of the iteration, but on getting to around $800$ iterations the equality between the between-chain and within-chain becomes so evident. This indicates that $1000$ iterations is sufficient to provide convergence for parameter $\lambda$. Likewise, the shrinking factor for the $\tau$ at the first $100$ iterations is $10$ which means the chains still need a lot of iterations to bring reduction in the shrinking factor. At about $600$ iterations, the shrinking factor is on the acceptable line of $1$.
\noindent Table \eqref{tab:6.3} shows some statistical inference to establish the convergence properties of the EP-MCMC parameters. The potential scale reduction faction of $\tau$ is $1.02$ which is still within the acceptable range of $1$. This can be interpreted as the chain represents the target distribution of the parameter $\tau$. Also, the PSRF for $\lambda$ is exactly $1$. The within-chain, between-chain variances, and pooled variance for $\tau$ are $W = 56.58$, $B/n = 3.1\times 10^{-4}$, and $\hat{V} = 56.84$ respectively. Similarly, the within-chain, between-chain variances, and pooled variance for $\lambda$ are $W = 1.21\times 10^{-8}$, $B/n = 1.14\times 10^{-11}$, and $\hat{V} = 1.21\times 10^{-8}$ respectively. It can be seen that according to the interpretation by \cite{gelmanandrubin1992a}, the equality of pooled variance $\hat{V}$ and $W$ for both parameters indicates the existence of convergence. 
\begin{figure}[H]
	\centering
	\begin{minipage}[b]{0.45\textwidth}
		\includegraphics[width = 2in, height=2in]{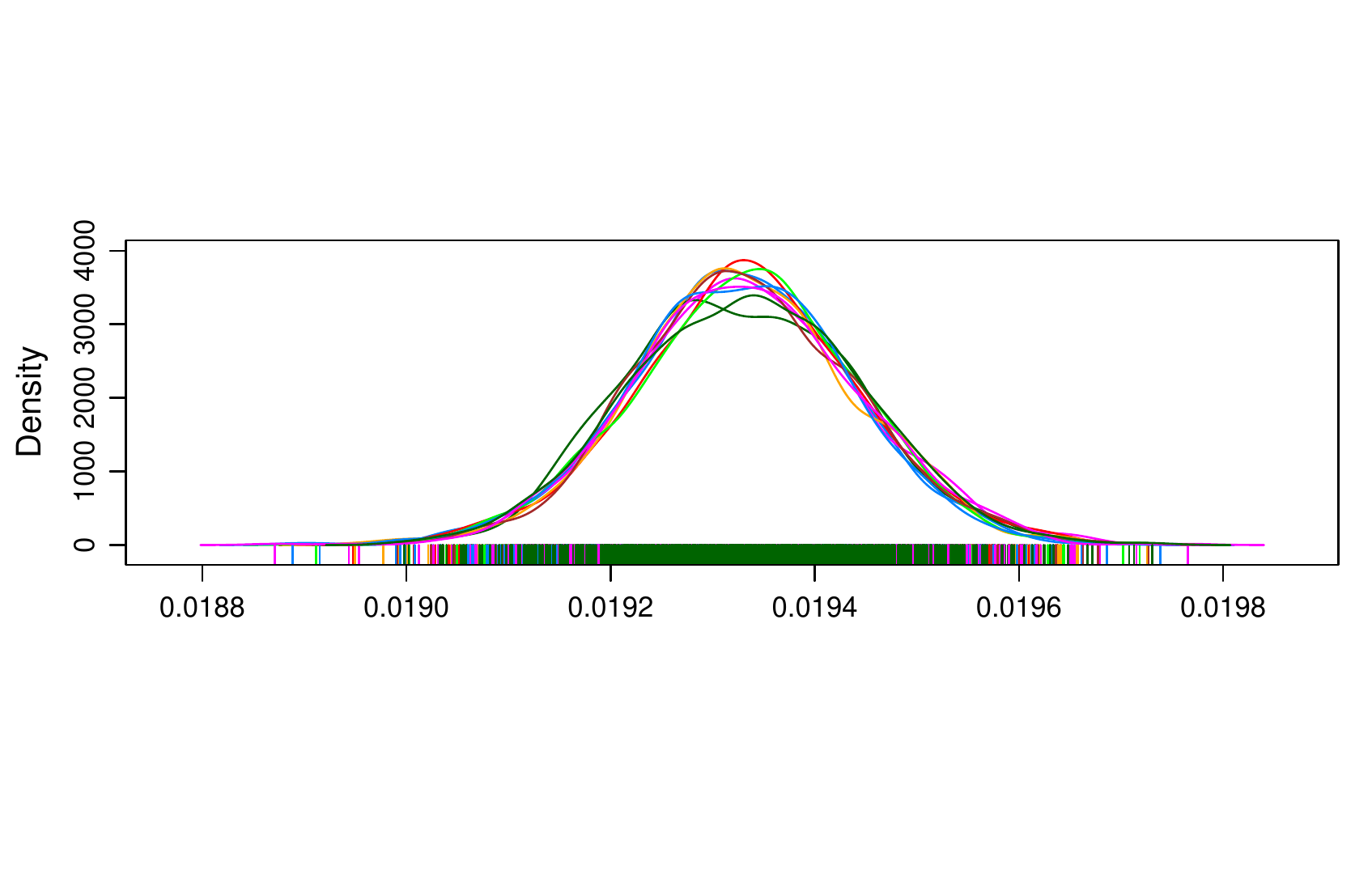}
		\centering\caption[Marginal posterior distribution of the parameter lambda for Mouse data]{Marginal posterior distribution of the parameter $\lambda$ for Mouse data. The tick horizontal bar signifies the posterior $95\%$ credible interval and the maximum value at $0.019$ is the posterior mean.}
		\label{fig:6.201}
	\end{minipage}
	\hfil
	\begin{minipage}[b]{0.45\textwidth}
		\includegraphics[width = 2.5in, height=2in]{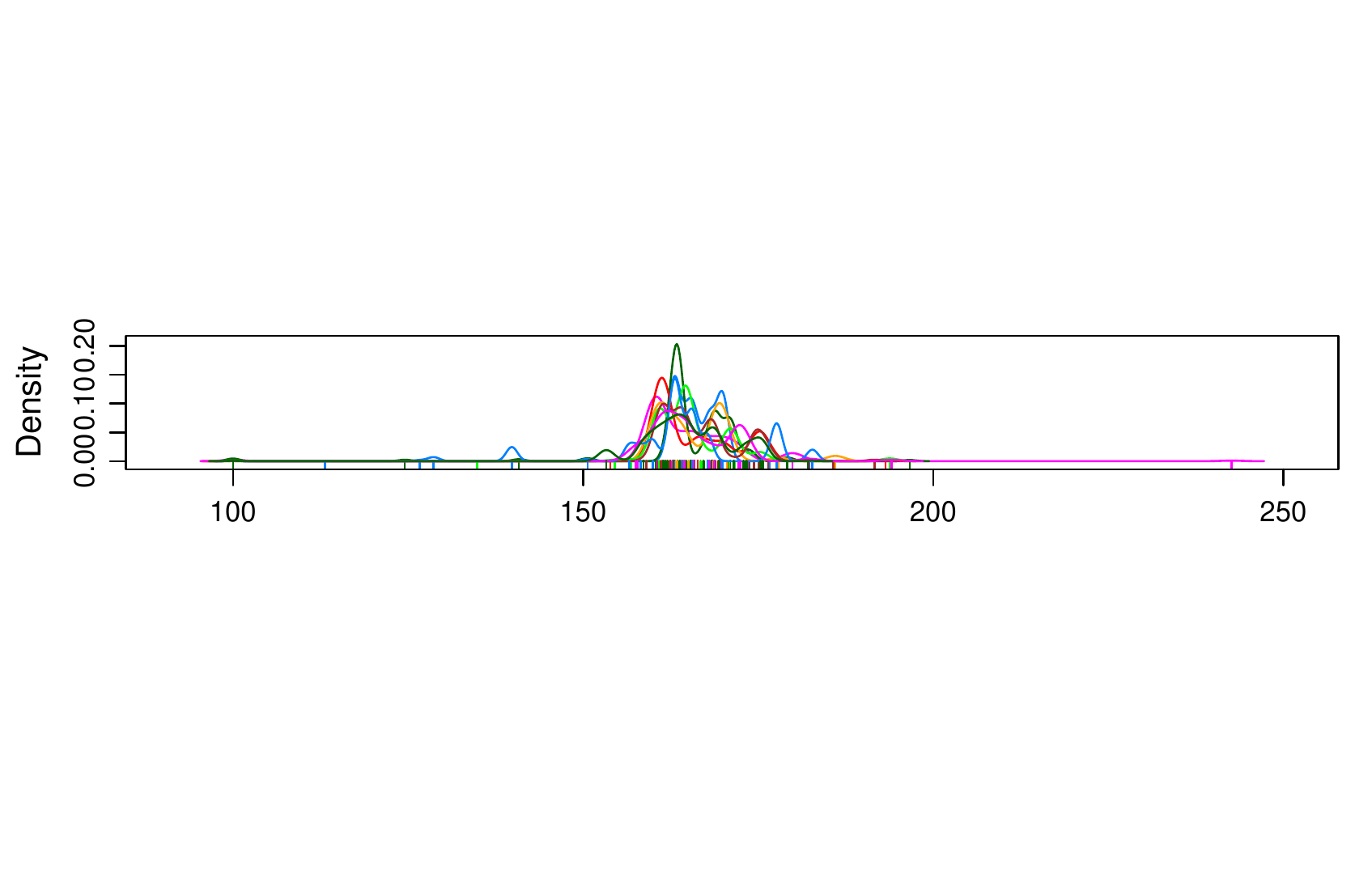}
		\centering\caption[Marginal posterior distribution of the parameter tau for Mouse data]{Marginal posterior distribution of the parameter $\tau$ for Mouse data. The tick horizontal bar signifies the posterior $95\%$ credible interval and the maximum value at about $166$ is the posterior mean.}
		\label{fig:6.211}
	\end{minipage}
\end{figure}
\begin{figure}[H]
	\centering
	\begin{minipage}[b]{0.45\textwidth}
		\includegraphics[width = 3in, height=2in]{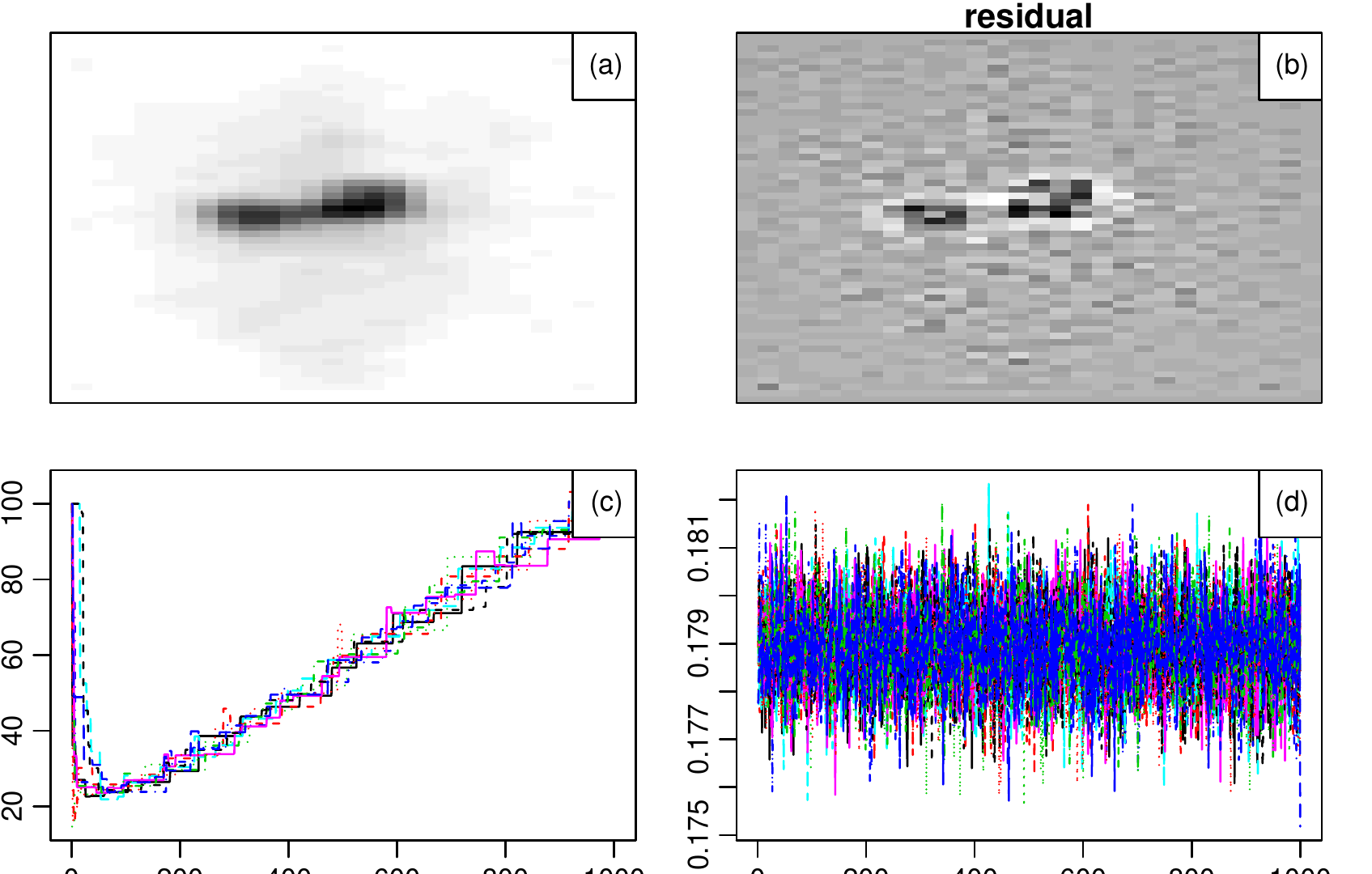}
		\centering\caption[DMSA results of SSEP, relative error, and the estimate of $\tau$, $\lambda$]{(a): dmsa\_1h\_mouse\_results from SEP, (b): relative error (c): estimate $\tau$ is $58.9$, (d): estimate $\lambda$ is $0.179$} 
		\label{fig:6.22}
	\end{minipage}
	\hfil
	\begin{minipage}[b]{0.45\textwidth}
		\includegraphics[width = 2.5in, height=1.98in]{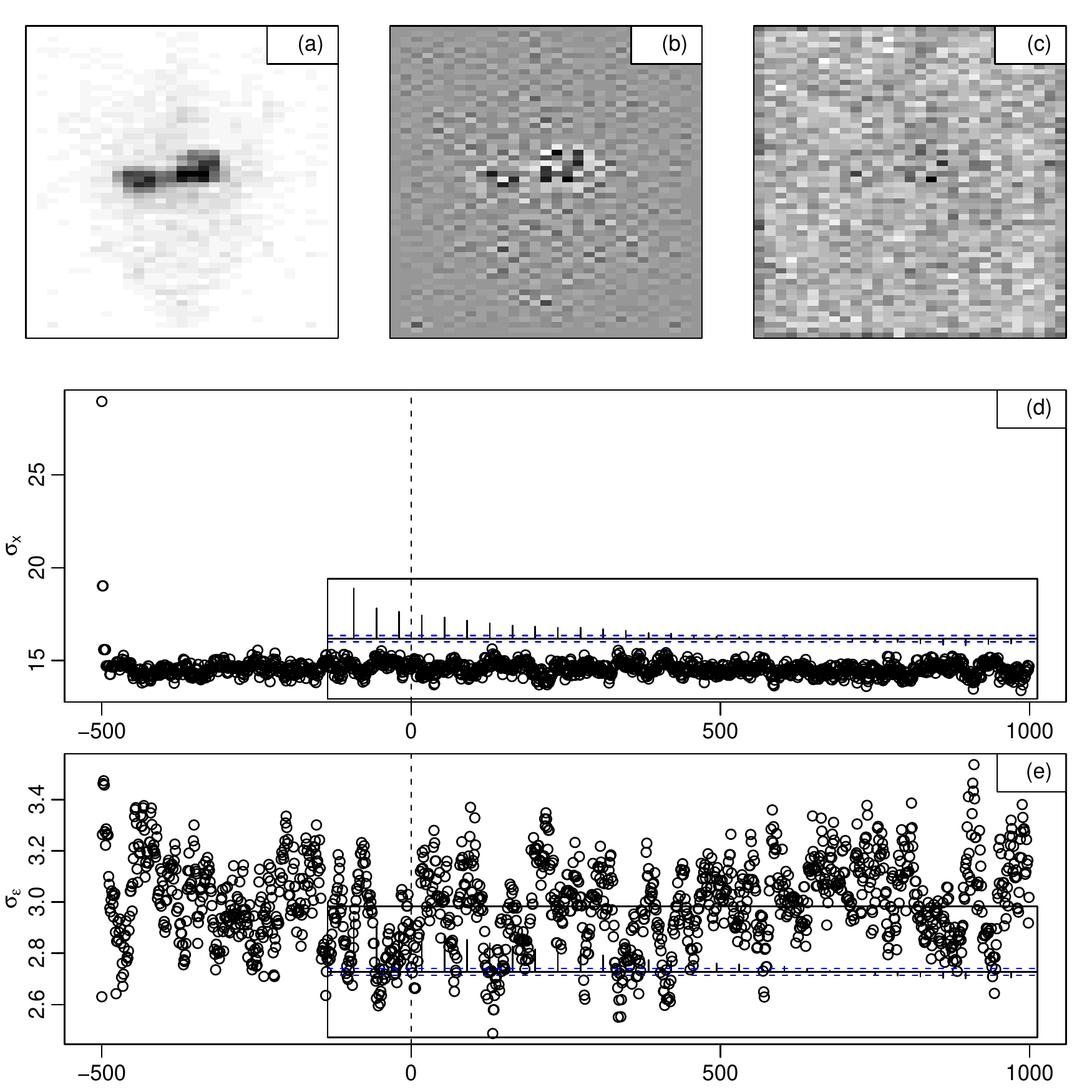}
		\caption[The reconstruction of DMSA image by MCMC]{(a): MCMC Dmsa (b)\&(c): error \& residual, (d)\&(e): the estimate of $\sigma_x$ \& $\sigma_\epsilon$ converges at $10$ and $3.0$}
		\label{fig:6.23}
	\end{minipage}
\end{figure}
\bigskip
\par \noindent Figure \eqref{fig:6.201} and Figure \eqref{fig:6.211} show the density of the marginal posterior distributions for both parameter $\lambda$ and $\tau$ respectively. In each plot, the density summarizes the posterior sample. The thick horizontal bar at the bottom shows the posterior $95\%$ credible interval and the maximum value is the posterior mean. The posterior estimate for $\lambda$ is $\hat{\lambda} = 0.019$, with the credible interval of $(0.01885, 0.0198)$. The posterior estimate for tau is $\hat{\tau} = 165.44$ with the credible interval of $(100, 250)$.
\subsubsection{Reconstruction result for DMSA Data}
\begin{figure}[H]
	\centering
	\begin{minipage}[b]{0.45\textwidth}
		\includegraphics[width=2.5in, height=2in]{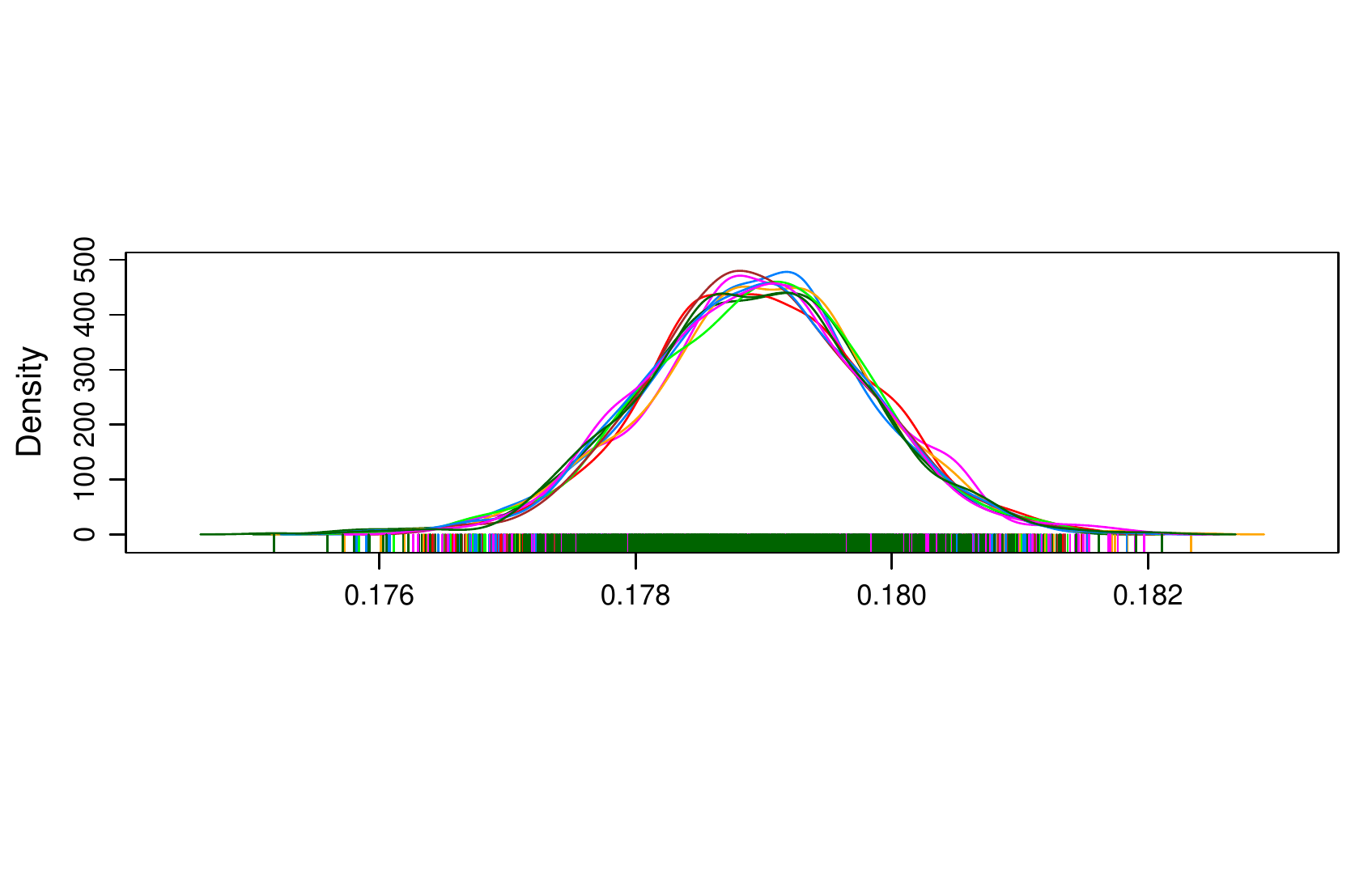}
		\centering\caption[Marginal posterior distribution of the parameter lambda for DMSA data]{Marginal posterior distribution of the parameter $\lambda$ for DMSA data. The tick horizontal bar signifies the posterior $95\%$ credible interval and the maximum value at $0.179$ is the posterior mean.}\label{fig:6.241}
	\end{minipage}
	\hfil
	\begin{minipage}[b]{0.45\textwidth}
		\includegraphics[width=2.5in]{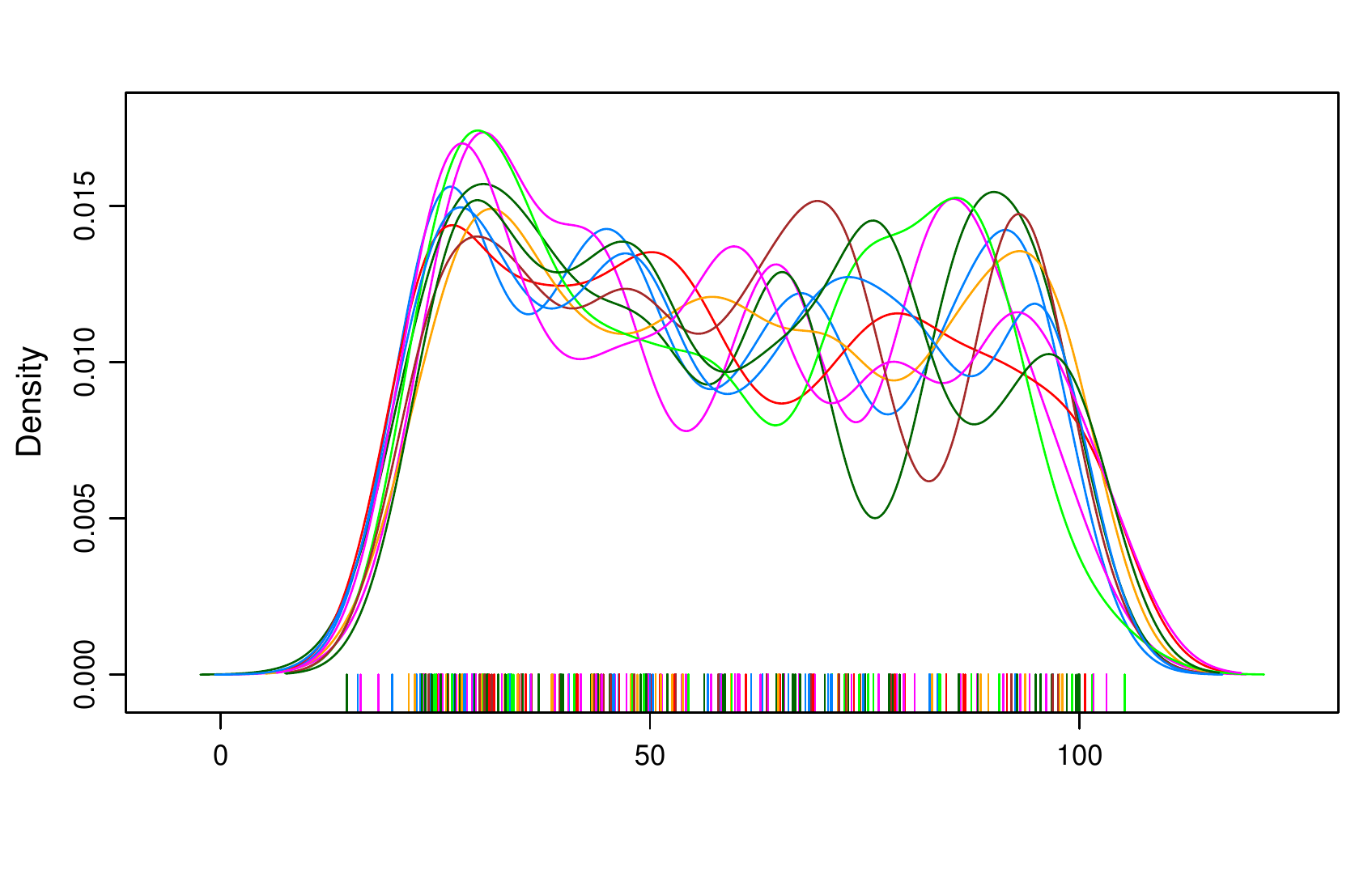}
		\centering\caption[Marginal posterior distribution of the parameter tau for DMSA data]{Marginal posterior distribution of the parameter $\tau$ for DMSA data. The tick horizontal bar signifies the posterior $95\%$ credible interval and the maximum value at about $59$ is the posterior mean.}	
		\label{fig:6.251}
	\end{minipage}
\end{figure}

Now, moving to the reconstruction of the mouse injected with DMSA. The results produced by splitting EP and MCMC are quite similar in Figure \eqref{fig:6.22}(a), Figure \eqref{fig:6.22}(b), Figure \eqref{fig:6.23}(a), and Figure \eqref{fig:6.23}(b) respectively. Given that we require clear and smooth image reconstruction, EP-ADMM produces a smooth image reconstruction with a clear and white background. In contrast, MCMC produces a white and clear background with sharp surface coupled with slightly rough edges. The residual errors shown in Figure \eqref{fig:6.22}(b) and Figure \eqref{fig:6.23}b comparing the estimates with the data, show similar pattern. Adjusting the value of $\alpha_\lambda$ and $\alpha_\tau$ in the hyperparameters $\lambda$ and $\tau$ respectively may produce random errors. In Figure \eqref{fig:6.22}(d), the mean estimate of the posterior for $\lambda$ across the $10$ replication is $0.179$, which can be seen in the trace-plot. Figure \eqref{fig:6.22}(c) does not show any pattern of convergence even after $1000$ iterations. This however may be due to the choice of proposal distribution. Figure \eqref{fig:6.241} and Figure \eqref{fig:6.251} show the density of the marginal posterior distributions for both parameter $\lambda$ and $\tau$ respectively. In each plot, the density summarizes the posterior sample.
 \begin{figure}[H]
	\centering
	\begin{minipage}[b]{0.45\textwidth}
		\includegraphics[width = 2.5in, height=1.5in]{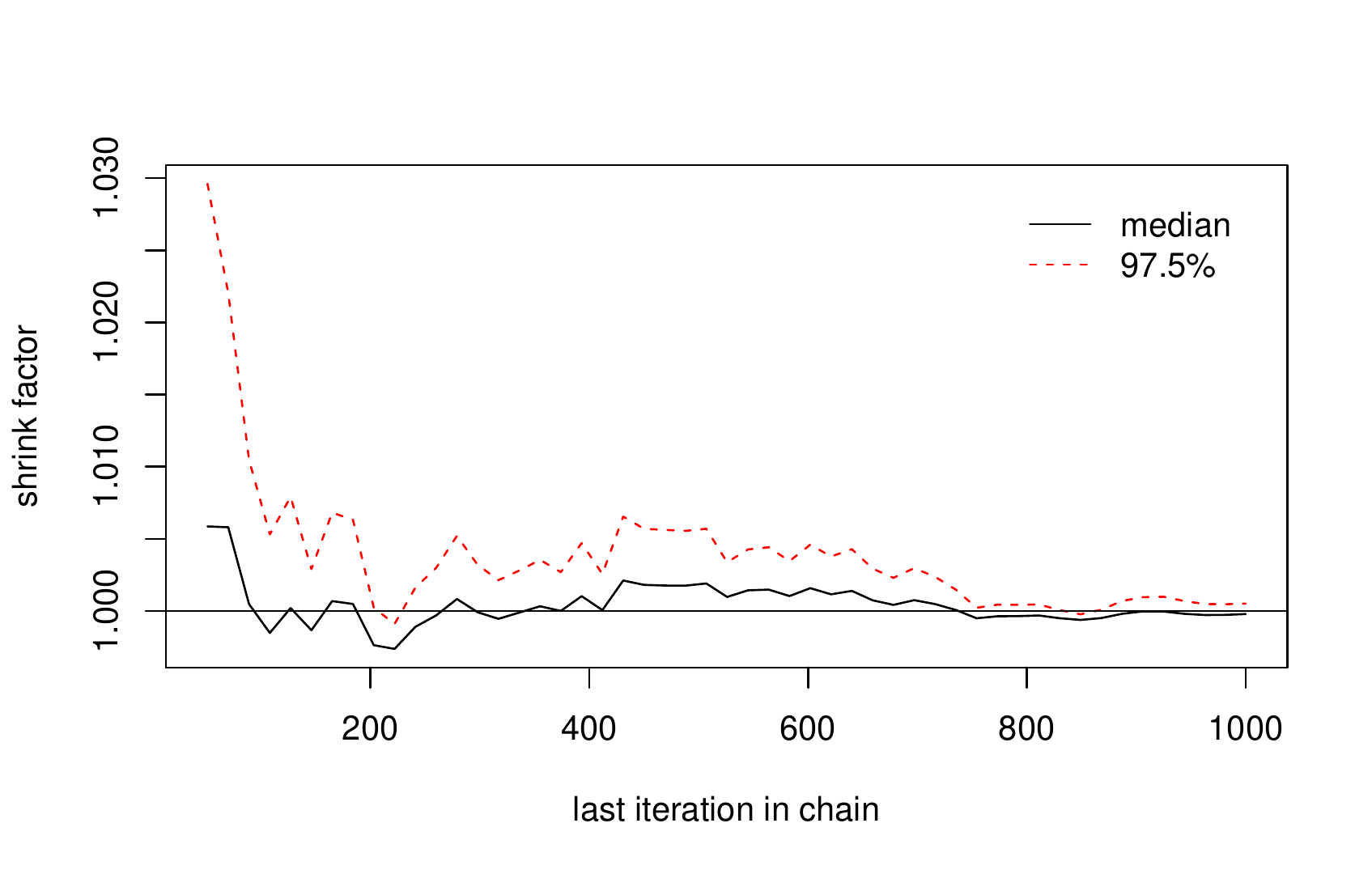}
		\centering\caption[Iterative psrf Plot for $\lambda$ on DMSA data]{Iterative PSRF Plot for $\lambda$ in Mouse image data (from $m = 10$ parallel sequence and $n = 1000$). After about $800$ iteration convergence starts till the end of the iteration.}
		\label{fig:24}
	\end{minipage}
	\hfil
	\begin{minipage}[b]{0.45\textwidth}
		\includegraphics[width = 2.5in, height=1.5in]{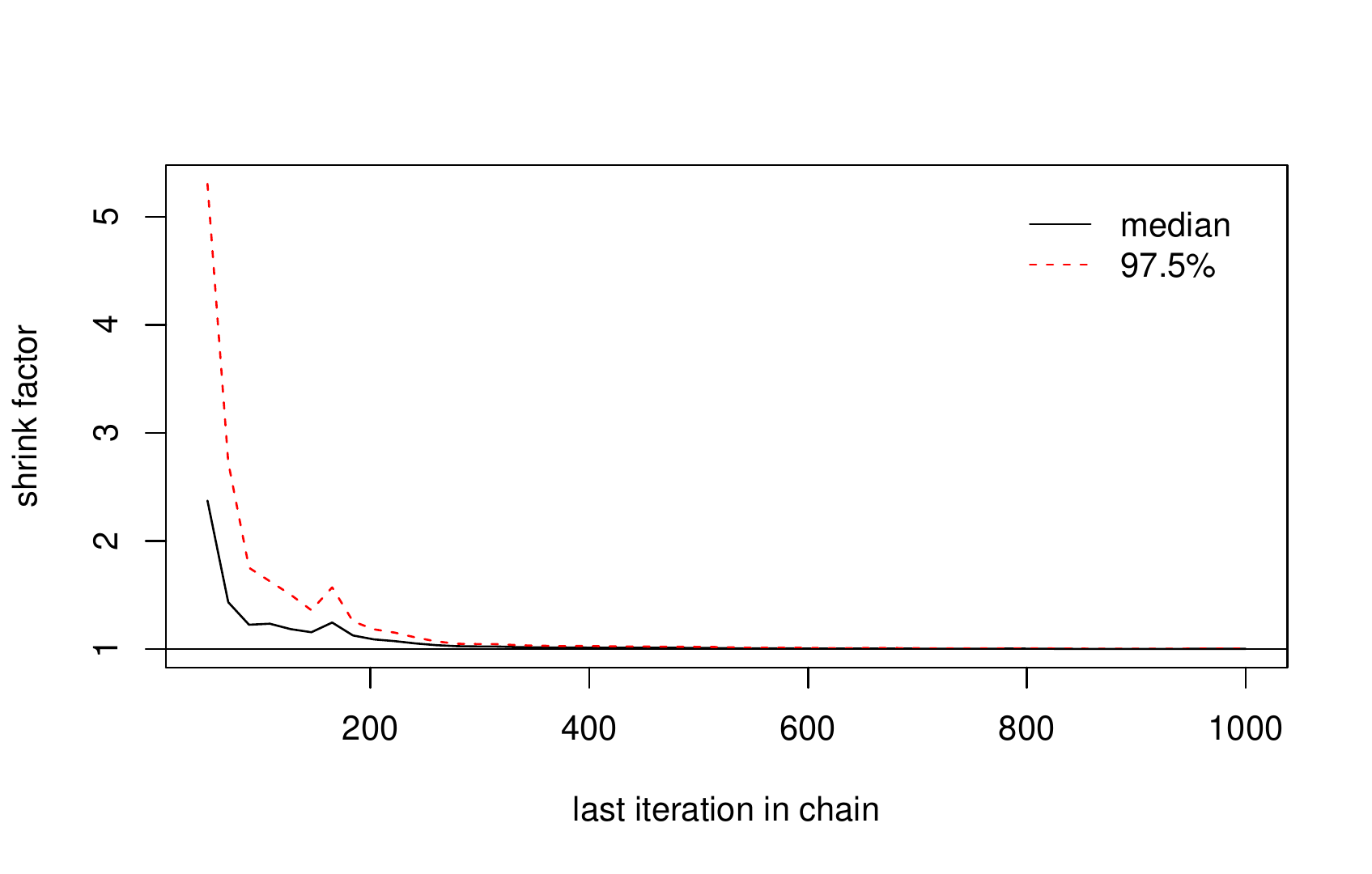}
		\caption[Iterative psrf Plot for $\tau$ on DMSA data]{Iterative PSRF Plot for $\tau$ in Mouse image data (from $m = 10$ parallel sequence and $n = 1000$). About $200$ iteration convergence starts till the end of the iteration.} 
		\label{fig:6.25}
	\end{minipage}
\end{figure}
\bigskip
\begin{table}[H]
	\centering\caption{The mean estimate of the ten EP-MCMC chains for prior $\tau$ and $\lambda$\\ \centering of DMSA data}
	\begin{tabular}{rrrrrrrrr}
		\hline
		parameter&Posterior-Mean&$\hat{R}$&$\hat{V}$&$W$&$B/n$&$\sigma_+^2$&\\
		\hline
		$\tau$&$58.82$&$1.00$&$59.89$&$59.87$&$0.75$&$59.89$&\\
		$\lambda$&$0.179$&$1.00$&$7.60e^{-07}$&$7.60e^{-07}$&$8.93e^{-10}$&$7.60e^{-07}$&\\
		\hline
	\end{tabular}
	\label{tab:3}
\end{table}
\bigskip
\noindent The thick horizontal bar at the bottom shows the posterior $95\%$ credible interval and the maximum value is the posterior mean. The posterior estimate for $\lambda$ is $\hat{\lambda} = 0.179$, with the credible interval of $(0.176, 0.182)$. The posterior estimate for tau is $\hat{\tau} = 58.82$ with the credible interval of $(1, 102)$.
\subsubsection{Reconstruction result of four Circles Data}
\begin{figure}[H]
	\begin{center}
		\includegraphics[height=1.5in]{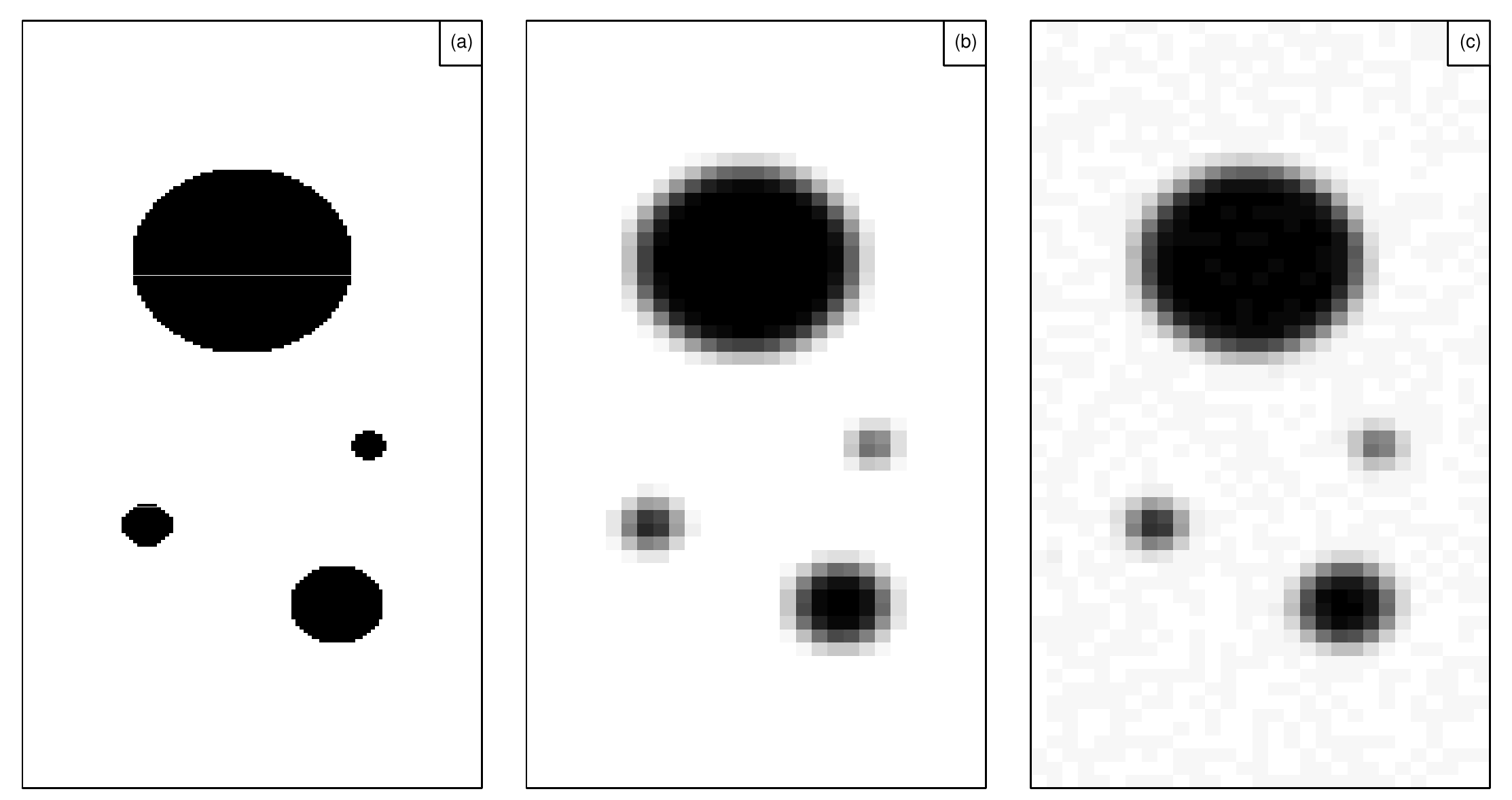}
		\caption[The original image of the circles]{$(a)$ Original image of circles to be reconstructed: $\mathcal{X}$, $(b)$ the mean: $\mathcal{G}\mathcal{X}$ and $(c)$ Noisy data:$\mathcal{Y}$}
		\label{fig:6.91}
	\end{center}
\end{figure}

\begin{figure}[H]
	\centering
	\begin{minipage}[b]{0.45\textwidth}
		\includegraphics[width = 3in, height=2in]{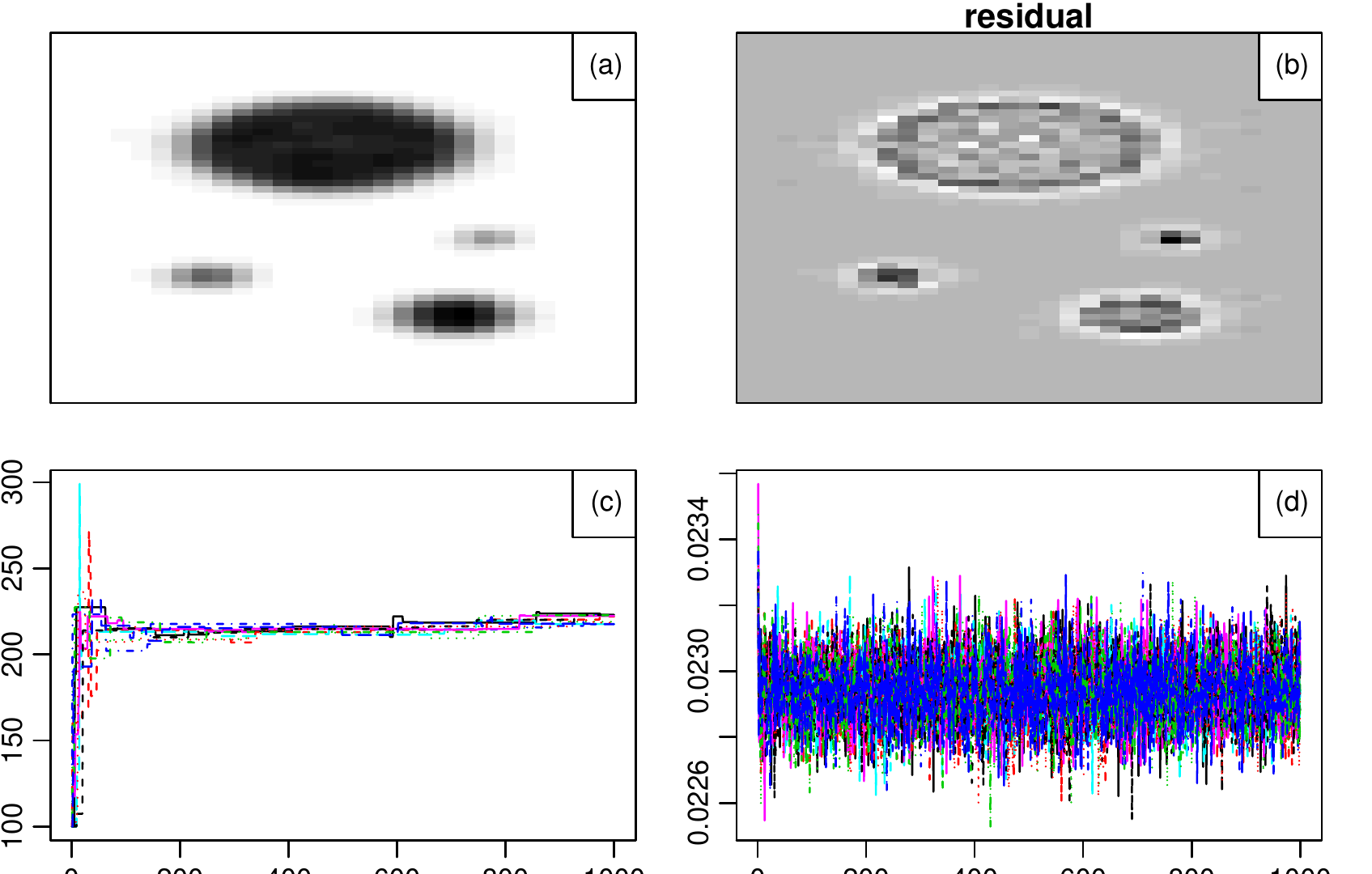}
		\centering\caption[The reconstruction image of four circles by EP-ADMM]{(a): The reconstruction image of four circles by EP-ADMM, (b): relative error (c): estimates $\tau$ converges at $200$ from EP-MCMC, (d): estimates $\lambda$ converges at $0.023$ from EP-MCMC} 
		\label{fig:6.26}
	\end{minipage}
	\hfil
	\begin{minipage}[b]{0.45\textwidth}
		\includegraphics[width = 2.5in, height=1.98in]{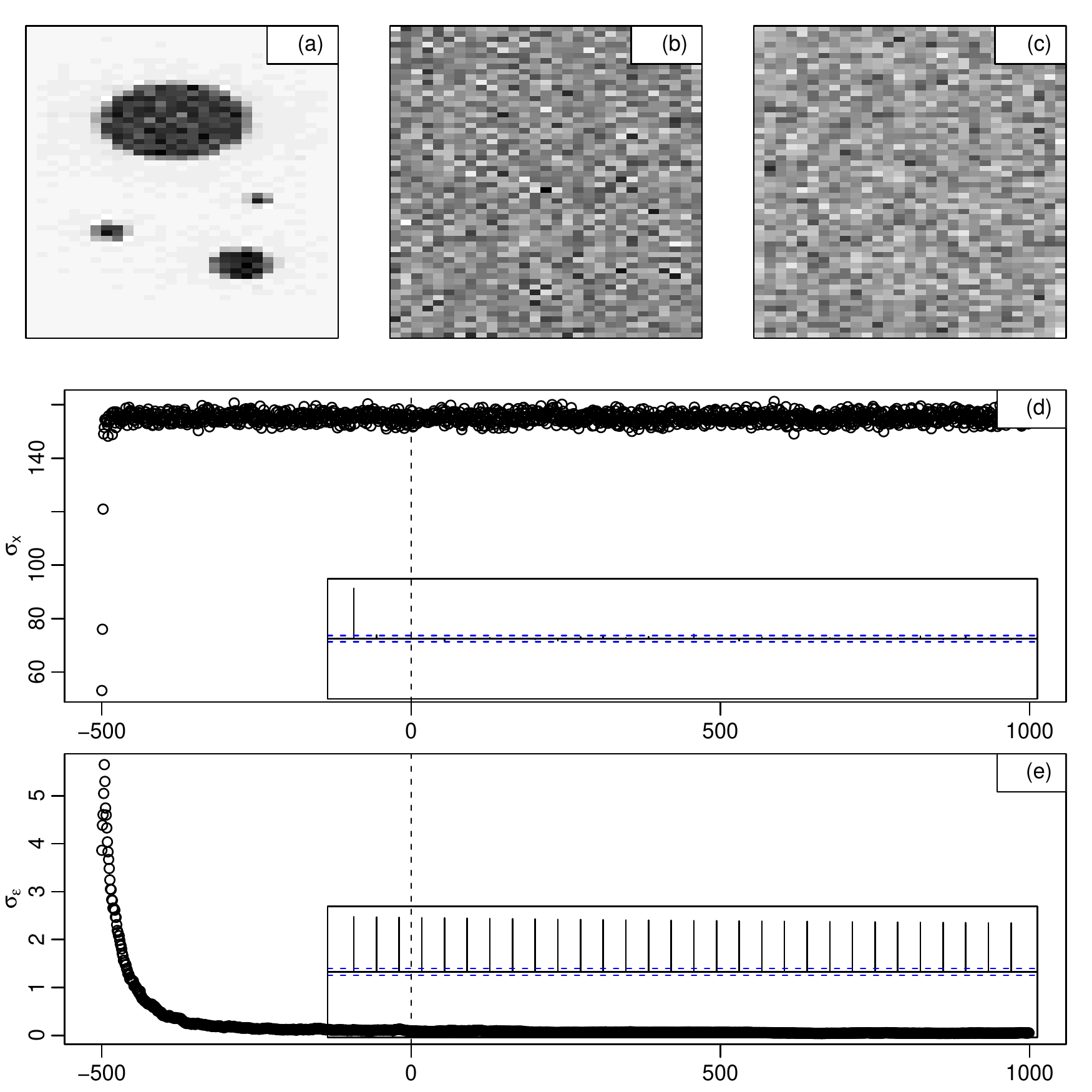}
		\caption[The reconstruction image of four circles by MCMC]{(a): The reconstruction image of four circles produced by MCMC (b)\&(c): error \& residual, (d): the estimate of $\sigma_x$ converges at $150$, (e): the estimates of $\sigma_\epsilon$ converges to about $0.5$} 
		\label{fig:6.27}
	\end{minipage}
\end{figure}

Finally, Figure \eqref{fig:6.91} shows the original image to be reconstructed. Figure \eqref{fig:6.26}(a) presents the reconstruction result for the image of four circles. It can be seen that EP-ADMM produces a clear and sharp reconstruction image, while MCMC produces a scaly and gray background as shown in Figure \eqref{fig:6.27}(a). Comparing the parameter estimates of MCMC with EP-MCMC, the parameter estimates of $\sigma_x$ in MCMC is $160$ while $\tau$ in EP-MCMC is about $200$, similarly the parameter estimates of $\sigma_\epsilon$ is about $0.5$ while that of EP-MCMC is $0.023$. The residual error shown in Figure \eqref{fig:6.26}(b) comparing the estimates with the data, shows a very clear pattern. Random residual as shown in Figure \eqref{fig:6.11}(b\&c) can be produced by using an appropriate values of the hyperparameters $\lambda$ and $\tau$. However, given our primary aim to produce a constant reconstruction, this remains a little concern and it can be further checked in the future. The computed potential scale reduction factors $\hat{R}$ for $\tau$ is $1.1$, upper C.I is $1.12$, and $\lambda$ is $1$ with upper C.I as $1$ are shown in Table \eqref{tab:4}.
\bigskip
\begin{figure}[H]
	\centering
	\begin{minipage}[b]{0.45\textwidth}
		\includegraphics[width = 3in, height=1.5in]{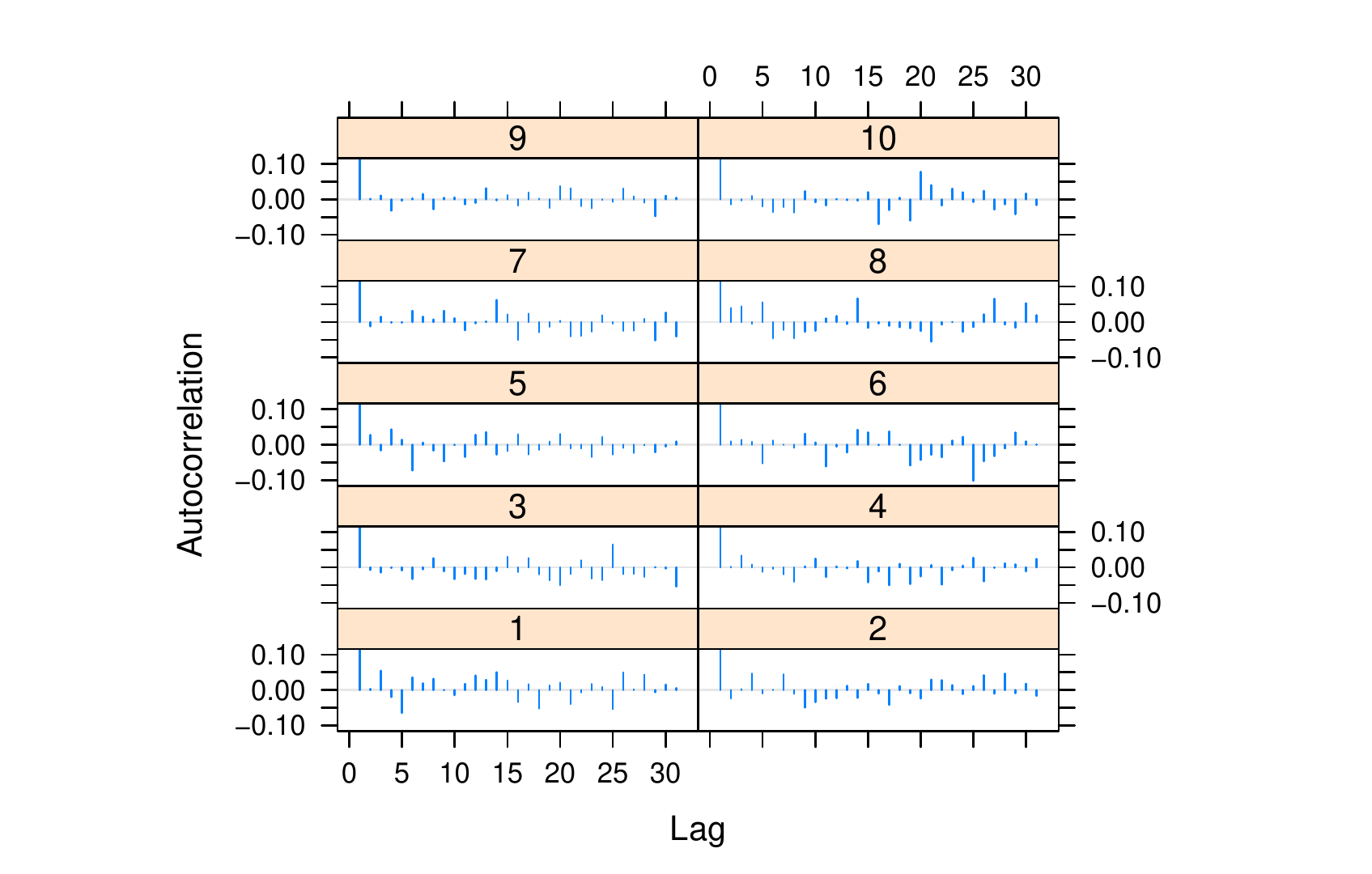}
		\centering\caption[The Autocorrelation plot of EP-MCMC for $\lambda$ on circles image.]{The Autocorrelation plot of the image of four circles; the EP-MCMC algorithm was run $10$ time each of length $1000$. The plot shows independence of chains at each replication.} 
		\label{fig:6.28}
	\end{minipage}
	\hfil
	\begin{minipage}[b]{0.45\textwidth}
		\includegraphics[width = 2.5in, height=1.71in]{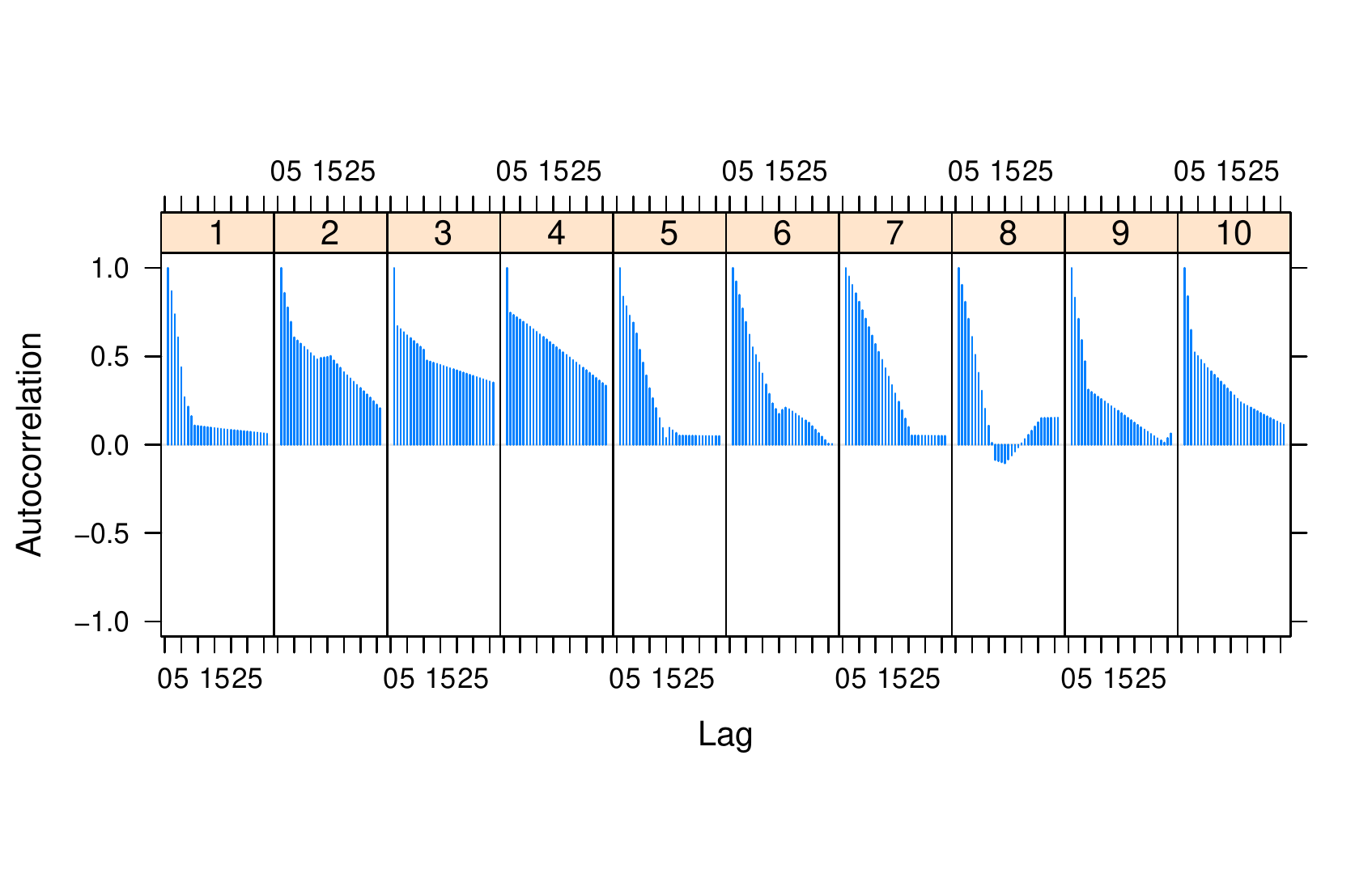}
		\caption[The Autocorrelation plot of EP-MCMC for $\tau$ on circles image.]{The plot shows a strong dependence of chains at first four replications but at the fifth and ninth there is no correlation and the last replication shows a decline in correlation} 
		\label{fig:6.29}
	\end{minipage}
\end{figure}
\bigskip
\par \noindent Here, the potential scale reduction factor for $\lambda$ is exactly $1$ which indicates strong convergence across the $10$ runs. Figure \eqref{fig:6.28} shows the correlation between the chains of $\lambda$ for each run. It can be seen that there is no autocorrection in the chains for each of the runs, this indicates that there is a convergence at every run which also contributes to potential scale reduction factor of $1$. Similarly, Figure \eqref{fig:6.29} shows an autocorrelation plot of $\tau$ for the $10$ runs at the initial stage of the runs. There is a high correlation among the chain of the first four runs, but at the fifth and ninth runs there is sharp reduction in the correlation while the tenth run shows a continuous decline in the correlation.
\begin{figure}[H]
	\centering
	\begin{minipage}[b]{0.45\textwidth}
		\includegraphics[width = 2.5in]{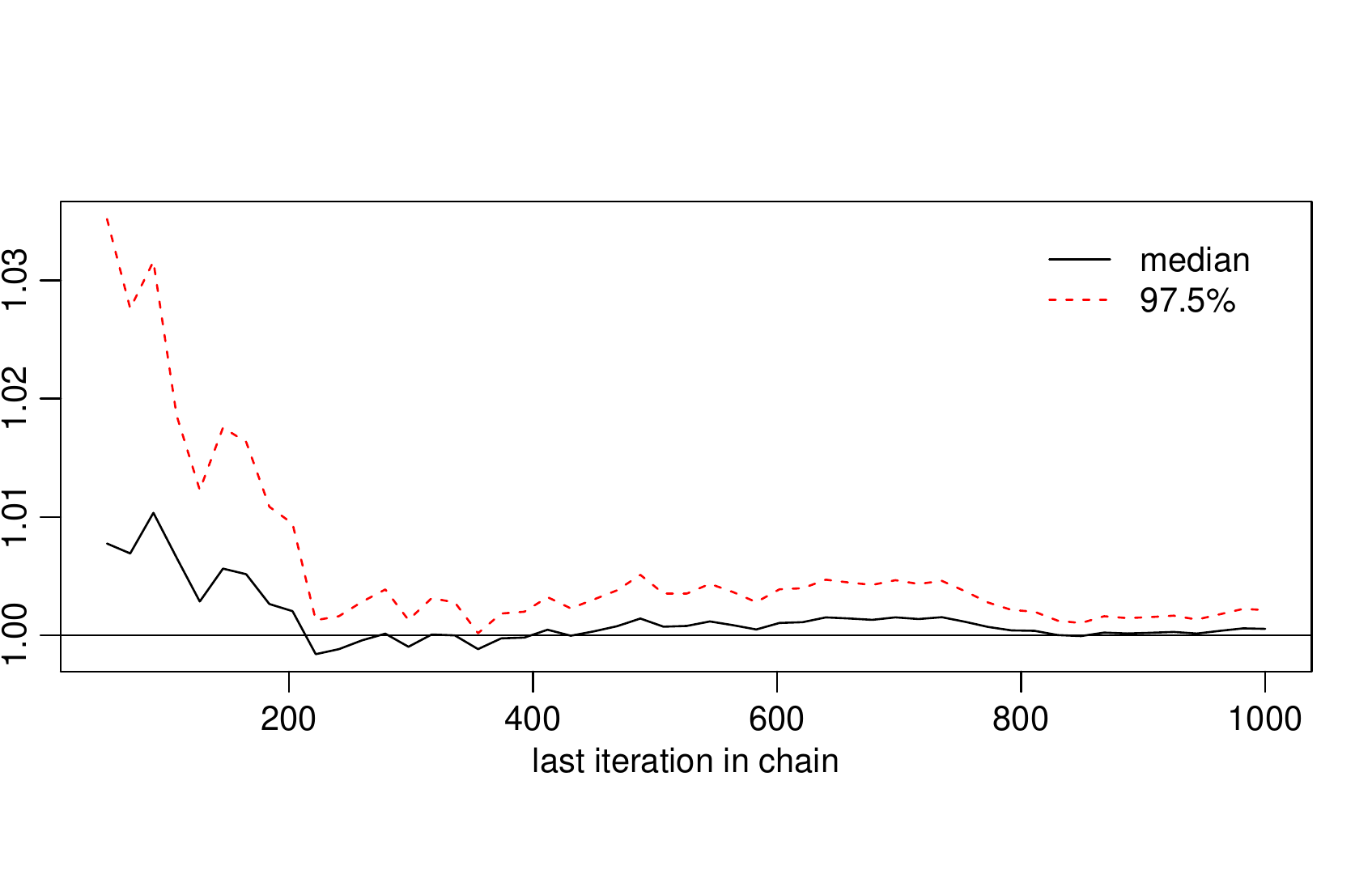}
		\centering\caption{Iterative PSFR Plot for $\lambda$ in image of four circle (from $m = 10$ parallel sequence and $n = 1000$). After about $200$ iteration convergence starts with a bump at the middle but got stabilized around $300$, till the end of the iteration.} 
		\label{fig:30}
	\end{minipage}
	\hfil
	\begin{minipage}[b]{0.45\textwidth}
		\includegraphics[width = 2.5in]{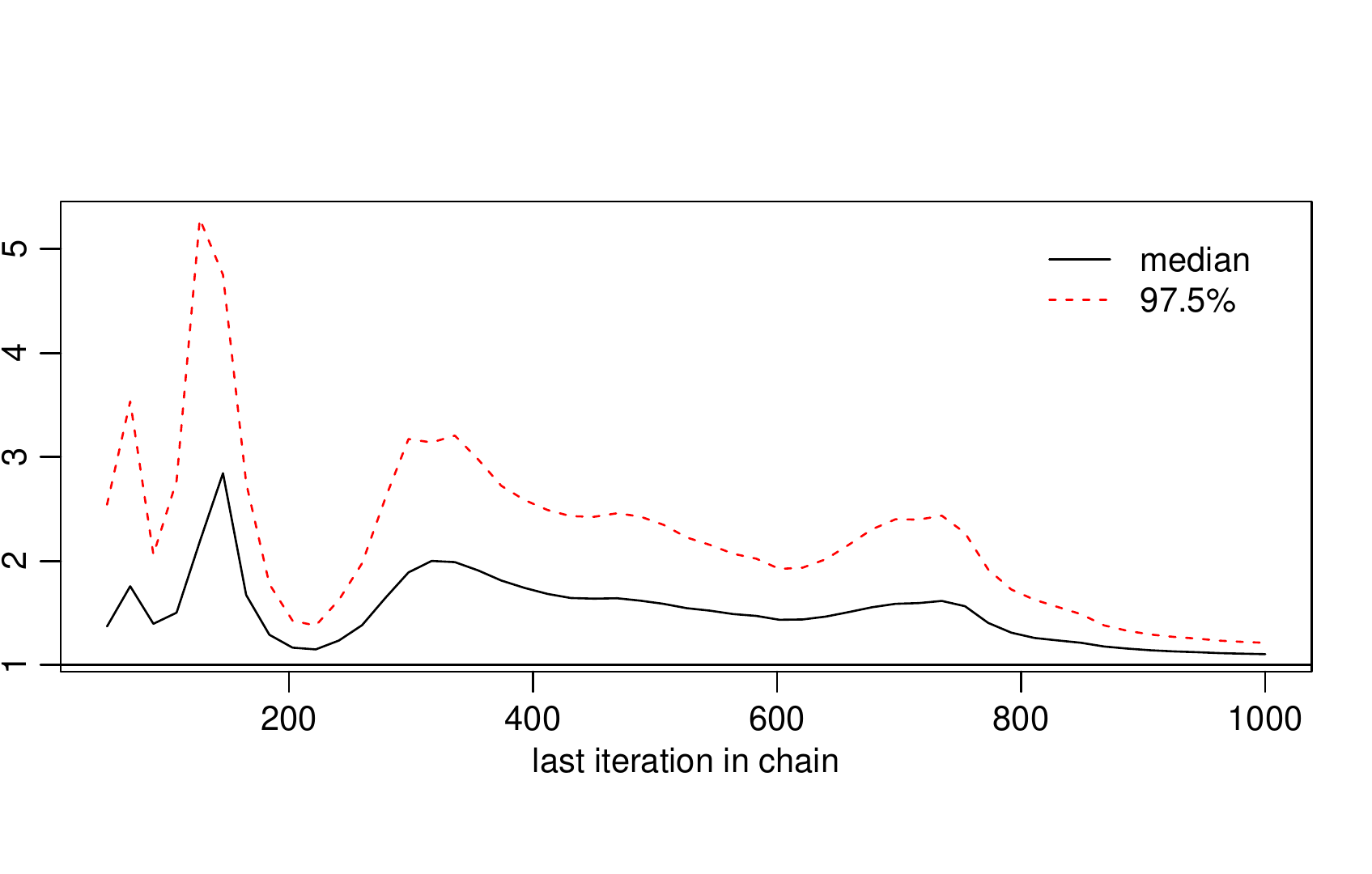}
		\caption{Iterative PSFR Plot for $\tau$ in image of four circles (from $m = 10$ parallel sequence and $n = 1000$). About $200$ iteration convergence started and diverged in the middle but later converged at about $800$ till the end of the iteration.} 
		\label{fig:31}
	\end{minipage}
\end{figure}
\bigskip
\begin{table}[H]
	\centering\caption{The mean estimate of the ten EP-MCMC chains for prior $\tau$ and $\lambda$\\ of Circles}
	\begin{tabular}{rrrrrrrrr}
		\hline
		parameter&Posterior-Mean&$\hat{R}$&$\hat{V}$&$W$&$B/n$&$\sigma_+^2$&&\\
		\hline
		$\tau$&$200$&$1.10$&$112.06$&$110$&$2.16$&$112.06$&&\\
		$\lambda$&$0.023$&$1.00$&$9.84e^{-09}$&$9e^{-09}$&$1.41e^{-11}$&$9.84e^{-09}$&&\\
		\hline
	\end{tabular}
	\label{tab:4}
\end{table}
\section{Conclusion and Future Direction}\label{sec:5}
This paper presented a hybrid algorithm called Splitting Expectation Propagation for approximating hierarchical Bayesian models. Our algorithm focused on the problems of instability and intractability in EP especially at the refinement stage where it often fails. Also, this paper generalizes EP to handle hierarchical Bayesian models which poses a problem due to rigorous moment matching especially when the prior and likelihood family distributions differ. The modification was employed from stochastic and deterministic standpoints. \par We adopted the technique by \cite{john2011vi} which uses Monte Carlo integration for the intractable integration of the EP. SEP was used for the image reconstruction and it competed well with the MCMC in that SEP produced well approximated parameter estimates. EP-ADMM was used to handle the image reconstruction and EP-MCMC was used for parameter estimation. According to the results presented EP-ADMM produced very clear, sharp and white background. EP-MCMC achieved low variances when compared to the ordinary MCMC. Splitting EP in these viewpoints is the umbrella over both EP-MC, EP-ADMM, and EP-MCMC. Splitting EP was able to solve the problems of intractability and inflexibility of EP to hierarchical Bayesian models. Splitting EP can therefore be seen as an alternative EP for image analysis in high dimensional space. \par The main limitation of EP-MC is that it required large number of samples to fulfill the central limit theorem due to the Monte Carlo integration. As a result of the large samples, EP-MC might be slow to converge. Moreover, the limitation of the EP-MCMC differs from the ordinary MCMC in that the inner loop EP-ADMM that contributes to the EP-MCMC for the hyperparameter is efficiently fast. On the contrary, ordinary MCMC for the prior $\mathcal{X}$ will be slow resulting in slow convergence in total. As a future direction, we look forward to handling the hyper-parameters with Variational Inference. This method would make perfect sense as the approximate posterior from the prior $\mathcal{X}$ will be used as the auxiliary posterior distribution factorized over both the hyper-parameters $\lambda$ and $\tau$ which can be seen as a Coordinate Ascent Variational Inference (CAVI) \cite{davidetal2018}.  

\section*{Appendix}
\subsection*{Full derivation of SEP}
\subsubsection*{Incorporating the priors into $q$}
We first incorporated priors on $\tau$ and $\lambda$ as factors. These factors have the same functional form as 

\begin{equation}
	q(\mathcal{X}, \lambda, \tau) = \bigg[\mathlarger \prod_{i=1}^m \mathlarger \prod_{j=1}^n \mathcal{N}(\mathcal{X}_{ij} \big | mx_{i,j}, vx_{i,j}) \bigg] \exp(\lambda|\alpha_\lambda) \exp(\tau|\alpha_\tau)
	\label{equ:F.1}
\end{equation}
The first update rules $\alpha_\lambda^{new}$ and $\alpha_\tau^{new}$ for $q$ is obtained by SSEP method which will be discussed in the algorithm below. The rest of the factors are sequentially incorporated into Equation (\ref{equ:F.1}) which is also updated in a similar manner. One difficulty encountered when applying the update rules in \cite{minka2001ep} is that the normalizer does not have a closed form. This brings about a uniqueness in SSEP which treats $Z$ as follows;
$$Z = \int \mathcal{N}(\mathcal{L}\mathcal{X}_{ij}\hspace{0.05cm}| \hspace{0.05cm}0, \hspace{0.05cm}\tau)\hspace{0.05cm}q(\mathcal{X},\lambda, \tau)\hspace{0.05cm}d\mathcal{X} \hspace{0.05cm}d \lambda \hspace{0.05cm}d\tau$$
\begin{equation}
	= \int \mathcal{N}(\mathcal{L}\mathcal{X}_{ij}\hspace{0.05cm}|0, \hspace{0.05cm}\tau)\hspace{0.05cm}\exp(\tau\hspace{0.05cm}|\hspace{0.05cm}\alpha_\tau)\hspace{0.05cm}d\mathcal{X}\hspace{0.05cm}d\tau
	\label{equ:F.2}
\end{equation}
In Equation (\ref{equ:F.2}), the integral involves $\tau$. We adopted the method by \cite{john2011vi}. The normalizing factor $Z$ in Equation (\ref{equ:F.2}) is solved for $\tau$ as follows,
\begin{equation}
	Z_\tau = \frac{1}{K}\sum_{k=1}^K \mathcal{N}(\mathcal{L}\mathcal{X}_{ij}\hspace{0.05cm}| \hspace{0.05cm}0,\hspace{0.05cm}\tau^{(k)})\hspace{0.05cm}\exp(\tau^{(k)}\hspace{0.05cm}|\hspace{0.05cm}\alpha_\tau),
	\notag
\end{equation}
$\tau$ is generated from exponential distribution while $\mathcal{X}$ is initialized by $\mathcal{Y}$ to compute $\mathcal{LX}$ which is in general the goal of this work. So the Monte Carlo integration of Equation (\ref{equ:F.2}) with respect to $\mathcal{X}$ becomes 
\begin{equation}
	Z_x = \frac{1}{K}\sum_{k=1}^K \mathcal{N}(\mathcal{Y}|\mathcal{GX}_{ij},\hspace{0.05cm}\lambda)\hspace{0.05cm}\mathcal{N}(\mathcal{X}^{(k)},mx_{ij},vx_{ij})\hspace{0.05cm}
	\notag
\end{equation}
Here, $\mathcal{X}$ is generated from any distribution student-t, Normal distribution, or Uniform distribution from which $\mathcal{LX}$ is recomputed.
Here, we write the exact likelihood that needs to be processed multiple times i.e $N \times D$ times, where $N$ and $D$ represent the number of rows and columns respectively. 
\begin{equation}
	p(L\mathcal{X}|\tau) = \prod_{i = 1}^m\prod_{j=1}^n \mathcal{N}(L\mathcal{X}_{ij} \big | 0, \tau)\label{equ:6.51}
\end{equation} 
To approximate the posterior with SSEP algorithm, the posterior distribution can be factorized as follows
\begin{equation}
	q(\mathcal{X}, \lambda, \tau) = \bigg[\prod_{i=1}^m\prod_{j=1}^n \mathcal{N}(\mathcal{X}_{ij} \big | mx_{i,j}, vx_{i,j}) \bigg] \exp(\lambda|\alpha_\lambda) \exp(\tau|\alpha_\tau)
	\notag
\end{equation}
The exact likelihood term is a joint distribution and is denoted as 
\begin{equation}
	t_{i,j}(\tau) = \mathcal{N}(L\mathcal{X}_{ij} \big | 0, \tau) 
	\notag
\end{equation}
Then, we approximate the exact likelihood term by choosing approximating term 
\begin{equation}
	\tilde{t}_{i,j}(\tau) = \exp(\tau|\tilde{\alpha}_{i,j}) \notag
\end{equation}
We choose term to refine by removing $\tilde{t}_{i,j}$ from $q$ to compute the cavity distribution on $\mathcal{X}_{i,j}$ and $\tau$ 
\begin{equation}
	q_{-i,j}(\tau) = \exp(\tau|\alpha_{-i,j})
	\notag
\end{equation}
\begin{equation}
	\alpha_{-i,j} = \alpha_{i,j} - \tilde{\alpha}_{i,j} \notag
\end{equation}
After this, we perform the moment matching between $q(\tau)$ and a normalizing version of $t_{i,j}(\tau)q_{-i,j}(\tau)$. It is computed as follows
$$\argmin_q KL(\tilde{p}_{i,j} \big|\big| q_{i,j})$$ 
where
\begin{equation}
	\tilde{p}_{i,j}(\tau) = \frac{t_{i,j}(\tau)q_{-i,j}(\tau)}{\int t_{i,j}(\tau)q_{-i,j}(\tau) d\tau}
	\label{equ:6.57}
\end{equation}
Following Equation (\ref{equ:6.57}), the integral contains an intractable terms resulting from $\nabla_\omega \log{Z}$
where 
\begin{equation}
	Z = \int_{\tau}t_{i,j}(\tau)q_{-i,j}(\tau)d\tau \label{equ:6.58}
\end{equation}
Our goal is to make a stochastic approximation of this gradient.
\begin{equation}
	\nabla_\omega \log{Z} = \frac{1}{Z} \nabla_\omega \int_{\tau} t_{i,j}(\tau)q_{-i,j}(\tau)d\tau
	\label{equ:6.59}
\end{equation}
\begin{equation}
	\nabla_\omega \log{Z} = \frac{1}{Z}  \mathlarger \int_{\tau} t_{i,j}(\tau)\nabla_\omega q_{-i,j}(\tau)d\tau
	\label{equ:6.60}
\end{equation}
Equation (\ref{equ:6.60}) can be written as
\begin{equation}
	\nabla_\omega \log{Z} = \frac{1}{Z} \mathlarger \int_{\tau} t_{i,j}(\tau) q_{-i,j}(\tau) \nabla_\omega \log q_{-i,j}(\tau)d\tau
	\label{equ:6.61}
\end{equation}
referring to Equation (\ref{equ:6.57}) as the posterior, we have Equation (\ref{equ:6.61}) as an expectation
\begin{equation}
	\nabla_\omega \log{Z} = E_{\hat{p}(\tau)}\big[\nabla_\omega \log q_{-i,j}(\tau)\big]
	\label{equ:6.62}
\end{equation}
we use the identity $\nabla_\omega q_{-i,j}(\tau)=q_{-i,j}(\tau) \nabla_\omega \log q_{-i,j}(\tau)$. We can stochastically approximate the expectation using Monte Carlo integration as follows;  
\begin{equation}
	\nabla_\omega \log{Z} = \frac{1}{K}\mathlarger \sum_{k=1}^{K} \nabla_\omega \log q_{-i,j}(\tau)
	\label{equ:6.63}
\end{equation}
Now, we substitute for $q_{-i,j}(\tau)$
\begin{equation}
	\nabla_\alpha \log{Z} = \frac{1}{K}\mathlarger \sum_{k=1}^{K} \nabla_\alpha \log \big(\exp(\tau|\alpha_{-i,j})\big)
	\label{equ:6.64}
\end{equation}
we compute the updates as follows;
\begin{equation}
	\nabla_{\alpha_{-ij}} \log{Z} = \frac{1}{K}\mathlarger \sum_{k=1}^{K} \big(\tau^{(k)}\big)
	\label{equ:6.65}
\end{equation}
The update rule is given in equation (\ref{equ:F.20}) below;
\begin{equation}
	\alpha_\tau^{new} \approx \alpha_{-ij} + v_{\alpha_{-ij}}\zeta_{\alpha_{-ij}}
	\label{equ:F.20}
\end{equation}
Where the mean $\frac{1}{K}\sum_{k=1}^{K}\tau^{(k)}$ is denoted as $\zeta_{\alpha_{-ij}}$ and variance as $v_{\alpha_{-ij}}$. 
\subsubsection*{Incorporating the likelihood factors into $q$}
Now, the $N \times D$ factors are sequentially incorporated for the likelihood in Equation (\ref{equ:6.51}). However, approaching this with conventional update rule in EP is difficult to compute because it requires integration of each likelihood factor. After incorporating all the factors in Equation (\ref{equ:6.51}), SSEP sequentially incorporates the $N \times D$ factors for the likelihood Equation (\ref{equ:6.51}) provided below in Equation (\ref{equ:6.67})
$$Z_\lambda = \int \mathcal{N}(\mathcal{Y}\hspace{0.05cm}| \hspace{0.05cm}\mathcal{G}\mathcal{X}, \hspace{0.05cm}\lambda)\hspace{0.05cm}q(\mathcal{X},\lambda, \tau)\hspace{0.05cm}d\mathcal{X} \hspace{0.05cm}d \lambda \hspace{0.05cm}d\tau$$
\begin{equation}
	= \int \mathcal{N}(\mathcal{Y}\hspace{0.05cm}| \hspace{0.05cm}\mathcal{G}\mathcal{X},\hspace{0.05cm}\lambda)\hspace{0.05cm}\exp(\lambda\hspace{0.05cm}|\hspace{0.05cm}\alpha_\lambda) d \lambda.
	\label{equ:6.67}
\end{equation}
Then the Monte Carlo approximation for Equation (\ref{equ:6.67}) is 
\begin{equation}
	\frac{1}{K}\sum_{k=1}^K \mathcal{N}(\mathcal{Y}|\mathcal{G}\mathcal{X},\lambda^{(k)})\exp(\lambda^{(k)}|\alpha_\lambda).
	\label{equ:6.68}
\end{equation}
Now we compute the gradient of the normalizing factor as follows;
\begin{equation}
	\nabla_{\alpha_\lambda} \log{Z} = \frac{1}{Z}  \mathlarger \int_\lambda t_{i,j}(\lambda)\nabla_{\alpha_\lambda} q_{-i,j}(\lambda) d \lambda
	\notag
\end{equation}
\begin{equation}
	\nabla_{\alpha_\lambda} \log{Z} = \frac{1}{Z}  \mathlarger \int_\lambda t_{i,j}(\lambda)q_{-i,j}(\lambda) \nabla_{\alpha_\lambda} \log q_{-i,j}(\lambda) d \lambda
	\notag
\end{equation}
\begin{equation}
	\nabla_{\alpha_\lambda} \log{Z} = \frac{1}{Z}  \mathlarger \int_\lambda \mathcal{N}(\mathcal{Y}|\mathcal{G}\mathcal{X},\lambda)q_{-i,j}(\lambda)\nabla_{\alpha_\lambda} \log \exp(\lambda|\alpha_\lambda) d \lambda
	\label{equ:6.71}
\end{equation}
Equation (\ref{equ:6.71}) can be approximated as follows;
\begin{equation}
	E_{\hat{p}}[\lambda] = \mathlarger \sum_{k=1}^{K} \nabla_{\alpha_\lambda} \log \exp(\lambda|\alpha_\lambda) 
	\notag
\end{equation}
Similarly, for $\lambda$ 
\begin{equation}
	\nabla_\omega \log{Z} = \frac{1}{K}\mathlarger \sum_{k=1}^{K} \nabla_\omega \log \big(\exp(\lambda|\alpha^{\lambda}_{-i,j})\big)
	\notag
\end{equation}
\begin{equation}
	\nabla_{\alpha^{\lambda}_{-ij}} \log{Z} = \frac{1}{K}\mathlarger \sum_{k=1}^{K} \big(\lambda^{(k)}\big)
	\notag
\end{equation}
We denote the mean $\frac{1}{K}\sum_{k=1}^{K}\lambda^{(k)}$ as $\zeta_{\lambda_{-ij}}$ and variance as $v_{\alpha^{\lambda}_{-ij}}$. The update rule for $\lambda$ is given in Equation (\ref{equ:F.29}) below;
\begin{equation}
	\alpha_\lambda^{new} \approx \alpha^{\lambda}_{-ij} + v_{\alpha^{\lambda}_{-ij}}\frac{1}{K}\mathlarger \sum_{k=1}^{K}\lambda^{(k)}
	\label{equ:F.29}
\end{equation}
The cavity of the prior over $\mathcal{X}$ is
\begin{equation}
	q_{-ij}(\mathcal{X}) = \mathcal{N}(x_{ij};m_{-ij},v_{-ij})
	\notag
\end{equation}
We choose the approximate posterior for $\mathcal{X}$ as 
\begin{equation}
	q(\mathcal{X}) = \mathcal{N}(x_{ij};m_x,v_x) \notag
\end{equation}
Then the approximate term is chosen from the Gaussian family
\begin{equation}
	\tilde{t}_{ij}(\mathcal{X}) = \mathcal{N}(\mathcal{X}_{ij};\tilde{m}_{ij},\tilde{v}_{ij})
	\notag
\end{equation}
In order to update the cavity distribution we remove $\tilde{t}_{ij}$ from $q$ and update its parameters as follows;
\begin{equation}
	v_{-i,j}^{-1} = vx_{i,j}^{-1} - \tilde{v}_{i,j}^{-1} \notag
\end{equation}  
\begin{equation}
	m_{-i,j} = mx_{i,j} + v_{i,j}\tilde{v}_{i,j}(mx_{i,j} - \tilde{m}_{i,j}) 
	\notag
\end{equation}
Our goal is to make a stochastic approximation of this gradient.
\begin{equation}
	\nabla_\omega \log{Z} = \frac{1}{Z} \nabla_\omega \int_{\mathcal{X}_{i,j}, \tau} t_{i,j}(L\mathcal{X}_{i,j}, \tau)q_{-i,j}(\mathcal{X}_{i,j}, \tau) d \mathcal{X}_{i,j}d\tau
	\label{equ:6.81}
\end{equation}
Replace $q_{-i,j}$ in Equation (\ref{equ:6.81}) with $q_{i,j} / \tilde{t}_{i,j}$ then it becomes 
\begin{equation}
	\nabla_\omega \log{Z} = \frac{1}{Z} \int_{\mathcal{X}_{i,j}, \tau} \frac{t_{i,j}(L\mathcal{X}_{i,j}, \tau)}{\tilde{t}_{i,j}(\mathcal{X}_{i,j}, \tau)}\nabla_\omega {q_{i,j}(\mathcal{X}_{i,j}, \tau)} d \mathcal{X}_{i,j}d\tau
	\notag
\end{equation}
\begin{equation}
	\nabla_\omega \log{Z} = \frac{1}{Z} \int_{\mathcal{X}_{i,j}, \tau} \frac{t_{i,j}(L\mathcal{X}_{i,j}, \tau)}{\tilde{t}_{i,j}(\mathcal{X}_{i,j}, \tau)}q_{i,j}(\mathcal{X}_{i,j}, \tau)\nabla_\omega \log{q_{i,j}(\mathcal{X}_{i,j}, \tau)} d \mathcal{X}_{i,j}d\tau
	\notag
\end{equation}
At this juncture, we proceed with the use of Monte Carlo Integration
\begin{equation}
	\frac{\delta_1}{\delta_0} = \sum_{n = 1}^N \frac{t_{i,j}(L\mathcal{X}_{i,j}, \tau)}{\tilde{t}_{i,j}(\mathcal{X}_{i,j}, \tau)}\nabla_\omega \log{q_{i,j}(\mathcal{X}_{i,j}, \tau)} \bigg / \sum_{n = 1}^N \frac{t_{i,j}(L\mathcal{X}_{i,j}, \tau)}{\tilde{t}_{i,j}(\mathcal{X}_{i,j}, \tau)}
	\notag
\end{equation}
when taking $\mathcal{X}$ into consideration then we fix $\tau$ and vice versa. So we update the parameter by taking the gradient step
\begin{equation}
	\omega^{t+1} = \omega^t + lr * \frac{\delta_1}{\delta_0} 
	\label{equ:6.85}
\end{equation}
\subsection*{Derivative of EP-ADMM}
We introduce the Alternating direction method of multiplier (ADMM) algorithm to update the approximate posterior parameters.
\begin{align}
	\begin{split}
		\text{minimize} \hspace*{0.2in}\text{KL}(\tilde{p}(x_{ij}||q(x)_{ij}))
		\notag
	\end{split} \\
	&\ \text{subject to}: m_{ij} \geq a; v_{ij} \geq b \notag
\end{align}
where $a$ and $b$ are constants. Then updating according to \cite{minka2001ep} is as follows;
\begin{equation}
	m_x = \nabla_{m_{-ij}} \log Z_x + \alpha + \rho(m_{-ij} - a) 
	\notag
\end{equation}
and 
\begin{equation}
	v_x = \nabla_{v_{-ij}} \log Z_x + \beta + \rho(v_{-ij} - b) 
	\notag
\end{equation}
where 
\begin{equation}
	q(x) = \mathcal{N}(\mathcal{X};m_x,v_x) \notag
\end{equation}
then the tilted distribution is 
\begin{equation}
	\tilde{p}(x_{ij}) = \frac{t_{ij}(x)q_{-ij}(x)}{\mathlarger \int_{\mathcal{X}_{ij}}t_{ij}(x)q_{-ij}(x) dx_{ij}}
	\notag
\end{equation}
Now the normalizing factor is 
\begin{equation}
	Z_x = \mathlarger \int_{x_{ij}} \mathcal{N}(\mathcal{Y}_{ij};G\mathcal{X}_{ij},\lambda)\mathcal{N}(\mathcal{X}_{ij};m_{-ij},v_{-ij}) dx_{ij}
	\notag
\end{equation}
By integrating out $x$, $Z_x$ becomes
\begin{equation}
	\mathlarger\int \exp\biggl\{-\frac{1}{2}\Big(y_{ij}-gx_{ij}\Big)^2 \lambda^{-1} + \Big(x_{ij}-m_{-ij}\Big)^2 v_{-ij}^{-1}\biggr\} dx_{ij}
	\notag
\end{equation}
Completing the squares, we have 
\begin{equation}
	-\frac{1}{2}\Bigg[x_{ij} - \frac{gy_{ij}\lambda^{-1} + m_{-ij}v_{-ij}^{-1}}{g^2\lambda^{-1} + v_{-ij}^{-1}}\Bigg]^2\bigg(g^2\lambda^{-1} + v_{-ij}^{-1}\bigg) + \frac{1}{2}\frac{\bigg(gy_{ij}\lambda^{-1} + m_{-ij}v_{-ij}^{-1}\bigg)^2}{g^2\lambda^{-1} + v_{-ij}^{-1}}
	\notag
\end{equation}
Then the normalizing factor of $x$ is
\begin{equation}
	Z_x = \mathcal{N}(y_{ij};m_y, v_y) \notag
\end{equation}
where 
\begin{equation}
	m_y = \frac{g\lambda^{-1}m_{-ij}v_{-ij}^{-1}}{\lambda^{-1} - g^2\lambda^{-2}\bigg(g^2\lambda^{-1}+v_{-ij}^{-1}\bigg)^{-1}} = gm_{-ij}
	\notag
\end{equation}
and 
\begin{equation}
	\frac{1}{\lambda^{-1} - g^2\lambda^{-2}\bigg(g^2\lambda^{-1} + v_{-ij}^{-1}\bigg)^{-1}} = v_{-ij}g^2 + \lambda 
	\notag
\end{equation}
then updating the parameters $m_x$ and $v_x$ of $q(\mathcal{X})$ as follows according to \cite{minka2001ep} with a method of multipliers, we have 
\begin{equation}
	m^{new}_x = m_{-ij} + v_{-ij} \frac{y_{ij} - gm_{-ij}}{v_{-ij}g^2 + \lambda}g + \alpha + \rho(m_i - a) 
	\notag
\end{equation}
\begin{equation}
	v_x^{new} = \frac{v_{-ij}\lambda}{v_{-ij}g^2 + \lambda} + \beta + \rho(v_i - b) \notag
\end{equation}
\begin{equation}
	\alpha^{new} = \alpha^k + \rho(m_x^{new} - a) \hspace*{0.2in} \text{and} \hspace{0.2in} \beta^{new} = \beta^k + \rho(v_x^{new} - b)
	\label{equ:6.99}
\end{equation}

\bibliographystyle{apalike}

\bibliography{biblio}

\end{document}